\documentclass[journal]{IEEEtran}
\usepackage{caption}
\usepackage{amsmath,amssymb,amsfonts}
\usepackage{graphicx}
\usepackage{textcomp}
\usepackage[pscoord]{eso-pic}
\usepackage{xcolor}
\usepackage{subfigure}
\usepackage{diagbox}
\usepackage{booktabs}

\usepackage{multirow}
\usepackage{algorithm}  
\usepackage{algpseudocode}  
\usepackage{amsmath}  
\usepackage{ifpdf}
\usepackage{cite}
\ifCLASSINFOpdf
\else
\fi
\usepackage{amsmath}
\usepackage{array}

\usepackage{xcolor}
\usepackage{xpatch}
 
\makeatletter
\def\changeBibColor#1{%
  \in@{#1}{XHuang2016:CDICS,chatterjee2017:tccad,Sukharev2018;tdmr,Gunes2015:arXiv,DeepXDE2019,Jin2021:DATE,diss1994,lagaris2000:tnn,lagar,IE1998:NN,,HUANG2016:intergration,chew2014:iccad}%  list of colored bib items
  \ifin@\color{black}\else\normalcolor\fi
}

\xpatchcmd\@bibitem
  {\item}
  {\changeBibColor{#1}\item}
  {}{\fail}
 
\xpatchcmd\@lbibitem
  {\item}
  {\changeBibColor{#2}\item}
  {}{\fail}
\makeatother

\begin{document}
% \title{STPINN: A Space-Time Physics-informed Neural Network for Temperature-Aware Electromigration Reliability Analysis}
\title{A Space-Time Neural Network for Analysis of Stress Evolution under DC Current Stressing}  

\author{Tianshu~Hou,
        Ngai~Wong,
        Quan~Chen,
        Zhigang~Ji,
        and~Hai-Bao~Chen
        % <-this % stops a space
\thanks{This work is supported in part by the National Key Research and Development Program of China under grant 2019YFB2205005, and in part by the Nature Science Foundation of China (NSFC) under No. 62034007. Corresponding author: Hai-Bao Chen.}
\thanks{T. Hou, Z. Ji and H. -B. Chen are with the Department of Micro/Nano Electronics, Shanghai Jiao Tong University. N. Wong is with the Department of Electrical and
Electronic Engineering, The University of Hong Kong. Q. Chen is with the School of Microelectronics, Southern University of Science and Technology.}}

\maketitle

\begin{abstract}
The electromigration (EM)-induced reliability issues in very large scale integration (VLSI) circuits have attracted increased attention due to the continuous technology scaling. Traditional EM models often lead to overly pessimistic prediction incompatible with the shrinking design margin in future technology nodes. Motivated by the latest success of neural networks in solving differential equations in physical problems, {\color{black}we propose a novel mesh-free model to compute EM-induced stress evolution in VLSI circuits.} The model utilizes a specifically crafted space-time physics-informed neural network (STPINN) as the solver for EM analysis. By coupling the physics-based EM analysis with dynamic temperature incorporating Joule heating and via effect, we can observe stress evolution along multi-segment interconnect trees under constant, time-dependent and space-time-dependent temperature during the void nucleation phase. 
% In particular, STPINN obviates the iterative solver required in conventional stress evolution solution and offers significant computational savings. 
{\color{black}The proposed STPINN method obviates the time discretization and meshing required in conventional numerical stress evolution analysis and offers significant computational savings.}
Numerical comparison with competing schemes demonstrates a $2\times \sim 52\times$ speedup with a satisfactory accuracy. 
% We further show tests extracted on IBM power grid, which demonstrates the scalability of the proposed model.
\end{abstract}


% \fi
\IEEEpeerreviewmaketitle

\section{Introduction}
Electromigration (EM) has become a major concern for nanometer VLSI designs due to the shrinking feature sizes and increased current density of copper damascene wire interconnects, especially for power grids carrying large unidirectional currents~\cite{Warnock-5981968}. 
% The 2013 International Semiconductor Technology Roadmap (ITRS) predicts that the life of each generation of metal wires will be halved and affected by variations of interconnect dimensions~\cite{ITRS}, shown in Fig.~\ref{fig:tn}. 
{\color{black}Consequently, EM verification is important in chip design sign-off.}
% \begin{figure}[htb]
% 	\centerline{\includegraphics[width=0.9\columnwidth]{tn.pdf}}
% 	\caption{The red curve shows the lifetime vs. technology node, the black curve shows trend of reduced critical void volume and the green curve shows the needed EM enhancement~\cite{ITRS}.}
% 	\label{fig:tn}
% \end{figure}
%(\textcolor{blue}{ better to %define all symbols in one %sentence: The well-known %empirical Black's model %\cite{Black1969} estimates the %mean-time-to-failure (MTTF) of %single metal wire based on the %equation $MTTF=Aj^{-n}\exp\{ %E_a/k_BT\}$, where $j,\ T,\ %k_B,\ E_a,\ n,\ A$ are the %current density, temperature, %Boltzmann’s constant and EM %activation energy, current %density exponent and empirical %constant, respectively.})
The well-known empirical Black's model~\cite{Black1969} estimates the mean-time-to-failure (MTTF) of single metal wire based on the equation $MTTF=Aj^{-n}\exp\{ E_a/k_BT\}$, where $j,\ T,\ k_B,\ E_a,\ n,\ A$ are the current density, temperature, Boltzmann’s constant, EM activation energy, current density exponent and empirical constant, respectively. It should be noted that $n$ and $E_a$ have typical values for copper metallization by enough stress tests. According to highly accelerated tests using Joule heating, Black's equation is employed for extrapolation to operating conditions, while both temperature and current density in stress conditions cannot be varied independently~\cite{Jvon2002:IRWFR}. Besides, Black's model and Blech's effect (immortal segment filter) mainly target single wires, which may lead to significant errors and excessive design guard band in real designs~\cite{RL2010:mr,LONG2019,zhang2017:JAP}. In practice, VLSI circuits are composed of multiple interleaved interconnects, whose stress evolution is interactive among wires and must be considered altogether~\cite{Mishra2013:DAC}.

Physics-based models have been proposed in EM-induced stress evolution assessment in recent years~\cite{XHuang2016:CDICS,Chen2019:VLSI,Sukharev2016:TDMR,chatterjee2017:tccad,chatterjee2018:TCAD}. These EM models focus on hydrostatic stress diffusion kinetics in continuously connected confined metal wires governed by coupled partial differential equations (PDEs), and on the accurate failure time estimation through critical stress. {\color{black}However, the mesh-based numerical methods such as the finite difference method (FDM) and the finite element method (FEM) are constrained by the discretization of space and time, which increase the number of variables to solve PDEs resulting from complex on-chip interconnect topologies~\cite{Wang2021}}. A voltage-based EM modeling and immortality check technique for general interconnect trees was proposed in~\cite{Sun2016:ACM}. In this work, if the largest stress of tree nodes is less than the critical stress, the tree is considered to be immortal. Although it is helpful for fast EM immortality check, the predicted failure time of wires can not be obtained.
In~\cite{XHuang2014:DAC} and~\cite{XHuang2016:CDICS}, compact models were developed to extend stress diffusion on the single metal wire to multi-segment interconnect structure through known steady-state stress distribution based on Korhonen's equation~\cite{Korhonen1993}. In~\cite{chen2016:CDICS}, Laplace transformation technology was applied with a complementary error basis function to find analytical solutions of stress evolution in multibranch interconnect trees during void nucleation. In~\cite{Chen2019:VLSI}, based on the accelerated separation of variables (ASOV), an analytical model was proposed to describe stress evolution with Gaussian elimination (GE), yet it can only deal with stress evolution under constant temperature.

Moreover, deep learning has demonstrated the ability to explore the underlying correlation of massive data in computational modeling of physical systems~\cite{cr2020:neur,justin2018:jcp,diss1994,lagaris2000:tnn}. %(\textcolor{blue}{what does it %mean? : In~\cite{IE1998:NN}, the %trial function was configured to %satisfy boundary conditions %(BCs) and initial conditions %(ICs) and employed a %one-hidden-layer neural network %as the nonlinear operation %component of trial function, %expressing differential, closed %and mesh-free solutions %describing PDE solution %approximations}), which offers %new insights for mathematical %equation solving. 
In~\cite{IE1998:NN}, the trial function is configured for satisfying boundary conditions (BCs) and initial conditions (ICs), and a single-hidden-layer neural network is employed as the nonlinear operation component of the trial function. This way offers a new insight for obtaining approximate solutions to PDEs. In~\cite{Lars2018:CoRR}, structural similarity between residual neural networks (ResNet) and PDE was observed to establish a new PDE-interpretation of convolutional neural networks (CNNs). Research in~\cite{ma2018:jcp} introduced hidden physics models (data-efficient learning machines) to extract patterns from high-dimensional data governed by time-dependent and nonlinear PDEs. In~\cite{ZANG2020:jcp}, weak adversarial network (WAN) was proposed for high-dimensional PDEs by leveraging their weak formulations. The primal and adversarial networks are set to minimize the converted objective function. Furthermore, multi-fidelity physics-informed neural network (MPINN) provides a composite structure to exploit the relationship between high-fidelity data and low-fidelity data of PDEs to obtain the PDE solutions~\cite{MENG2020:jcp}. 
{\color{black} Inspired by these works, a physics-constrained deep learning scheme based solver was proposed for analyzing electric potential and electric fields in \cite{Jin2021:DATE}, which provides new insight for applying deep learning in EM assessment.}
{\color{black}In this paper, we propose a space-time conversion based multi-channel network motivated by the physics-informed neural network (PINN)~\cite{PINN2019:Journal}, which encodes PDEs respecting given laws of physics into the neural networks for learning tasks.
The proposed network, called space-time physics-informed neural network (STPINN), aims at calculating stress evolution in multi-segment interconnect trees during the void nucleation phase under complex temperature conditions.} It should be noted that the space-time in the proposed STPINN method means that the diffusivities depend on both location and time. 
{\color{black}The proposed STPINN method is a mesh-free approach that can obtain the continuous and differentiable stress solution without a mesh generation since STPINN applies automatic differentiation (AD) \cite{Gunes2015:arXiv} for all the derivative operations to enforce the strong form of the stress evolution equation \cite{DeepXDE2019}. Mesh is needed to be created in numerical methods such as FEM and FDM for the discretization in the space domain.}
The main contributions of this paper are:
\begin{itemize}
  \item [$\bullet$] 
{\color{black}We propose the STPINN architecture for the first time for solving the mesh-free stress evolution problem in EM-based reliability analysis without requiring a prior knowledge of the relevant datasets.}
  \item [$\bullet$] 
We extend the proposed STPINN method from analyzing EM-based reliability of single wire to interconnects with multi-segments. 
{\color{black}STPINN can be used to solve the equations describing stress evolution on multi-segment interconnect trees under space-time related diffusivity.}

% {The new method can be used to efficiently process the dynamic stress flux problem with the space-time related diffusivity embedded in the stress evolution equation.}   
  \item [$\bullet$]
We consider the effects of the dynamic temperature incorporating Joule heating and the via effect in the proposed STPINN for calculating the EM-induced stress evolution during the void nucleation phase.
  \item [$\bullet$]
We compare the proposed method against the FEM and the compact analytical model. The proposed model is $2\times \sim 52\times$ faster than existing methods, with mean relative errors $<1.22\%$ vs FEM and $<0.44\%$ vs the analytical model. 
\end{itemize}

% We show that the proposed STPINN can solve PDEs with more challenging configurations: i) irregular equation parameters such as tiny diffusion coefficient and extensive temporal range; ii) discontinuous flux due to different current densities in neighboring segments; iii) space-time related diffusivity due to temperature distribution. 
% We show that the new method analyze stress development under three different temperature conditions: constant temperature, dynamic temperature and dynamic temperature incorporating Joule heating and via effect to illustrate the impact of temperature condition on EM-induced stress evolution.

% We compare the proposed method against FEM-based COMSOL\cite{Comsol} and compact analytical solution model\cite{Chen2017:TDMR}. The proposed model is $2\times \sim 52\times$ faster than existing methods, with mean relative errors $<1.22\%$ vs FEM and $<0.44\%$ vs the analytical model.    

The rest of the paper is organized as follows. {\color{black}Section~\ref{2} describes the physics-based Korhonen's equation for EM modeling and the thermal model considering Joule heat, via effect and time-varying temperature.} In Section~\ref{3}, we introduce the STPINN and the corresponding preprocessing procedure. Experiments are presented in Section~\ref{4} for the configured multi-segment interconnect trees in analyzing stress evolution under different thermal models, followed by conclusion in Section~\ref{5}.
% , we extract n-segment interconnects from the power grid benchmarks and employ the proposed method in analyzing stress evolution under dynamic temperature condition. 

\section{BACKGROUND OF PHYSICS-BASED EM MODELING}\label{2}
The physics-based stress evolution model governed by Korhonen's equation demonstrates its reliability in stress evolution estimation compared to empirical methods. The space-time related temperature model shows the temperature distribution through the heat conduction modeling. {\color{black} In this section, we will introduce the general EM modeling and 
the thermal model used in EM analysis.}
% the Joule heat effects on reliability in metal interconnects. Also, we will discuss the non-uniform distribution of temperature caused by heat conduction and the effects on EM analysis.}

%(Black' equation and Blech effect).
%{\color{black}In this section, we will introduce the general EM modeling and the Joule-heat conduction incorporating via effect model in metal interconnects. The physics-based stress evolution model governed by Korhonen's equation demonstrates its reliability in stress evolution estimation compared to experience-based methods. 
%(Black' equation and Blech effect).

\subsection{Stress Evolution Model}\label{stress evo sec}
\begin{figure}[t]
	\centerline{\includegraphics[width=0.8\columnwidth]{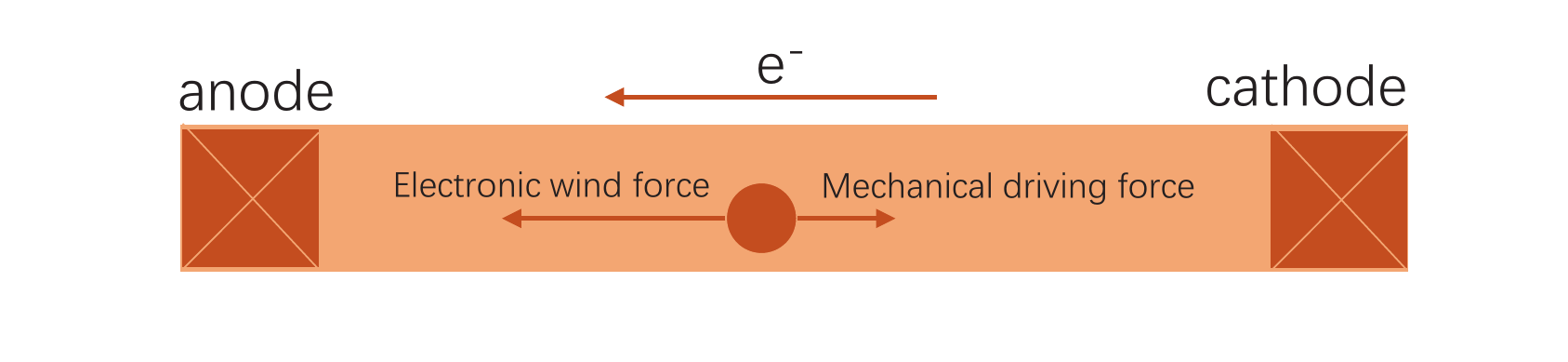}}
	\caption{Atomic forces on single segment wire due to high-density current.}
	\label{fig:em}
\end{figure}

% EM is the physical migration of metal atoms in the direction of electrical field with the momentum exchange, 
{\color{black}EM phenomenon results from a combined action of the momentum exchange with the conducting electrons and the electrostatic force.} 
The electrons and metal atoms collide within the high-density current. During collision, the metal atoms are driven by the electronic wind force and the opposite mechanical driving force, shown in Fig.~\ref{fig:em}.
{\color{black}As a result, when the metal line is embedded into the rigid confinement, the atoms are subject to tensile stress at the cathode and compressive stress at the anode, causing depletion and accumulation of volume changes, respectively.} Void nucleation can be determined when tensile stress exceeds the critical stress, called as $\sigma_{crit}$. It is worth mentioning that void nucleation will first occur at the cathode node due to the atomic migration. 

Physics-based model of Korhonen $\emph{et al.}$~\cite{Korhonen1993} governs hydrostatic stress evolution during void nucleation phase with PDEs. Let $\sigma(x,t)$ denote the stress evolution at location $x$ and time $t$. {\color{black} The stress evolution along a single metal wire in one-dimension takes the form:}  
\begin{equation}\label{eq:Korhonen's PDE}
\begin{aligned}
&{\rm PDE}:\frac{\partial\sigma(x,t)}{\partial t}=\frac{\partial}{\partial x}\Big[\kappa\Big(\frac{\partial \sigma(x,t)}{\partial x}+G\Big)\Big],x\in{\color{black}\mathcal{L}},\\
&{\rm BC}:\kappa\Big(\frac{\partial \sigma(x,t)}{\partial x}+G\Big)=0,x\in{\color{black}\mathcal{B}},\\
&{\rm IC}:\sigma(x,0)=\sigma_T,x\in {\color{black}\mathcal{L}},\\
\end{aligned}
\end{equation}
where $t$ satisfies $t\in(0,+\infty)$. {\color{black}The notations $\mathcal{L}$ and $\mathcal{B}$ represent the set of points within the metal wire and the collection of points at the blocked terminals.} The notations $\kappa=D_aB\Omega/(kT)$ and $B$ are stress diffusivity and effective bulk related to line geometry, especially width, aspect and grain morphology~\cite{Hau:2000MRS}. Whereas $k$ and $T$ are the Boltzmann's constant and the absolute temperature, respectively. The effective atomic diffusion coefficient is expressed as:
\begin{equation}\label{eq:da}
    D_a=D_0\exp\Big(-\frac{E_a}{kT}\Big).
\end{equation}
{\color{black}Here, $D_0$ is the self-diffusion coefficient and $G=|Z^*|e\rho j/\Omega$ is the EM driving force}. The atomic lattice 
volume, current density, metal resistivity, the effective charge number and activation energy are denoted by $\Omega$, $j$, $\rho$, $Z^*$ and $E_a$, respectively. {\color{black}The notations and their values in this paper are summarized in Table~\ref{tab:parameters}.}

\begin{table}[t]
	\centering
	\caption{{\color{black}Description and typical value of parameters in the experiments.}}
	\setlength{\tabcolsep}{2.5mm}{
		\begin{tabular}{|c|c|c|}
			% 		\toprule
			\hline
			\textbf{Parameter} & \textbf{Value} & \textbf{Description} \\ \hline
			\textbf{$k$} & $1.38\times 10^{-23}J/K$ & Boltzmann constant \\ \hline
			\textbf{$e$} & $1.6\times10^{-19}C$ & Electric charge \\ \hline
			\textbf{$Z^*$} & $10$ & Effective valence charge \\ \hline
			\textbf{$E_a$} & $1.1eV$ & Activation energy \\ \hline
			\textbf{$B$} & $1\times10^{11}Pa$ & Effective bulk \\ \hline
			\textbf{$D_0$} & $5.2\times10^{-5}m^2/s$ & Self-diffusion coefficient \\ \hline
			\textbf{$\rho$} & $3\times10^{-8}\Omega\cdot m$ & Resistivity of Cu \\ \hline
			\textbf{$\Omega$} & $8.78\times10^{-30}m^3$ & Atomic volume \\ \hline
			\textbf{$\sigma_{crit}$} & $4\times10^{8}Pa$ & Critical stress \\ \hline
			% 		\bottomrule
	\end{tabular}}
	\label{tab:parameters}
\end{table}

In \eqref{eq:Korhonen's PDE}, the stress evolution $\sigma(x,t)$ is simultaneously described by Korhonen's PDEs, zero-flux BCs and ICs. Here, $\sigma_T$ is the pre-existing stress along wire. In our work, we assume $\sigma_T$ to be zero. During void nucleation phase, atomic flux is blocked at boundary of wire geometry in corresponding dimensions and the atomic flux~\cite{Korhonen1993} is given by:
\begin{equation}\label{eq:atomic flux}
J(x,t)=\frac{D_aC\Omega}{ kT}\Big(\frac{\partial \sigma(x,t)}{\partial x}+G\Big),
\end{equation}
where $C$ represents the number of metal atoms per unit volume.
In multi-segment topology, the value of stress and atomic flux should be continuous at the interior junction nodes of adjacent segments. For an interconnect tree with multi segments, Korhonen's equations can be extended to the following form to describe the stress evolution $\sigma_i(x,t)$ along the $i$-th segment:
% \begin{equation}\label{eq:extendedKorhonen's PDEs}\small
% \begin{aligned}
% {\rm PDE}:
% &\frac{\partial\sigma_1(x,t)}{\partial t}=\frac{\partial}{\partial     x}\Big[\kappa_1\Big(\frac{\partial\sigma_1(x,t)}{\partial     x}+G_1\Big)\Big],x\in{\color{black}\mathcal{L}_1},\\
% {\rm PDE}:
% &\frac{\partial\sigma_2(x,t)}{\partial t}=\frac{\partial}{\partial     x}\Big[\kappa_2\Big(\frac{\partial\sigma_2(x,t)}{\partial    x}+G_2\Big)\Big],x\in{\color{black}\mathcal{L}_2},\\
% {\rm BC:}
% &\kappa_1\Big(\frac{\partial \sigma_1(x,t)}{\partial   x}+G_1\Big)=0,x\in{\color{black}\mathcal{B}},\\
% {\rm BC:}
% &\kappa_2\Big(\frac{\partial \sigma_2(x,t)}{\partial    x}+G_2\Big)=0,x\in{\color{black}\mathcal{B}},\\
% {\rm {\color{black}BC}}:
% &\kappa_1\Big(\frac{\partial \sigma_1(x,t)}{\partial    x}+G_1\Big)=\kappa_2\Big(\frac{\partial     \sigma_2(x,t)}{\partial x}+G_2\Big),x\in{\color{black}\mathcal{I}},\\
% {\rm {\color{black}BC}}:
% &\sigma_1(x,t)=\sigma_2(x,t),x\in{\color{black}\mathcal{I}},\\
% {\rm IC}:
% &\sigma_1(x,0)=0,x\in{\color{black}\mathcal{L}_1},\\
% {\rm IC}:
% &\sigma_2(x,0)=0,x\in{\color{black}\mathcal{L}_2}.\\
% \end{aligned}
% \end{equation} 
{\color{black}\begin{equation}\label{eq:extendedKorhonen's PDEs}
\begin{aligned}
{\rm PDE}:
&\frac{\partial\sigma_i(x,t)}{\partial t}=\frac{\partial}{\partial     x}\Big[\kappa_i\Big(\frac{\partial\sigma_i(x,t)}{\partial    x}+G_i\Big)\Big],x\in \mathcal{L}_i,\\
{\rm BC}:
&\kappa_i\Big(\frac{\partial \sigma_i(x,t)}{\partial    x}+G_i\Big)=0,x\in\mathcal{B},\\
{\rm BC}:
&\sigma_{i_1}(x,t)=\cdots =\sigma_{i_p}(x,t),x\in\mathcal{I},\\
{\rm BC}:
&\sum_{j\in \mathcal{N}_i}\kappa_j\Big(\frac{\partial\sigma_j(x,t)}{\partial x}+G_j\Big)\cdot n_j=0,x\in\mathcal{I},\\
{\rm IC}:
&\sigma_i(x,0)=0,x\in{\mathcal{L}_i}.\\
\end{aligned}
\end{equation}
% where $t$ satisfies $t\in(0,+\infty)$. 
% while $\partial\Omega_j$ and $\partial\Omega$ are junction area and boundary of interconnect trees, respectively. $\Omega_i$ denotes the entire geometric space except $\partial\Omega_j$ and $\partial\Omega$ on the $i$-th segment. 
We employ $\mathcal{I}$ to describe the collection of coordinates of interior junction nodes. The collection of adjacent segments of each interior junction node on the $i$-th segment is defined as $\mathcal{N}_i=\{i_1,\cdots,i_p\}$. The notation $n_j$ represents the unit inward normal direction of the interior junction node on the neighbouring $j$-th segment, of which the value ($1$ or $-1$) is determined by the direction of electron propagation~\cite{chen2016:CDICS}.} The BCs at both interior junction nodes of each segment should be satisfied. It is remarked that the impacts of constant, time-varying, and space-time dependent temperatures are quantified with $\kappa_i$, leading to three different types of PDEs for the stress evolution model.

%We employ $\mathcal{I}$ to describe the collection of coordinates of interior junction nodes. The collection of neighbour segments of each interior junction node on the $i$-th segment is defined as $\mathcal{N}_i=\{i_1,\cdots,i_p\}$. The notation $n_j$ represents the unit inward normal direction of the interior junction node on the neighbouring $j$-th segment, of which value is $+1$ for left and lower direction and $-1$ for right and upper direction. BCs at both interior junction nodes of each segment should be satisfied.}
%It should be noted that the impacts of constant, time-varying, and space-time dependent temperature are quantified with $\kappa_i$, which derives the stress evolution model to three different types of PDEs.

\subsection{Space-time Related Temperature Model}\label{condition}
% \begin{figure}[t]
% 	\centerline{\includegraphics[width=0.9\columnwidth]{temttf.pdf}}
% 	\caption{TTF distribution of interconnects stressed at $250^\circ C$–$350^\circ C$~\cite{zhang2017:JAP}. }
% 	\label{fig:temttf}
% \end{figure}

\begin{figure}[t]
	\centerline{\includegraphics[width=0.7\columnwidth]{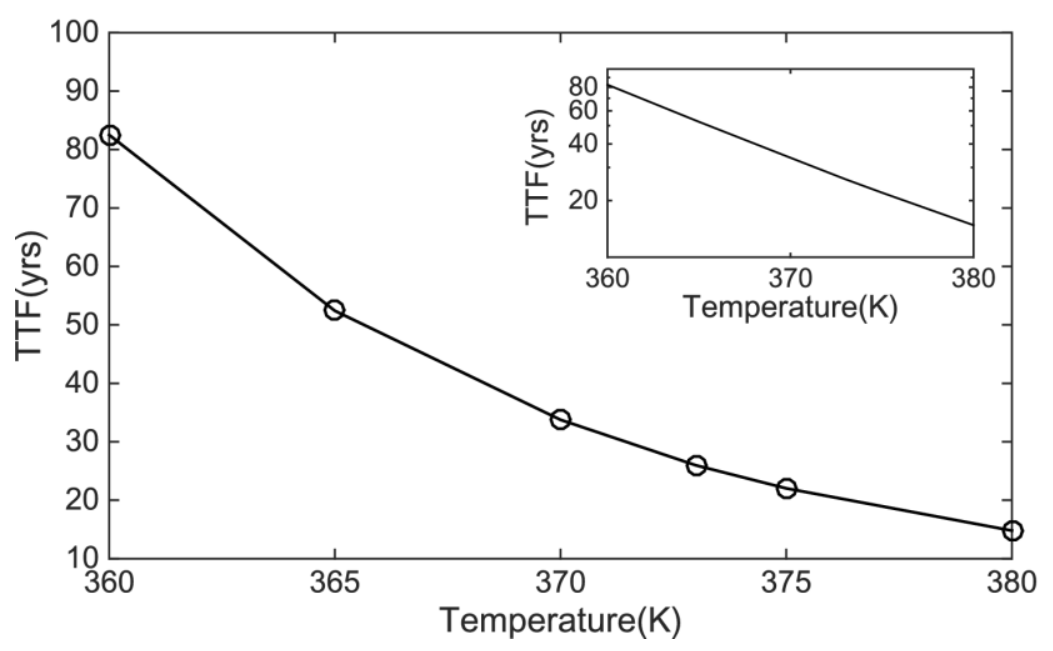}}
	\caption{{\color{black}Effect of temperature on TTF~\cite{XHuang2016:CDICS}.} }
	\label{fig:temttf}
\end{figure}

In EM reliability analysis, temperature is one critical factor affecting the EM failure process. {\color{black}The temperature variation was taken into account by \cite{Sukharev2018;tdmr,HUANG2016:intergration,chew2014:iccad} in EM analysis.} {\color{black}In Fig.~\ref{fig:temttf}, the simulation \cite{XHuang2016:CDICS} reveals that the temperature rising will lead to an early failure of the full-chip.} 
{\color{black}The temperature distribution of the interconnect trees depends on the gate switching activity and the quality of heat dissipation. The total thermal power dissipated can be represented as a sum of the average power dissipated by Joule heating of the metal branches and the average heating dissipated by the underlying logic due to active switching and leakage currents \cite{Sukharev2018;tdmr}.}
% In addition to the relationship between temperature and time, the high-density current flow within the interconnect metal wire will generate Joule heating, which affects the temperature distribution on the wire. This implies the temperature is not the same at different locations.
% which is often ignored in EM reliability analysis. 

{\color{black}In this section, we review the Joule heat model \cite{thermal} and combine the model with time-varying temperature to customize the space-time related temperature model for EM analysis. In the metal wire with high-density current, the non-uniform temperature rising is generated due to the Joule heat effect.}
%(\textcolor{blue}{ what does it %mean?: During Joule-heat %conduction procedure on metal %wire, higher temperature %estimation of interconnects will %be obtained due to the %negligence of efficient heat %dissipation paths}). 
{\color{black}It should be noted that during Joule heat conduction, the estimated temperature will be higher than the practical temperature if the heat dissipation paths are ignored.
The impact of low-k dielectrics such as ${\rm SiO_2}$ on the interconnect temperature was reported in \cite{lowk2000}.
In copper processes, vias serve as more efficient heat dissipation paths since they have much better thermal conductivity than the low-k dielectrics \cite{lowk1996}. Thus, the efficient heat dissipation of vias, called via effect, should be considered in temperature estimation during Joule heat conduction.
% The influence of the via effect is more critical because low-k dielectrics such as ${\rm SiO_2}$ have lower thermal conductivity than vias. 
% For a metal wire, two terminals are connected to the underlying layer through vias with negligible heat loss. 
Given the length of metal wire, the space-time related temperature distribution considering via effect follows \cite{thermal} as:
\begin{equation}\small
\label{eq:thermal}
T(x,t)=T_0(t)+\frac{j^2_{r\!m\!s}\rho L_H^2}{k_M}\Big(1-\frac{\cosh(\frac{x}{L_H})}{\cosh(\frac{L}{2L_H})}\Big),x\in[-\frac{L}{2},\frac{L}{2}],
\end{equation}
where $L,\ \rho,\ k_M,\ j_{r\!m\!s}$ are the length, the resistivity, the thermal conductivity and the uniform root-mean-square current density of the metal wire, respectively. 
The time-varying model $T_0(t)$ represents the temperature on the underlying layer and characterizes the impacts of the gate switching activity and the quality of heat dissipation.
The notation $L_H$ denotes the healing length \cite{thermal} and it can be further explained in detail as follows:
\begin{equation}\label{eq:s}
\begin{aligned}
&L_H=\left[\frac{k_MH_{I\!L\!D}}{k_{I\!L\!D}}\left(\frac{1}{s}\right)\right]^{\frac{1}{2}},\\
&s=\Big(\frac{w}{t_{I\!L\!D}}\Big[\frac{1}{2}\ln\Big(\frac{w+d}{w}\Big)+\frac{t_{I\!L\!D}-\frac{d}{2}}{w+d}\Big]\Big)^{-1}.
\end{aligned}
\end{equation} 
% In a metal wire, it is supposed {\color{black}\cite{thermal}} that heat only flows downwards the silicon substrate attached to the heat sink, which is governed by 
% \begin{equation}\label{eq:heat equation}
% \frac{d^2T}{dx^2}-\frac{T-T_0(t)}{L^2_H}=-\frac{\rho j_{rms}^2}{k_M},
% L_H=\Big[\frac{k_MHt_{I\!L\!D}}{k_{I\!L\!D}}\Big(\frac{1}{s}\Big)\Big]^{\frac{1}{2}},
% \end{equation}
Here, $s$ is the heat spreading factor, and the notations $d$ and $w$ are the spacing and width of the metal lines. The notations $t_{I\!L\!D},\ k_{I\!L\!D}$ are the thickness and the conductivity of Inter Layer Dielectric (ILD), which separates the metal wire and underlying layer. The temperature on the multi-segment interconnect tree is assumed to be a combination of the non-uniform distributed temperatures on each segment calculated by \eqref{eq:thermal}.
For the node connecting to more than one segments, we suppose that there is negligible heat loss between the node and the underlying layer \cite{thermal}. Based on \eqref{eq:thermal}, the temperature of the node over time can be expressed as:
\begin{equation}
T(\pm \frac{L}{2},t)=T_0(t).
\end{equation}
}
In this work, we consider three temperature scenarios: I) constant temperature over space and time; II) time-varying temperature; III) space-time related temperature considering Joule heat spreading, via effect and time-varying temperature on the underlying layer. Combining with the EM modeling, the equation \eqref{eq:da} shows the relationship between the effective atomic diffusion coefficient $D_a$ and the temperature $T$, which induces the correlation of the diffusivity $\kappa$ and the space-time variables.

\section{STPINN-BASED EM RELIABILITY ANALYSIS }\label{3}
In this section, we first modify the PINN structure in EM analysis under constant thermal condition, and then extrapolate the PINN method to a novel STPINN-based EM reliability analysis model for accurate mesh-free stress evolution on multi-segment interconnects during void nucleation phase under different temperature conditions.
\begin{figure*}[t]
	\centerline{\includegraphics[width=1.5\columnwidth]{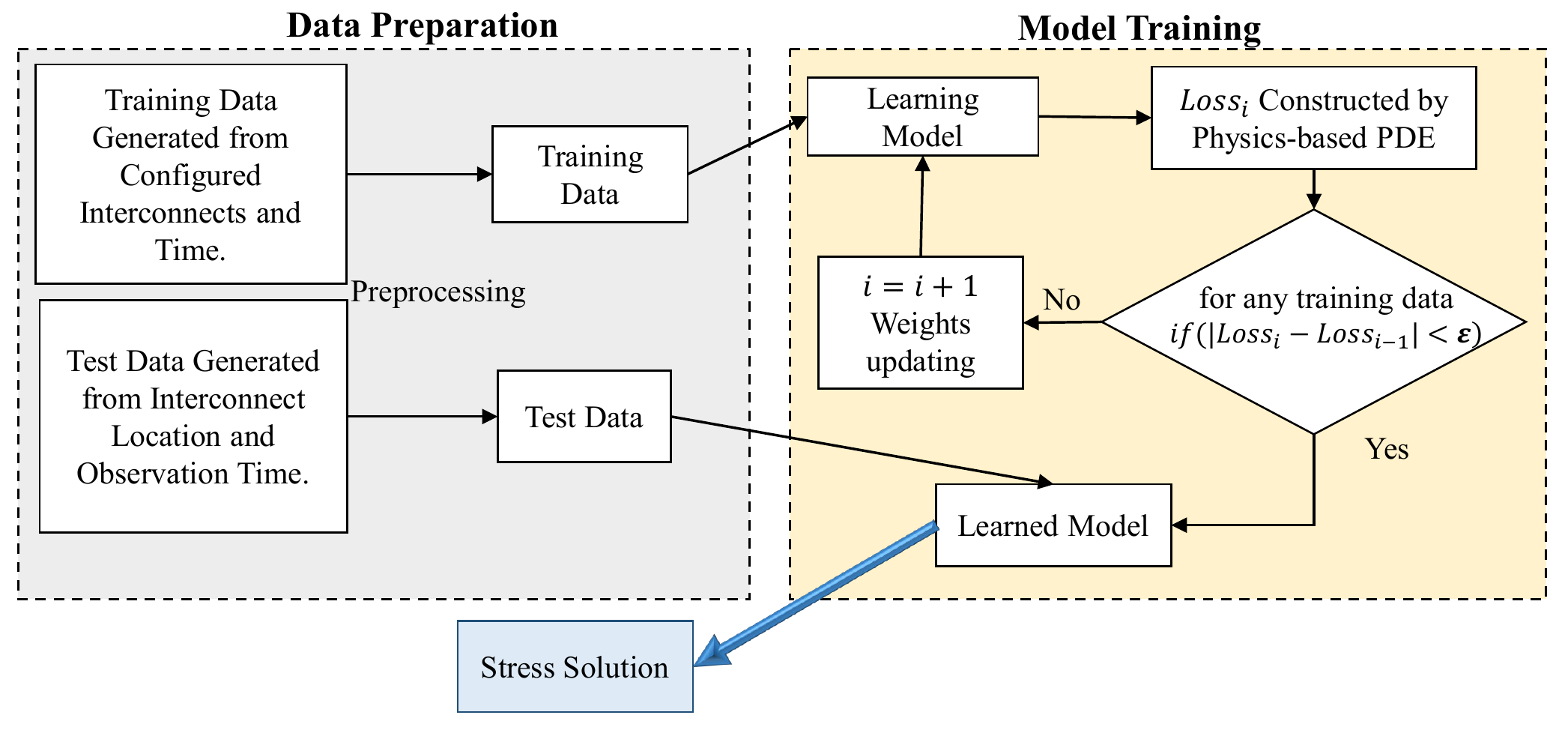}}
	\caption{{\color{black}The flowchart of learning based EM reliability analysis model.}}
	\label{fig:flowchart}
\end{figure*}
Fig.~\ref{fig:flowchart} shows an overview flowchart of the learning-based EM model which consists of data preparation, model training and stress solution. {\color{black}The training data is sampled from the configured geometry and time range, and then preprocessed to regular parameters for 
% the unsupervised 
learning}. During the training phase, the network weights are continually updated until the objective function converges to the minimum. The learned model first obtains normalized stress by the needed location and time (test data), then restores the normalized stress through scaling to achieve the EM stress evolution along the configured interconnect tree.  
More details in the flowchart will be discussed.

\subsection{A Modified Physics-informed Neural Network}\label{PINN}
Physics-informed Neural Networks (PINNs) have demonstrated the strong expressive powers in solving forward and inverse PDE problems through their nonlinear learning capabilities. Application in learning velocity and pressure fields by partially observing space-time visualizations of a passive scalar~\cite{raissi2018} relies on both data and physics-based information described by PDEs~\cite{PPINN2019:arxiv}. We now enhance and customize the PINN method for solving the interconnect reliability problem.

To fit into EM-based reliability problem, we first transform Korhonen's stress evolution equation~\eqref{eq:Korhonen's PDE} to the following form:
%for stress ${\sigma}(x,t)$ to the %following form:}
%We first extend PINN in analyzing mesh-free EM-induced stress evolution through PDEs governed by extended Korhonen's equations with extremely low diffusion, large parameters and truncated convection term at junctions. 
\begin{equation}\label{eq:parabolic}
{\sigma}_t+\mathcal{N}[{\sigma}]=0,
\end{equation}
{\color{black}where $\mathcal{N}[\sigma]=\partial[\kappa(\partial\sigma/\partial x+G)]/\partial x$ and ${\sigma}_{t}=\partial\sigma/\partial t$ represent the nonlinear differential operator and the first derivative of ${\sigma}$ with respect to temporal variable $t$, respectively.} We also define the diffusion operator $f$ which is employed to substitute the left side of \eqref{eq:parabolic} 
%to satisfy the equation condition, 
and $f$ is expected to be zero. 
% It should be noted that all the derivative operations are applied by automatic differentiation through chain rule for mesh-free solutions instead of iterative solution on a meshed grid.

\begin{figure}[t]
	\centerline{\includegraphics[width=0.8\columnwidth]{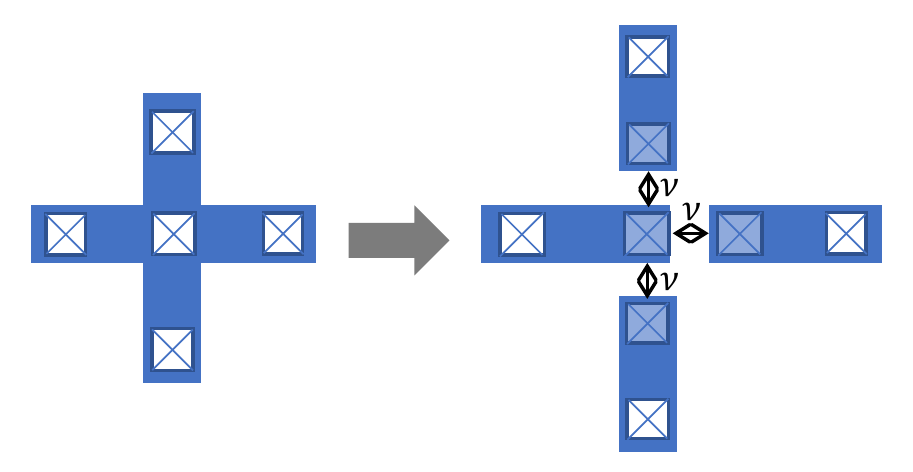}}
	\caption{{\color{black}Schematic diagram of virtual distance for calculating the derivatives of stress evolution with respect to the location at the interior junction node. The blue nodes (right) are overlapping at the interior junction (left).} }
	\label{fig:vd}
\end{figure}

% {\color{black}It should be noted that all the derivative operations are applied by automatic differentiation through chain rule, and thus PINNs are mesh-free \cite{DeepXDE2019}.}
{\color{black}It has been shown in \cite{HORNIK1989:NN} that Feed-forward Neural Network (FNN) is a useful tool for function approximation with the practical benefits of modeling highly nonlinear functions. The structure of FNN is intuitive and straightforward to understand and manipulate.
Thus, we use the FNN to approximate the solution ${\sigma}$, and take spatial-temporal information $(x,t)$ as inputs.}
% Thus, we use the FNN $\mathcal{F}(x,t;\alpha)$ with weights $\alpha$ to approximate the solution ${\sigma}$, and take spatial-temporal information $(x,t)$ as inputs.} 
% BCs, ICs and PDEs correct the network by updating weights $\alpha$ to minimize the objective function. 
% It is worth mentioning that the effective atomic diffusion coefficient is a piece-wise function over space in EM analysis so that $\mathcal{N}[\sigma]$ becomes a discontinuous function. Therefore, the stress solution $\sigma$ is hard to be approximated by the multi-order differentiable activation function. 
{\color{black}It is worth mentioning that the EM driving force is not the same in different segments due to various current densities within each segment. According to 
the boundary conditions \eqref{eq:extendedKorhonen's PDEs} corresponding to the interior junction node constraints, derivatives of stress evolution with respect to the location at the interior junction node in different directions are different.} Therefore, the stress solution $\sigma$ of the whole interconnect tree is non-differentiable at the interior junction nodes and is hard to be approximated by the multi-order differentiable activation function.
%To mitigate the problem, 
% In order to overcome this problem, we introduce virtual distance $\nu$ in the spatial domain as an input and set solution-flux continuity constraint at the boundary of virtual distance for smoothing the discontinuous stress and flux at junctions in interconnect trees. 
{\color{black}In order to overcome this problem, we insert a virtual distance $\nu$ at the intersection of adjacent segments and set the boundary conditions
corresponding to the interior junction node constraints at the terminals of the virtual distance, which are the overlapping points at the intersection node of the interconnect structure, shown in Fig.~\ref{fig:vd}.} {\color{black} 
It should be noted that derivatives of the multi-order differentiable output of PINN in different directions with respect to location need to be equal at any time. However, this is incompatible with the interior junction node constraints so that we cannot directly construct the objective function without introducing a virtual distance at the interior junction node. The virtual distance smooths the stress evolution and helps to construct the objective function for network training in our proposed STPINN method.
% we employ AD for calculating the spatial gradient of the output of PINN and each output can only get one spatial gradient at the corresponding location. Specifically, as the PINN takes locations and time instances as inputs, the spatial gradient of the output at the interior junction node in any adjacent segment which is calculated by AD is equal. This is incompatible with the interior junction node constraints so that we cannot directly construct the objective function without the insertion of virtual distance. 
The virtual distance is set to be $\nu=0.5 \mu m$ in our experiment and the objective function is finally defined as follows:}
% The objective function is finally defined as:
\begin{equation}
\label{eq:MSEloss}
MSE=MSE_f+MSE_b+MSE_i+MSE_c.
\vspace{-0.2cm}
\end{equation}
In the equation~\eqref{eq:MSEloss}, $MSE_f$ represents the mean-square error of diffusion operator $f$ and describes the relationship between the nonlinear operator term $\mathcal{N}[{\sigma}]$ and the temporal term $\sigma_t$. 
{\color{black}The losses $MSE_b$, $MSE_c$ and $MSE_i$ correspond to BCs at terminals, BCs at interior junction nodes and ICs, respectively.}
% Losses $MSE_b$, $MSE_c$ and $MSE_i$ correspond to BCs and ICs and solution-flux continuity constraints at junctions through virtual distance.
When solving the stress using the modified PINN and the proposed STPINN, each term on the right side of~\eqref{eq:MSEloss} is defined as follows:
%Details for each term in %\eqref{eq:MSEloss} follow:
% \begin{equation}\small
% \label{eq:MSE}
% \begin{aligned}
% &MSE_f = \frac{1}{N_f}\sum^{N_f}_{i=1}\Big|\frac{\partial {\sigma}(x,t_i
% 	)}{\partial t}-{\color{black}{\kappa(x,t_i)}}\frac{\partial^2{\sigma}(x,t_i)}{\partial x^2}\Big|^2, x\in{\color{black}\mathcal{L}_1}\cup{\color{black}\mathcal{L}_2}\\
% 	%\\&\qquad\qquad\qquad\qquad\qquad\qquad\qquad\qquad\  x\in{\color{black}\mathcal{L}_1}\cup{\color{black}\mathcal{L}_2},\\
% &MSE_b = \frac{1}{N_b}\sum^{N_b}_{i=1}\Big|\frac{{
% 		\partial\sigma}(x,t_i)}{\partial x}+G_b\Big|^2,x\in{\color{black}\mathcal{B}},\\
% &MSE_i = \frac{1}{N_0}\sum^{N_0}_{i=1}\Big|{\sigma
% }(x,0)-\sigma_T\Big|^2,x\in{\color{black}\mathcal{L}_1}\cup{\color{black}\mathcal{L}_2}, \\
% &MSE_c=\frac{1}{N_c}\sum^{N_c}_{i=1}\Big(\Big|{\sigma}
% (x_l,t_i)-{\sigma}(x_r,t_i)\Big|^2+\Big|\kappa(x_l,t_i)\\&\quad\qquad\times\big(\!\frac{{
% 		\partial\sigma}(x_l,t_i)}{\partial x}+G_l\big)-\kappa(x_r,t_i)\big(\frac{{
% 		\partial\sigma}(x_r,t_i)}{\partial x}+G_r\big)\Big|^2\Big),\\ %\\&  x_l\in %{\color{black}\mathcal{I}%},x_r\in{\color{black}\ma%thcal{I}',} 
% \end{aligned}
% \end{equation}
{\color{black} 
\begin{equation}\small
\label{eq:MSE}
\begin{aligned}
&MSE_f = \frac{1}{N_f}\sum^{N_f}_{l=1}\Big|\frac{\partial {\sigma}(x_l,t_l
	)}{\partial t}\!-\!\frac{\partial}{\partial x}\Big[{\kappa(x_l,t_l)}\Big(\frac{\partial{\sigma}(x_l,t_l)}{\partial x}\!+\!G_l\Big)\Big]\Big|^2\!, \\
&MSE_b = \frac{1}{N_b}\sum^{N_b}_{l=1}\Big|\kappa(x,t_l)\Big(\frac{{
		\partial\sigma}(x,t_l)}{\partial x}+G_b\Big)\Big|^2,x\in{\mathcal{B}'},\\
&MSE_i = \frac{1}{N_0}\sum^{N_0}_{l=1}\Big|{\sigma
}(x_l,0)\Big|^2,x_l\in\mathcal{L}', \\
&MSE_c=\frac{1}{N_c}\sum^{N_c}_{l=1}\sum_{m=1}^q\Big(\sum_{x_k\in \mathcal{V}_m}\Big|{\sigma}
(x_k,t_l)-{\sigma}(c_m,t_l)\Big|^2\\&\qquad\qquad\ +\Big|\sum_{x_j\in \mathcal{P}_m}\kappa(x_j,t_l)\Big(\frac{{\partial\sigma}(x_j,t_l)}{\partial x}+G_j\Big)\cdot n_j\Big|^2\Big),\\
% \\&\qquad\qquad\ -\kappa(x_j^+,t_i)\big(\frac{{
% 		\partial\sigma}(x_j^+,t_i)}{\partial x}+G_j^+\big)\Big|^2\Big),\\ %\\&  x_l\in %{\color{black}\mathcal{I}%},x_r\in{\color{black}\ma%thcal{I}',} 
\end{aligned}
\end{equation}
% where ${\Omega} = \{x:x_1\cap x_2,\;x_1\in{\Omega}_1,\;x_2\in{\Omega}_{2\nu}\}$ and ${\Omega}_{2\nu}=\{v:x+\nu,\;x\in{\Omega}_2\}$.
where $x_l\in\mathcal{L}'$ in $MSE_f$. Notations $\mathcal{L}'$ and $\mathcal{B}'$ are the set of points within all segments and at blocked terminals, respectively, extracted from interconnect trees inserted with virtual distances at interior junction nodes. It is supposed that there are $q$ interior junction nodes in the interconnect tree. We denote the collection of overlapping nodes describing the $m$-th interior junction node located at $c_m$ by $\mathcal{P}_m=\{c_m,\mathcal{V}_m\}$ where $\mathcal{V}_m=\{c_{m1},\cdots,c_{mp}\}$ represents the collection of coordinates of the neighbouring nodes at virtual distances connected with the $m$-th interior junction node. The notations $G_l,\ G_b,\ G_j$ are the EM driving forces on the segment corresponding to $x_l$, the blocked segments, and the segment connected with node $x_j$, respectively.
% According to the interconnect structure shown in Fig.~\ref{fig:vd}, we define the collection points of left and lower terminal coordination of the virtual distance in $\mathcal{V}_m$ as $\mathcal{I}_m^-=\{a_1^-,\cdots,a_p^-\}$ and the collection points of right and upper terminal coordination as $\mathcal{I}_m^+=\{a_1^+,\cdots,a_p^+\}$, where $|a_i^-a_i^+|=\nu\ (i=1,\cdots,p)$ and $|a_i^-a_i^+|$ represents the distance between points $a_i^-, a_i^+$. We define $x_j^-\in \mathcal{I}_m^-$, $x_j^+\in \mathcal{I}_m^+$ and $G_b,\ G_j^-,\ G_j^+$ are the EM driving forces on the blocked segments and segments consisting node $x_j^-$ and $x_j^+$, respectively.
}
The notation $t_l$ denotes the time point sampled randomly within the configured time range. \textcolor{black}{For the PINN analysis, we first obtain the temperature $T(x,t)$ according to the thermal model \eqref{eq:thermal} and then calculate the diffusivity $\kappa(x,t)=D_aB\Omega/(kT(x,t))$ through the obtained temperature at a certain time and location. The mean-square error $MSE_f$ can be calculated by the equation \eqref{eq:MSE} through collocation points randomly sampled in space and time domains and the corresponding diffusivity $\kappa(x_l,t_l)$. After that, we can calculate the objective function $MSE$ by the losses $MSE_f$, $MSE_b$, $MSE_c$, and $MSE_i$, which is used for the PINN training.}
% {\color{black} and $\kappa(x_i,t_i)$ is the stress diffusivity depending on the space-time related temperature.}
% The losses $MSE_i$, $MSE_b$ and $MSE_c$ correspond to the initial, boundary and solution-flux continuity constraint condition, and $MSE_f$ enforces the structure imposed by Korhonen's equation \eqref{eq:Korhonen's PDE} at a finite set of sampling points.

% In~\cite{PINN2019:Journal},  Latin Hypercube Sampling (LHS)~\cite{LHS1987} is applied to generate spatial-temporal input data. Based on observation of the transformed diffusion coefficient and time range, Logarithmic Isometric Sampling (LIS) demonstrates its rationality in sampling, since the gradient is relatively steep near the zero point.
% %which can be noticed in Fig.~\ref{fig:sample}. 
% In case tests, spatial information is sampled by LHS and temporal data are sampled by both LIS and LHS.

% {\color{black}The PINN-based EM stress evolution model is composed of data preparation, model training with a PINN network.} 
The PINN model works as a solver for predicting the stress evolution with differently configured currents flowing in multiple interconnected segments.
%The model works as a solver to predict the stress evolution in days or years, where different-configured currents flow in multiple interconnected segments.
% The above-mentioned configuration is an effective PINN expansion for solving the stress evolution equation with a constant temperature. 
Compared with general supervised neural network methods, training data preparation of PINN omits data labeling since the objective function is only related to the sampled location and time point. However, one potential limitation of the PINN-based model for calculating EM-based stress evolution stems from the learning ability and approximate capacity of the network structure for solving PDE with different diffusivities. It is required to improve the network structure to cope with different diffusivities such that we can obtain an exact stress solution under complex temperature conditions.

%However, one potential limitation of the PINN-based model in EM stress evolution estimation stems from the learning ability and approximate capacity of the network structure for PDE solving with different diffusivities. It is necessary to improve a network structure suitable for different diffusivities so that we can obtain stress evolution under complex temperature conditions.

%It can be observed in Section~\ref{experiment} that PINN show sufficient performance in stress evolution analysis under constant temperature.
%\begin{figure}[htb]
%	\centerline{\includegraphics[width=0.6\columnwidth]{example.pdf}}
%	\caption{Structure of neural network based model for Korhonen's equations.}
%	\label{fig:sample}
%\end{figure}
%\subsection{STPINN based EM Analysis under Complex Temperature}
% \begin{figure}[t]
% 	\centerline{\includegraphics[width=0.8\columnwidth]{gradient.pdf}}
% 	\caption{{\color{black}The absolute gradient value along with the training iterations.}}
% 	\label{fig:lgradient}
% \end{figure}
\begin{figure}[t]
	\centering 
	\subfigure[]{
		\includegraphics[width=0.465\columnwidth]{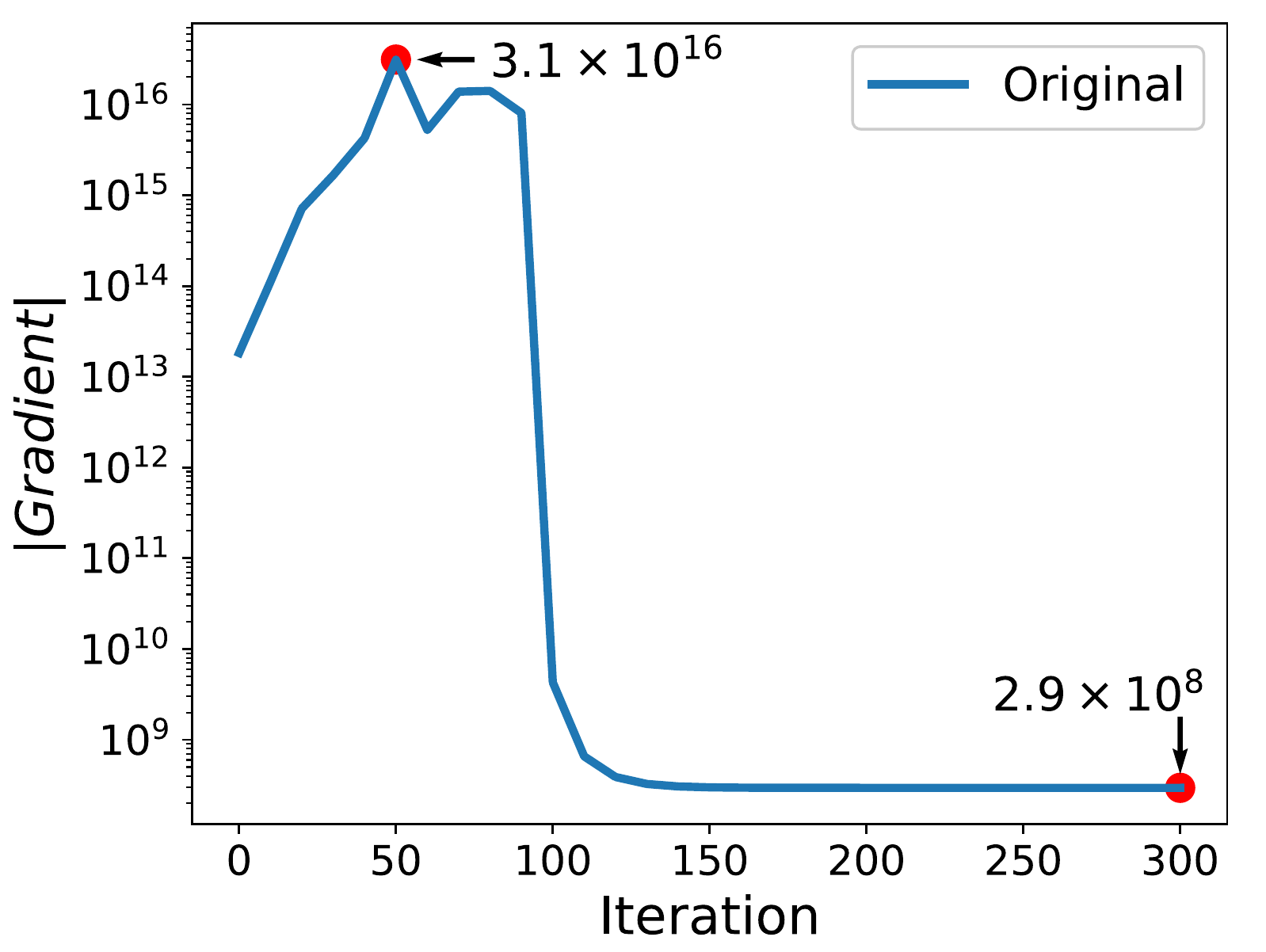}
		\label{fig:lgradient}} 
	\subfigure[]{
		\includegraphics[width=0.465\columnwidth]{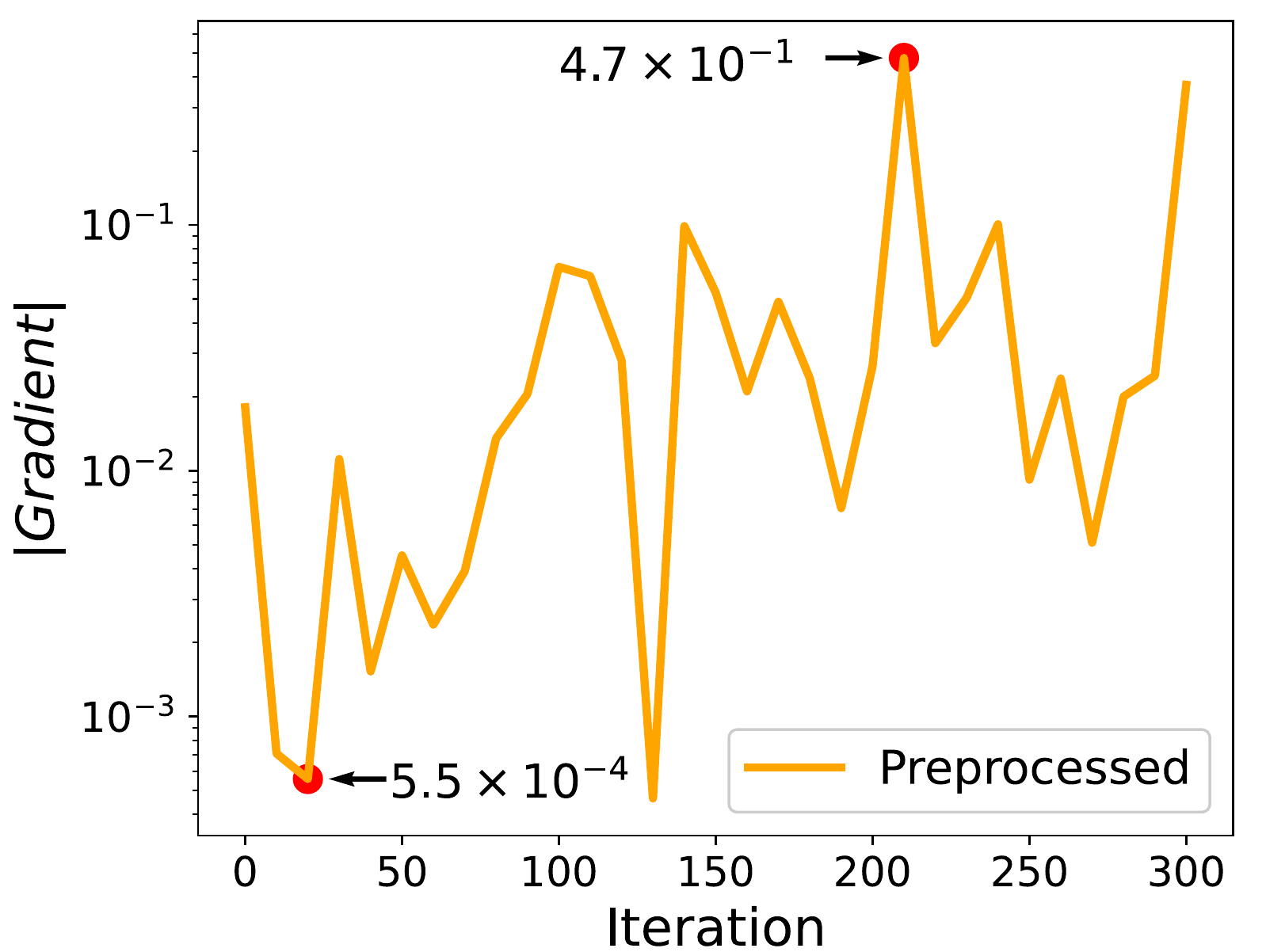}
		\label{fig:pgradient}}
	\caption{{\color{black}The absolute value of (a) original gradient and (b) gradient after preprocessing over training iterations.}}
	\label{fig:gradient}
\end{figure}

{\color{black}
\subsection{Gradient Computation}
% Neural networks have been applied in mesh-free-based solution of physical systems as a universal function approximator~\cite{HORNIK1989:NN}. 
Before employing the neural network based method for stress assessment, we first analyze the gradient of neural networks during the training phase. We use an FNN consisting of two input units, one hidden layer with $H$ $\tanh$ units, and one linear output unit to approximate the stress evolution. The input vector is defined as $a_j(j=1,2)$ comprising of the coordinate of the location and the time instance. The output is $N=\sum_{i=1}^Hv_io(z_i)$ where $z_i=\sum_{j=1}^2w_{ij}a_{j}+u_i$. Here, $u_i$ and $w_{ij}$ denote the bias of hidden unit $i$ and the weight from the input unit $j$ to the hidden layer unit $i$. The notation $v_i$ denotes the weight from the hidden unit $i$ to the output and $o(z)$ is the $\tanh$ activation function. Then, the derivative term of the stress evolution in the objective function follows:
\begin{equation}\label{eq:derivative}
\frac{\partial^k N}{\partial a_j^k} = \sum_{i=1}^Hv_iw_{ij}^ko_i^{(k)},\\
\end{equation}
where $o_i=o(z_i)$ and $o^{(k)}$ is the $k$-th order derivative of the $\tanh$ function. Thus, the gradient of each derivative term in \eqref{eq:derivative} can be obtained as:
\begin{equation}\label{eq:gradient}
    \begin{aligned}
    &\frac{\partial^k N}{\partial a_j^k\partial v_i}=w_{ij}^ko_i^{(k)},\\
    &\frac{\partial^k N}{\partial a_j^k\partial u_i}=v_iw_{ij}^ko_i^{(k+1)},\\
    &\frac{\partial^k N}{\partial a_j^k\partial w_{ij}}=a_jv_iw_{ij}^ko_i^{(k+1)}
    +v_ikw_{ij}^{k-1}o_i^{(k)}.\\
    \end{aligned}
\end{equation}
For the parameters in the objective function, the magnitudes of $x,\ t$ and the EM driving force are close to $1\times10^{-5},\ 1\times10^{8},\ 1\times10^{13}$, and the value of $\kappa$ in Korhonen's function is about $1.4136\times 10^{-18}$ under the temperature $T=350K$. The gradient of the objective function with respect to the network parameters can be defined based on \eqref{eq:MSEloss}, \eqref{eq:MSE} and \eqref{eq:gradient}. Fig.~\ref{fig:lgradient} shows the absolute gradient value $|Gradient|$ along with the training iterations. It can be observed that the maximum value of $|Gradient|$ is about $3.1\times 10^{16}$ and the value of $|Gradient|$ after convergence is about $2.9\times 10^8$ with minor fluctuations. The learned model can not provide accurate stress solutions through the large gradient. The constraints in Korhonen's equation will not be satisfied if we directly normalize the parameters. To deal with the problem, we develop a linear preprocessing method for EM model.
}

\subsection{Preprocessing and Data Preparation}\label{secpreprocess}
% Neural networks have been applied in mesh-free-based solution of physical systems as a universal function approximator~\cite{HORNIK1989:NN}. 

% {\color{black}Specifically, due to the tiny diffusion coefficient and large parameters in Korhonen's equations, the training gradient will become extremely huge if the original parameters are sent into the model directly. }

% Specifically, due to the tiny diffusion coefficient and large parameters provided by Korhonen's equations, the gradient will explode when all the original parameters are sent into the model directly. 
{\color{black}In this section, we propose a preprocessing method for EM analysis to handle the gradient problem.} We first unfold the PDE, BC and IC in~\eqref{eq:Korhonen's PDE} to Euler format and perform linear transformation:

%Specifically, due to the tiny diffusion coefficient and large parameters provided by Korhonen's equations, the gradient will explode when all the original parameters are sent into the model directly. {\color{black}We propose a neural network based preprocessing method for EM analysis to handle the gradient explosion problem.} We first unfold the PDE, BC and IC in \eqref{eq:Korhonen's PDE} to Euler format and perform linear transformation:

%, which is given by:
%	\begin{equation}\label{Eulerorigin}
%	\begin{aligned}
%{\rm PDE}:&\frac{\sigma^{m+1} - 	\sigma^{m}}{\tau}-\kappa\Big(\frac{\sigma_{j-1}-2\sigma_j+\sigma_{j+1}}{h^2}\Big)=O(h^2)-O(\tau),\\
%&m=0,1,\cdots,M,j=1,2,\cdots,N-1,\\
%{\rm BC}:&\frac{\sigma_{1}-\sigma_{0}}{h}+O(h)=-G,\\
%&\frac{\sigma_{N}-\sigma_{N-1}}{h}+O(h)=-G.\\
%	\end{aligned}
%\end{equation}
\begin{equation}\label{Euler}
\begin{aligned}
{\rm PDE}:&\frac{\hat{\sigma}^{m+1} - 	\hat{\sigma}^{m}}{\hat{\tau}}-\frac{\omega_x^2}{\omega_t}\kappa\Big(\frac{\hat{		\sigma}^{m}_{j-1}-2\hat{\sigma}^{m}_j+\hat{\sigma}^{m}_{j+1}}{\hat{h}^2}\Big)=0,\\
&m=0,1,\cdots,M,j=1,2,\cdots,N-1,\\
{\rm BC}:&\frac{\hat{\sigma}^{m}_{1}-\hat{\sigma}^{m}_{0}}{\hat{h}}=-\hat{G}_1,
\frac{\hat{\sigma}^{m}_{N}-\hat{\sigma}^{m}_{N-1}}{\hat{h}}=-\hat{G}_2,\\
{\rm IC}:&\hat{\sigma}^0=\omega_{\sigma}\sigma(x,0).
\end{aligned}
\end{equation}
We denote $\tau$, $h$ as the time and length interval for time and space discretizing, and construct numerical relationship $\hat{\sigma}=\omega_{\sigma}\sigma$, $\hat{h}=\omega_xh$, $\hat{\tau}=\omega_t\tau$ and $\hat{G}_i=\omega_{\sigma}/\omega_xG_i(i=1,2)$ for the linear scale transformations on $h$, $\tau$, $\sigma$, $G_i$. Here,  $\omega_x, \omega_t, \omega_{\sigma}$ represent scaling factors of the length, the time and the stress $\sigma$, respectively. Number of mesh grids in the space domain $N$ is related to the transformed length interval $\hat{h}$ with the relationship $N\hat{h}=\omega_xL$, where $L$ is the total length of wire. The temporal domain is configured to $t\in[0,T_s]$. The notation $T_s$, selected for simulating the reliability assessment time range, satisfies the relationship $M\hat{\tau}=\omega_tT_s$. In addition, the scaling factor of the stress $\omega_{\sigma}$ is unrelated to PDEs and BCs for its elimination in both sides of the formulas.
Let $\sigma(x,t,\kappa)$ and 
$\hat{\sigma}(x,t,\kappa)$ be the solutions of original and normalized expressions. Then, the solution of stress evolution equation satisfies:
\begin{equation}\label{preprocess}
\sigma(x,t,\kappa)=\omega_{\sigma}\hat{\sigma}(\omega_x
x,\omega_tt,\frac{\omega_x^2}{\omega_t}\kappa).
\end{equation}
The inputs of networks are converted to the regular parameters consisting of variables $\omega_xx$, $\omega_tt$. Correspondingly, the diffusion coefficient in both the stress evolution equation and the objective function is rewritten as $\omega_x^2/\omega_t\kappa$. {\color{black}Fig.~\ref{fig:pgradient} shows the absolute value of the gradient after preprocessing along with training iterations. It can be observed that $|Gradient|$ fluctuates between $5.5\times10^{-4}$ and $4.7\times10^{-1}$. The learned model can achieve satisfactory accuracy after employing the preprocessing scheme.} This procedure plays an important role in the application of learning based method for solving stress evolution equation in EM reliability analysis.

%We denote $\tau$, $h$ as the time and length interval for time and space discretizing, and construct numerical relationship $\hat{\sigma}=\omega_{\sigma}\sigma$, $\hat{h}=\omega_xh$, $\hat{\tau}=\omega_t\tau$ and $\hat{G}_i=\omega_{\sigma}/\omega_xG_i(i=1,2)$ for the linear scale transformations on $h$, $\tau$, $\sigma$, $G_i$. Here,  $\omega_x, \omega_t, \omega_{\sigma}$ represent scaling factors of length, time and solution, respectively. Number of mesh grids in the space domain $N$ is related to the transformed length interval $\hat{h}$ with the relationship $N\hat{h}=\omega_xL$, where $L$ is the total length of wire. The temporal domain is configured to $t\in[0,T_s]$. Notation $T_s$, selected beyond the reliability assessment time range, satisfies the relationship $M\hat{\tau}=\omega_tT_s$. In addition, the scaling factor of solution $\omega_{\sigma}$ is unrelated to PDEs and BCs for its elimination in both sides of the formulas. 
%As a result, let $\sigma(x,t,\kappa)$ and $\hat{\sigma}(x,t,\kappa)$ be the solutions of original and normalized expression, the solution of Korhonen's equation satisfies:
%\begin{equation}\label{preprocess}
%\sigma(x,t,\kappa)=\omega_{\sigma}\hat{\sigma}(\omega_x
%x,\omega_tt,\frac{\omega_x^2}{\omega_t}\kappa).
%\end{equation}
%Here, inputs of networks are converted to regular parameters consisting of variables $\omega_xx$, $\omega_tt$, and $\omega_x^2/\omega_t\kappa$ is the new diffusion coefficient in the objective function. Note that the above preprocessing procedure is a pivotal operation in data preparation.
% , which  

For the original data preparation of  preprocessing on solving PDEs, Latin Hypercube Sampling (LHS)~\cite{LHS1987} is applied to generate spatial-temporal input data in the PINN method~\cite{PINN2019:Journal}. Based on observation of the transformed diffusion coefficient and time range in the stress evolution equation for VLSI reliability analysis, Logarithmic Isometric Sampling (LIS) demonstrates its rationality in sampling since the gradient is relatively steep near the zero point. In test cases on neural network based model, we sample the spatial location by LHS and the time point by both LIS and LHS, and then generate the training data and test data by preprocessing the collection of spatial location and time points, shown in the data preparation procedure of Fig.~\ref{fig:flowchart}.

%For the original data preparation before preprocessing, Latin Hypercube Sampling (LHS)~\cite{LHS1987} is applied to generate spatial-temporal input data in the PINN method~\cite{PINN2019:Journal}. Based on observation of the transformed diffusion coefficient and time range, Logarithmic Isometric Sampling (LIS) demonstrates its rationality in sampling, since the gradient is relatively steep near the zero point.
%which can be noticed in Fig.~\ref{fig:sample}. 
%In case tests, spatial information is sampled by 

%In test cases, we first sample the spatial location by LHS and the time point by LIS and LHS, and then generate the training data and test data by preprocessing the collection of spatial location and time point, shown in data preparation of Fig.~\ref{fig:flowchart}.

\subsection{STPINN}\label{stpinnsec}
\begin{figure*}[htb]
	\centerline{\includegraphics[width=2.0\columnwidth]{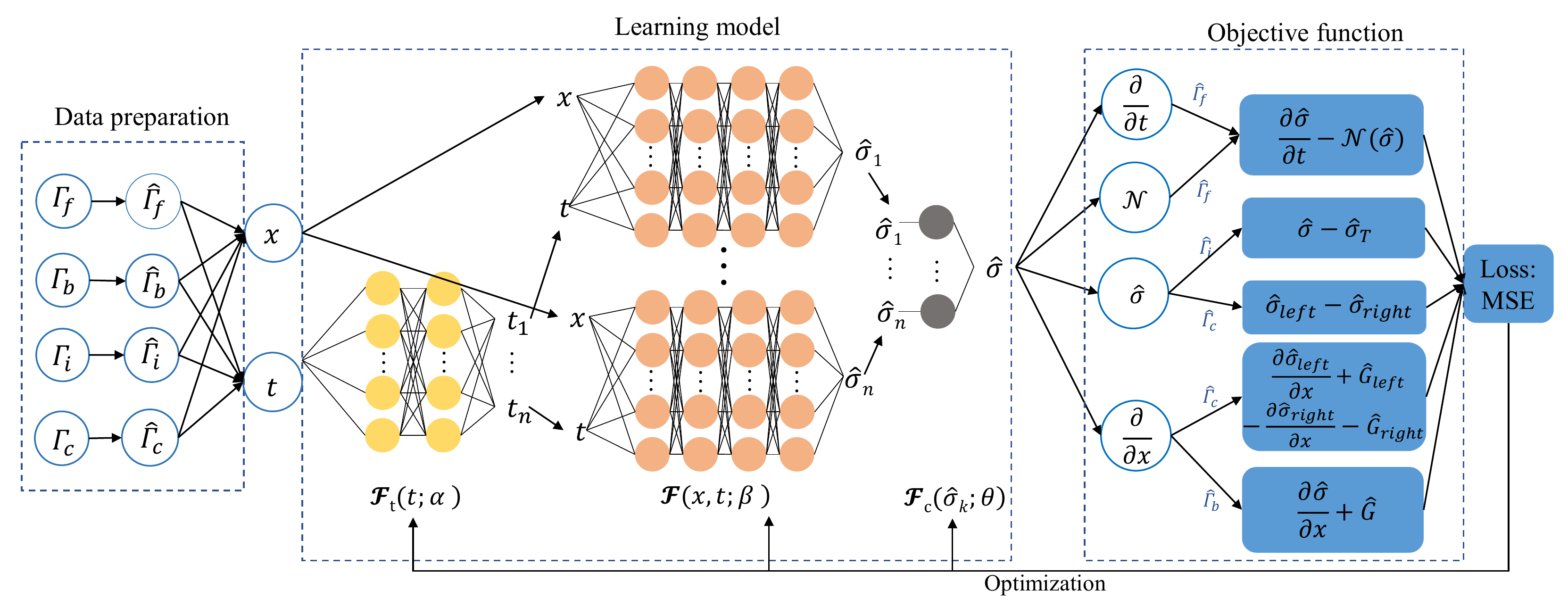}}
	\caption{Structure of STPINN: Network $\mathcal{F}_t$ is an FNN for multi-transformation in time domain, network $\mathcal{F}$ supplies solutions of PDEs with basis diffusivity and $\mathcal{F}_c$ is a fully connected layer to operate mapping conversion in space domain. By combining the parallel components, learning ability for non-constant diffusivity PDEs is strengthened by neural networks. }
	\label{fig:pdenn}
\end{figure*}
%In this section, we propose an unsupervised neural network structure called STPINN (space-time physics-informed neural network), introducing new space and time variables for PDEs with non-constant diffusivity. The non-constant diffusivity $\kappa$ is determined by the dynamic temperature model and thermal model \eqref{eq:thermal}.However, our experiments show that PINN has poor training performance in stress assessment with non-constant diffusivity. To improve the training performance, we first unfold the diffusion function \eqref{eq:Korhonen's PDE} with diffusivity $\kappa(x,t)$ in Euler format:
{\color{black}It is reported that the PINN method can be used to solve some types of PDEs (such as constant diffusivity) \cite{PINN2019:Journal}. However, the accuracy of the PINN method depends on the complexity of the diffusivities. The PINN method cannot well capture the stress evolution process when the temperature varies with respect to space and time. To solve the problem, 
we propose the STPINN by developing new space and time variables for PDEs with temperature dependent diffusivity.}
%The non-constant diffusivity $\kappa$ is related to a temporal and spatial temperature. 
% Our experiments show that PINN has unsatisfactory training performance in stress assessment with non-constant diffusivity.
{\color{black}We suppose the stress $\sigma(x,t)$ can be divided to $n$ solutions of different diffusion process on various spatial-temporal space, which can be expressed as follows:
% we first separate the stress  $\sigma(x,t,\kappa(x,t))$ in \eqref{eq:Korhonen's PDE} into $n$ components, which can be expressed as follows:
% we first rewrite the diffusion equation \eqref{eq:Korhonen's PDE} with diffusivity $\kappa(x,t)$ in Euler format:}
% \begin{equation}\label{eq:eulerk}
% \begin{aligned}
% \frac{{\sigma}^{m+1} -{\sigma}^{m}}{{\tau}}&-\frac{\kappa_{j-\frac{1}{2}}{		\sigma}^{m}_{j-1}-(\kappa_{j+\frac{1}{2}}+\kappa_{j-\frac{1}{2}}){\sigma}^{m}_{j}+\kappa_{j+\frac{1}{2}}{\sigma}^{m}_{j+1}}{{h}^2}\\
% &-\frac{\kappa_{j+\frac{1}{2}}-\kappa_{j-\frac{1}{2}}}{{h}}{G}=0.
% \end{aligned}
% \end{equation} 
%Then, we separate the diffusion expression by the separation %of variables method, which can be expressed as 
% \textcolor{black}{We also separate the solution $\sigma(x,t,\kappa(x,t))$ into $n$ components, which can be expressed as follows:}
% \begin{equation}\label{eq:separation}
% \kappa(x,t)=\sum_{i=0}^nA_i(t)B_i(x)E(x,t),
% \end{equation}
\begin{equation}\label{eq:separation}
\sigma(x,t)\approx\sum_{k=0}^nu(\hat{x}_k,\hat{t}_k,E_k(\hat{x}_k,\hat{t}_k)),
\end{equation}
%where $E(x,t)$, $A_i(t)$ and $B_i(x)$ are the unknown basis diffusion coefficient, mapping operations on temporal and spatial variables in PDE with diffusivity $\kappa(x,t)$. Specifically, the separation procedure can be expressed in Euler format based on \eqref{eq:eulerk}:
where $u(\hat{x}_k,\hat{t}_k,E_k(\hat{x}_k,\hat{t}_k))$ is the solution of the diffusion process with location $\hat{x}_k$, time $\hat{t}_k$ and diffusivity $E_k(\hat{x}_k,\hat{t}_k)$. The stress evolution satisfies $\sigma(x,t)=u(x,t,\kappa(x,t))$. It should be noted that $\hat{x}_k$, $\hat{t}_k$ and $E_k(\hat{x}_k,\hat{t}_k)$ are new spatial, temporal transformed variables and the unknown basis diffusion coefficient in the $k$-th component of the stress, respectively. We define $u(\hat{x}_k,\hat{t}_k,E_k(\hat{x}_k,\hat{t}_k))=u_k$ and rewrite the diffusion equation of $u_k$ as follows:
\begin{equation}\label{eq:euler}
    \frac{\partial u_k}{\partial t}\frac{\partial t}{\partial \hat{t}_k}=\frac{\partial}{\partial x}\frac{\partial x}{\partial \hat{x}_k}\Big[E_k(\hat{x}_k,\hat{t}_k)\Big(\frac{\partial u_k}{\partial x}\frac{\partial x}{\partial \hat{x}_k}+G\Big)\Big].
\end{equation}
We define $\partial t/\partial \hat{t}_k=\gamma_k(t),\ \partial x/\partial \hat{x}_k=\xi_k(x)$ and introduce a new temporal variable $t_k$ which satisfies the following equation:
\begin{equation}\label{eq:t}
\frac{\partial t_k}{\partial t}=\frac{\xi^2_k(x)}{\gamma_k(t)}.\\
\end{equation}
Moreover, we define the notation $\hat{u}_k=u(x,t_k,E_k(\hat{x}_k,\hat{t}_k))$ which follows the diffusion equation:
% and thus \eqref{eq:euler} yields:
% \begin{equation}\label{eq:euler}
% \begin{aligned}
% \frac{{\sigma}_e^{m+1} -{\sigma}_e^{m}}{\gamma_{i,m}{\tau}}&-\frac{e_{j-\frac{1}{2}}{		\sigma}^{m}_{j-1}-(e_{j+\frac{1}{2}}+e_{j-\frac{1}{2}}){\sigma}^{m}_{j}+e_{j+\frac{1}{2}}{\sigma}^{m}_{j+1}}{(\frac{1}{\theta_i}{h})^2}\\
% &-\frac{e_{j+\frac{1}{2}}-e_{j-\frac{1}{2}}}{\frac{1}{\theta_i}{h}}{\hat{G}}=0.
% \end{aligned}
% \end{equation} 
% Here, $e$ is the discrete value of the basis diffusivity $E(x,t)$ in \eqref{eq:separation}. The $i$-th scaling factors in the spatial and temporal domain are set to be $1/\theta_i$ and $\gamma_{i,m}$, where $m$ represents time-discrete position. The function $\hat{G}$ is linear related to the reciprocal of spatial scale $\theta_i$ according to BC, defined as $\hat{G} = \theta_i G$. 
% % In order to reduce the complexity of network training, the operation on spatial variables has the same impact when employing a linear scale expansion on approximate solutions with factor $\theta_i$. 
% We transform \eqref{eq:euler} to the following form in case of the hypothesis $\tau\rightarrow 0,\ h\rightarrow 0$:
% We define the $i$-th component of solution as $\hat{\sigma}_i(x,t)=\sigma_i(x,t,E(x,t))$, which satisfies the following expression when :
\begin{equation}\label{eq:euler1}
\frac{\partial \hat{u}_k}{\partial t_k}=\frac{\partial}{\partial x}\Big[E_k(\hat{x}_k,\hat{t}_k)\Big(\frac{\partial \hat{u}_k}{\partial x}+G\Big)\Big].\\
\end{equation} 
According to \eqref{eq:euler}, \eqref{eq:t} and \eqref{eq:euler1}, the $k$-th component $u_k$ can be calculated as follows:
\begin{equation}\label{eq:x}
   u_k = \xi_k(x) \hat{u}_k.
\end{equation}
In \eqref{eq:euler1}, we define $E_k(\hat{x}_k,\hat{t}_k)=E(x,t_k)$, where $E(x,t_k)$ is an unknown basis diffusion coefficient. In this way, the shared parameters can be employed to obtain $\hat{u_k}$ through the inputs $x$ and $t_k$. It can be observed in \eqref{eq:t} that the transformation from $t$ to $t_k$ is related to the transformation from $x$ to $\hat{x}$. For simplify, we suppose that a linear transformation is executed on $x$, which satisfies $\xi_k(x)=\theta_k$.
%of each component solution
% \begin{align}
%     &t_i = \gamma_i(t)\times t,\label{eq:t}\\
%     &\sigma_i(x,t) = \theta_i \hat{\sigma}_i(x,t_i).\label{eq:x}
% \end{align}
% In \eqref{eq:t}, $\gamma_i(t)$ is the continuous form of $\gamma_{i,m}$ used for the nonlinear temporal conversion, and the linear spatial-dependent conversion of stress is shown in \eqref{eq:x}. It can be observed that the spatial conversion is a scaling process for the $i$-th component of the stress $\hat{\sigma}_i$ with the scale factor $\theta_i$. 
Thus, we can derive the stress evolution solution $\sigma(x,t)$ in \eqref{eq:Korhonen's PDE} as:%
\begin{equation}\label{eq:sigma}
\begin{aligned}
\sigma(x,t)&\approx \sum_{k=0}^{n}\theta_k u(x,t_k,E(x,t_k)),\\
t_k&=\int_0^t\theta_k^2/\gamma_k(t')dt'.
\end{aligned}
    \end{equation}
The proposed STPINN method is motivated by the spectral method in the mathematical intuition, which expands the solution approximately into a finite series expansion of a smooth function. In this way, it is supposed that the solution of the stress evolution equation can be divided into several smooth solutions respecting different sub-diffusion processes in different spatial-temporal spaces.}
% where $\hat{\sigma}_i(x,t)$ is the approximate expression of the PDE solution with the basis diffusivity $E(x,t)$, and $\gamma_{i,m}\times t$ is the nonlinear transformation of time variable $t$.

{\color{black}Based on \eqref{eq:sigma}, we build a network consisting of neural network-based linear and nonlinear operators for variable transformation in the proposed STPINN architecture.} The proposed method comprises three connected neural networks with multi channels, as shown in Fig.~\ref{fig:pdenn}. The first network $\mathcal{F}_t(t;\alpha)$ is employed as the nonlinear temporal transformation operator in \eqref{eq:t}, which converts a time variable $t$ into $n$ new sub variables $t_k(k=1,2,\cdots, n)$. The second FNN $\mathcal{F}(x,t;\beta)$ obtains the mesh-free solutions of PDEs with the diffusivity $E(x,t_k)$. Several couples of outputs $\hat{\sigma}_k(k=1,2,\cdots,n)$ perceived by $\mathcal{F}(x,t;\beta)$ are connected to the fully connected layer $\mathcal{F}_c(\hat{\sigma}_k;\theta)$, which operates spatial conversion in \eqref{eq:x}. {\color{black}The original data are sampled through the method mentioned in Section~\ref{secpreprocess} into $\Gamma_f,\ \Gamma_i,\ \Gamma_b,\ \Gamma_c$ and then they are preprocessed into $\hat{\Gamma}_f,\ \hat{\Gamma}_i,\ \hat{\Gamma}_b,\ \hat{\Gamma}_c$, which are collocation points for the diffusion operator, ICs, BCs at terminals and BCs at interior junction nodes.}
% which are defined by the diffusion operator, BCs, ICs and the solution-flux continuity constraint in the objective function, respectively.

\begin{figure}[t]
	\centerline{\includegraphics[width=0.7\columnwidth]{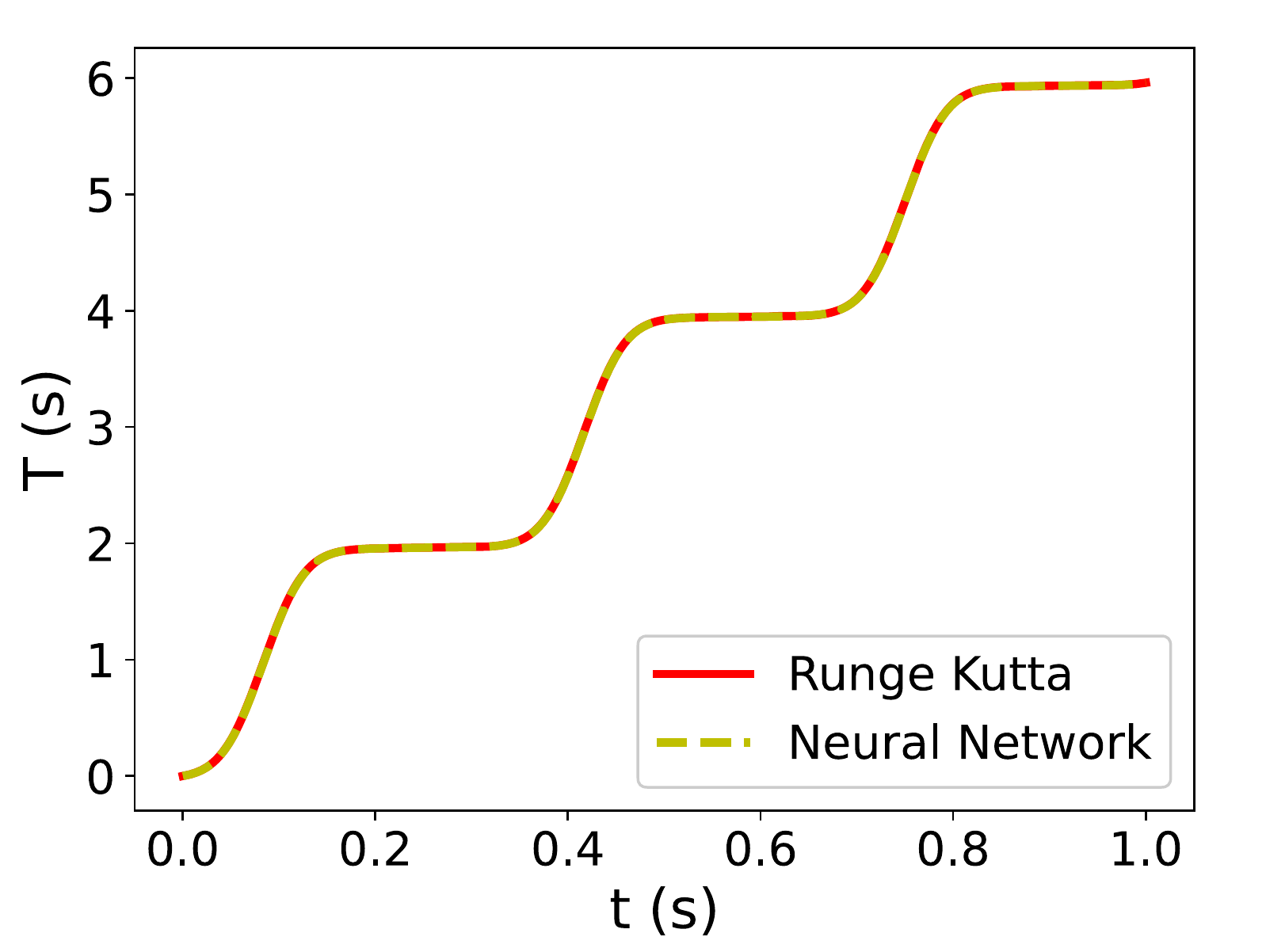}}
	\caption{The comparison between Runge Kutta and the neural network based method.}
	\label{fig:inte}
\end{figure}

For the number of STPINN channels $n$, we choose different $n$ according to different diffusion coefficient types. For instance, a one-channel STPINN  (1-STPINN) has sufficient capacity for solving the PDE with time-dependent diffusivity $\kappa(t)$ in Case II, since no spatial-dependent transformation is caused by the non-constant diffusivity. We multiply both sides of the PDE in \eqref{eq:Korhonen's PDE} by $\partial t/\partial T'$, where $T'=\int_0^t(\kappa(t')/\kappa_0) dt'$. The transformation follows:
\begin{equation}\label{eq:conditionbex}
\frac{\partial\sigma}{\partial t}\frac{\partial t}{\partial T'}=\frac{\partial}{\partial x}[\kappa(t)\frac{\partial t}{\partial T'}(\frac{\partial \sigma}{\partial x}+G)],x\in{\color{black}\mathcal{L}}.
\end{equation}
Then the PDE with new temporal variable takes the form:
\begin{equation}\label{eq:1stpinn}
\frac{\partial\sigma}{\partial T'}=\frac{\partial}{\partial x}[\kappa_0(\frac{\partial \sigma}{\partial x}+G)],x\in{\color{black}\mathcal{L}}.
\end{equation}
Here, the time-related diffusivity PDE can be converted to a new one with constant diffusivity $\kappa_0$ through the transformation on temporal variable~\cite{Ashlee2014}. 
% Furthermore, compared with the numerical integration based dynamic EM model in~\cite{Chen2017:TDMR}, a neural network-based solver consumes less computation time by ignoring iterative procedure. 
% {\color{black} Furthermore, compared with the numerical integration based dynamic EM model in~\cite{Chen2017:TDMR}, the proposed method can obtain the value of transformed variable $T'$ without setting time-stepping.}
Fig.~\ref{fig:inte} shows the comparison between the results of the transformed variable $T'$ obtained by the neural network based solver and the numerical Runge Kutta method. It can be seen from Fig.~\ref{fig:inte} that the results by the neural network method fit well with the numerical method within 0.2\% relative error. {\color{black} The neural network based solver consumes 0.0003s for calculating $T'$ and the Runge Kutta method costs 0.002s for calculating $T'$ in 100 time steps.}

The objective function is imposed by \eqref{eq:MSEloss}, \eqref{eq:MSE} through the training data sets $\hat{\Gamma}_f,\ \hat{\Gamma}_b,\ \hat{\Gamma}_i,\ \hat{\Gamma}_c$, as shown in Fig.~\ref{fig:pdenn}. The outputs of the network are restored based on stress scaling in preprocessing. Hence, STPINN enhances the ability of the neural network to learn approximate solution of complex PDEs. 
{\color{black}It should be noted that the structure of STPINN is not just adding neurons and hidden layers over PINN. The proposed STPINN method establishes a constraint for approximating each sub-diffusion process and combining the solutions together to achieve accurate approximation of the stress evolution under complex diffusivities. The proposed STPINN method with multi-channels employs the same MLP to solve each sub-diffusion process, which decreases the demand of the number of trainable weights.}

{\color{black}In order to employ STPINN for EM stochasticity assessment, an atomic diffusivity $D_a^*$ can be obtained for the interconnect tree in each Monte Carlo iteration. 
For Case I of constant temperature analysis, the stress evolution $\sigma^*(x,t)$ with $D_a^*$ can be obtained by:
\begin{equation}\label{eq:emmc1}
\sigma^*(x,t)=\sigma(x,\frac{D_a^*}{D_a}t),
\end{equation}
which can be derived by:
\begin{equation}
    \frac{\partial\sigma^*(x,t)}{\partial t}\frac{\partial t}{\partial t'} = \frac{\partial}{\partial x}\Big[\frac{D_a^*}{D_a}\frac{D_a\Omega}{kT}(\frac{\partial \sigma^*(x,t)}{\partial x}+G)\Big].
\end{equation}
In \eqref{eq:emmc1}, $D_a$ is a determined atomic diffusivity at a constant temperature $T$ used for neural network training and $\sigma(x,t)$ is the stress solution under $D_a$.
For Case II of time-varying temperature, the stress evolution $\sigma^*(x,t)$ with $D_a^*$ can be obtained through 1-STPINN as:
\begin{equation}\label{eq:emmc}
    \sigma^*(x,t) = \mathcal{F}(x,\frac{D_a^*}{D_a}\mathcal{F}_t(t;\alpha);\beta).
\end{equation}
In \eqref{eq:emmc}, $\mathcal{F}(x,t;\beta)$ is trained for the stress evolution analysis with the determined atomic diffusivity $D_a$.
In this way, the proposed STPINN may capture the randomness in EM degradation by employing the trained STPINN model several times under time-varying temperature. 
}

In the next section, experiments will demonstrate the effectiveness of the STPINN structure in solving stress evolution equation under complex temperature.
% PDEs with non-constant diffusivity caused by the non-constant temperature in EM modeling.

% In the next section, experiments will demonstrate the effectiveness of the STPINN structure in solving PDEs with non-constant diffusivity caused by the non-constant temperature in EM modeling.

\section{Experimental Results}\label{4}
\subsection{Experiment Setup}\label{sec:es}
%\begin{figure}[htb]
%	\centering 
%	\subfigure[]{
%		\includegraphics[width=0.55\columnwidth]{casetemp.pdf}
%		\label{fig:case12}}
%	\subfigure[]{
%		\includegraphics[width=0.6\columnwidth]{temp4pi.pdf}
%		\label{fig:case3}} 
%	\caption{Temperature configuration under three different cases: (a) Temperature profile under constant temperature (Case I) and dynamic temperature (Case II), (b) Temperature profile along Cu wires with $20 \mu m$ and $30 \mu m$ via separation under dynamic temperature (Case III).}
%	\label{fig:temp}
%\end{figure}

%In this section, we first demonstrate the temperature effects under three thermal cases, stated in Section~\ref{condition}. The proposed model is compared with the FEM-based COMSOL~\cite{Comsol} in Cases I-III and compact analytical solution method~\cite{Chen2017:TDMR} in Cases I-II. {\color{black}Then, we employ the STPINN in multi-segment interconnect for extension}. The inference model of STPINN is implemented in Python 3.6.2 with Tensorflow 1.12.0, and the compact analytical solution method is implemented in Python 3.6.2. The experiments are carried out on a 3.0-GHz Windows machine with 8-GB of RAM.
%Then we will discuss the impact of network structure configuration on accuracy and time consumption trade-off among different methods. 

We demonstrate the thermal effects on interconnect wires with three cases stated in Section~\ref{condition}. The proposed model is compared with the FEM~\cite{Comsol} and the compact analytical method~\cite{Chen2017:TDMR}. Experimental results will show that the proposed STPINN method can be also applicable to interconnect wires with multi-segments. The proposed STPINN method is implemented in Python 3.6.2 with Tensorflow 1.12.0, and the compact analytical method is implemented in Python 3.6.2. The experiments are carried out on a 3.0-GHz PC with 8-GB of RAM.

%In the training progress, we first perform Adam optimization for $5000$ iterations and then use L-BFGS for the subsequent training. Initial learning rate is $0.001$ together with Xavier's initialization method, and numbers of training data are set to be $N_f=2000,\ N_b=2000,\ N_0=500$ and $N_c=2000$. Scaling factors $\omega_{\sigma},\ \omega_x$ and $\omega_t$ are set to be $1\times 10^{-9},\ 1\times 10^5$ and $1\times 10^{-7}$. The second-order derivable function $\tanh$ is used as the activation function. Geometric and current density configuration of a two-segment interconnect tree is set as:

\begin{figure}[t]
	\centering 
		\includegraphics[width=0.7\columnwidth]{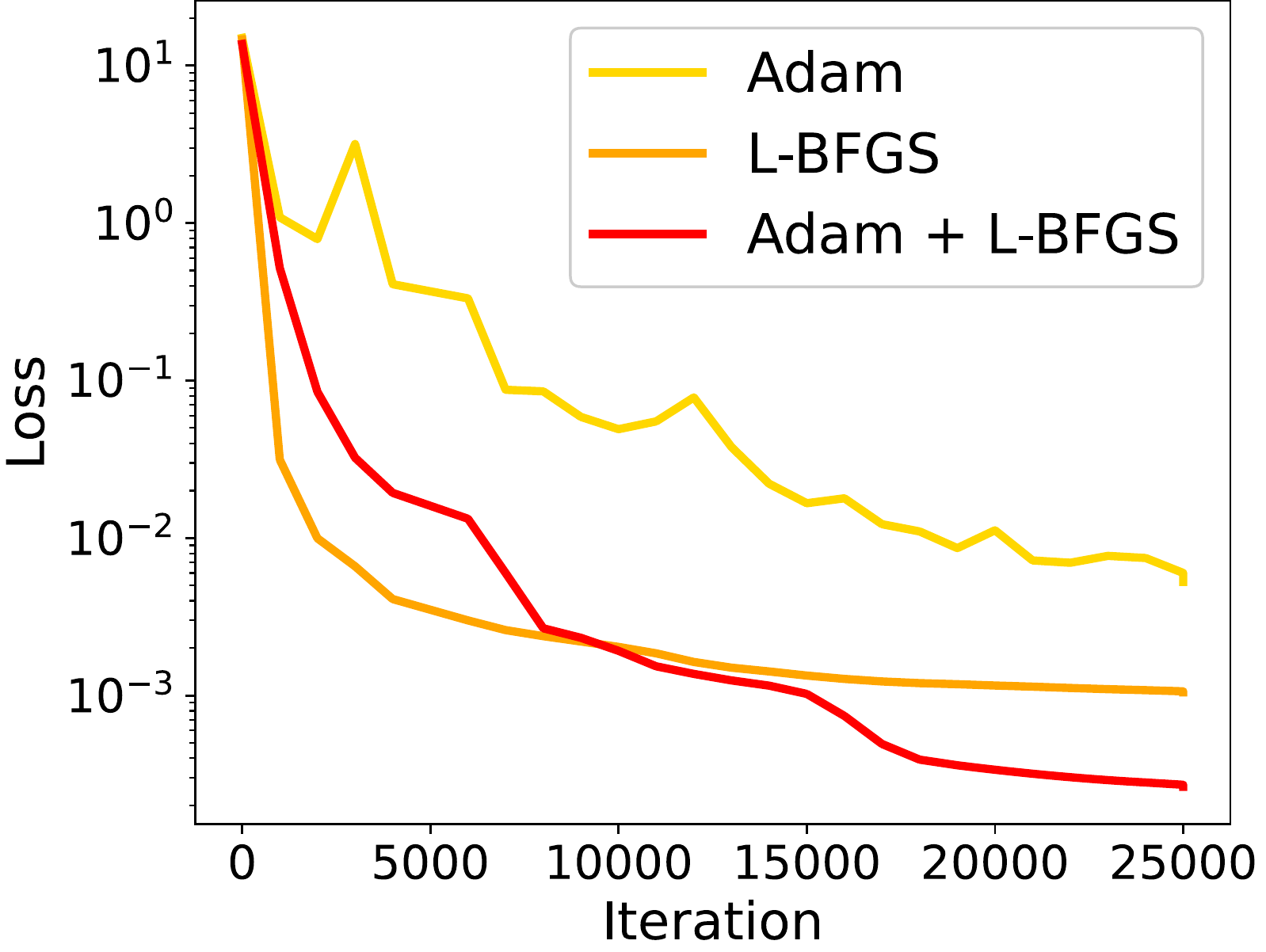}
	\caption{{\color{black}Comparison of loss over the number of iterations by Adam, L-BFGS and the combination of Adam and L-BFGS.}}
	\label{fig:training}
\end{figure}

In the training process, we first perform Adam optimization \cite{Adam2014} for 5k iterations and then use L-BFGS \cite{LBFGS1995} for the subsequent training. {\color{black}We remark that the second-order derivative based L-BFGS can achieve better accuracy with less iterations than the first-order derivative based Adam \cite{DeepXDE2019}. Our experiments show that the combination of Adam and L-BFGS can achieve better training performance, shown in Fig.~\ref{fig:training}. During the training phase, we use the Adam algorithm to calculate the initial weights of L-BFGS for avoiding stacking at a bad local minimum.}
The initial learning rate of Adam is set to be $0.001$ together with Xavier's initialization method, and the numbers of training data are set to be $N_f=25000,\ N_b=N_c=1000,\ N_0=500$, respectively. The scaling factors $\omega_{\sigma},\ \omega_x$ and $\omega_t$ are set to be $1\times 10^{-9},\ 1\times 10^5$ and $1\times 10^{-7}$, respectively. 
% The second-order derivable function $\tanh$ is used as the activation function. 
{\color{black} Since there are multi-order differential operations in the objective and optimization functions, a multi-order differentiable activation function is required for the nonlinear transformation in the proposed STPINN method. In this work, we choose $\tanh$ as the activation function since the experimental results show that the $\tanh$ function has better approximating ability in the EM model than the sigmoid function.}
Geometric and current density configuration of interconnect wire with two segments is set as follows:
\begin{equation}\small
\begin{aligned}
&{\color{black}\mathcal{L}_1}:\{x:0< x < 20\mu m\},\ j_1=4\times 10^{10}A/m^2,\\
&{\color{black}\mathcal{L}_2}:\{x:20\mu m< x < 50\mu m\},\ j_2=-1\times10^{10}A/m^2.
\end{aligned}
\end{equation}

\textcolor{black}{For the comparisons of FEM, PINN, and STPINN, the typical values of the parameters used for calculating the stress evolution of this two-segment structure are shown in Table \ref{tab:parameters}. The original self-diffusion coefficient $D_0$ is to keep the same in this two-segment structure. The effective atomic diffusion coefficient $D_a$ is calculated by the equation \eqref{eq:da}.}

\subsection{Experimental results}\label{experiment}
%In order to discover stress evolution under different temperature conditions, transient stress profiles at certain times under cases I-III are compared in Fig.~\ref{fig:ktkxt}, showing good agreements between the proposed model and COMSOL.
\begin{figure}[t]
	\centering 
	\includegraphics[width=0.7\columnwidth]{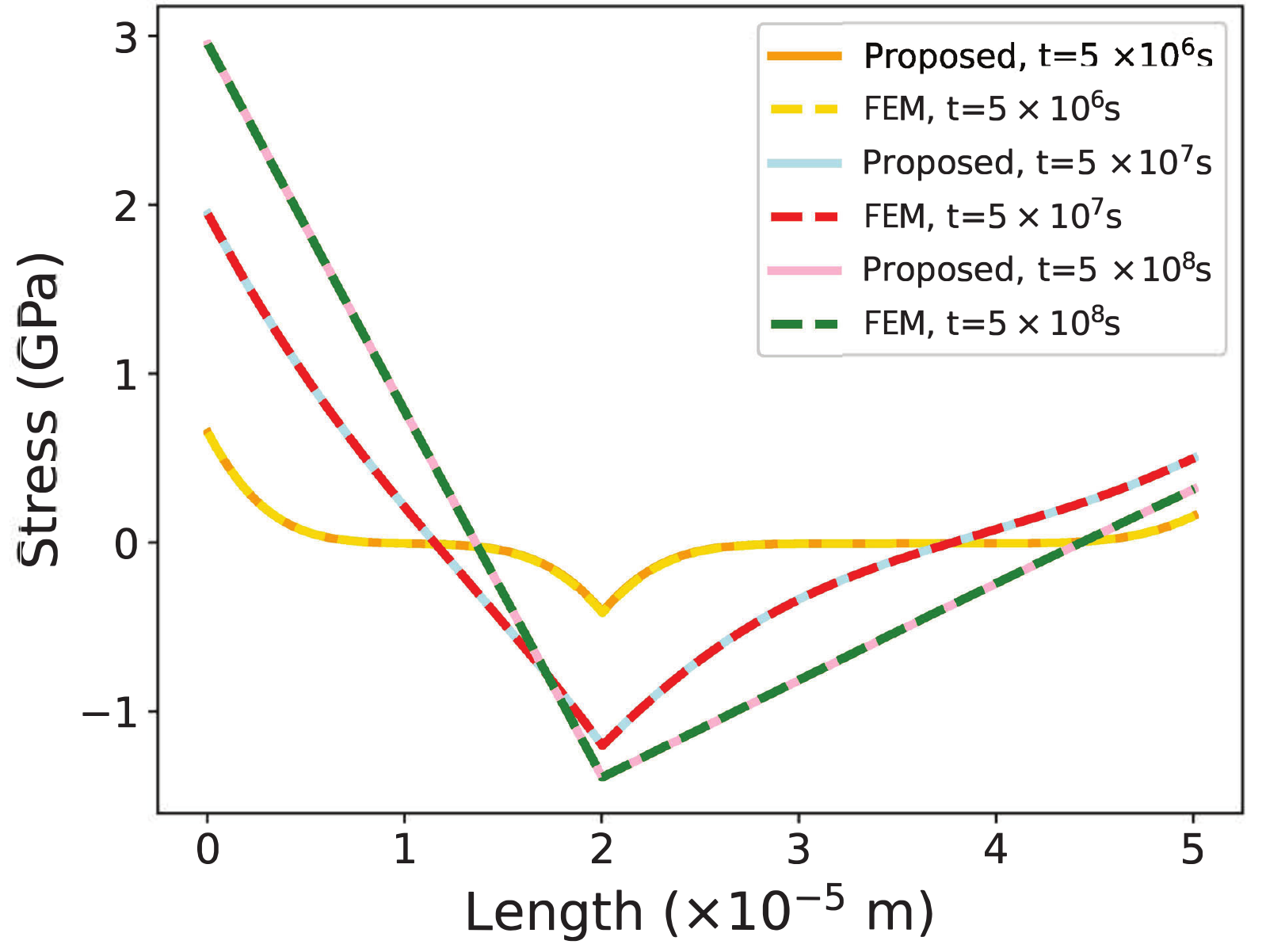}
	\caption{Comparisons of EM stress development under Case I between the proposed model and FEM along two-segment interconnect tree at different times.}
	\label{fig:consacc}
\end{figure}

	\begin{figure}[t]
		\centering 
		\includegraphics[width=0.8\columnwidth]{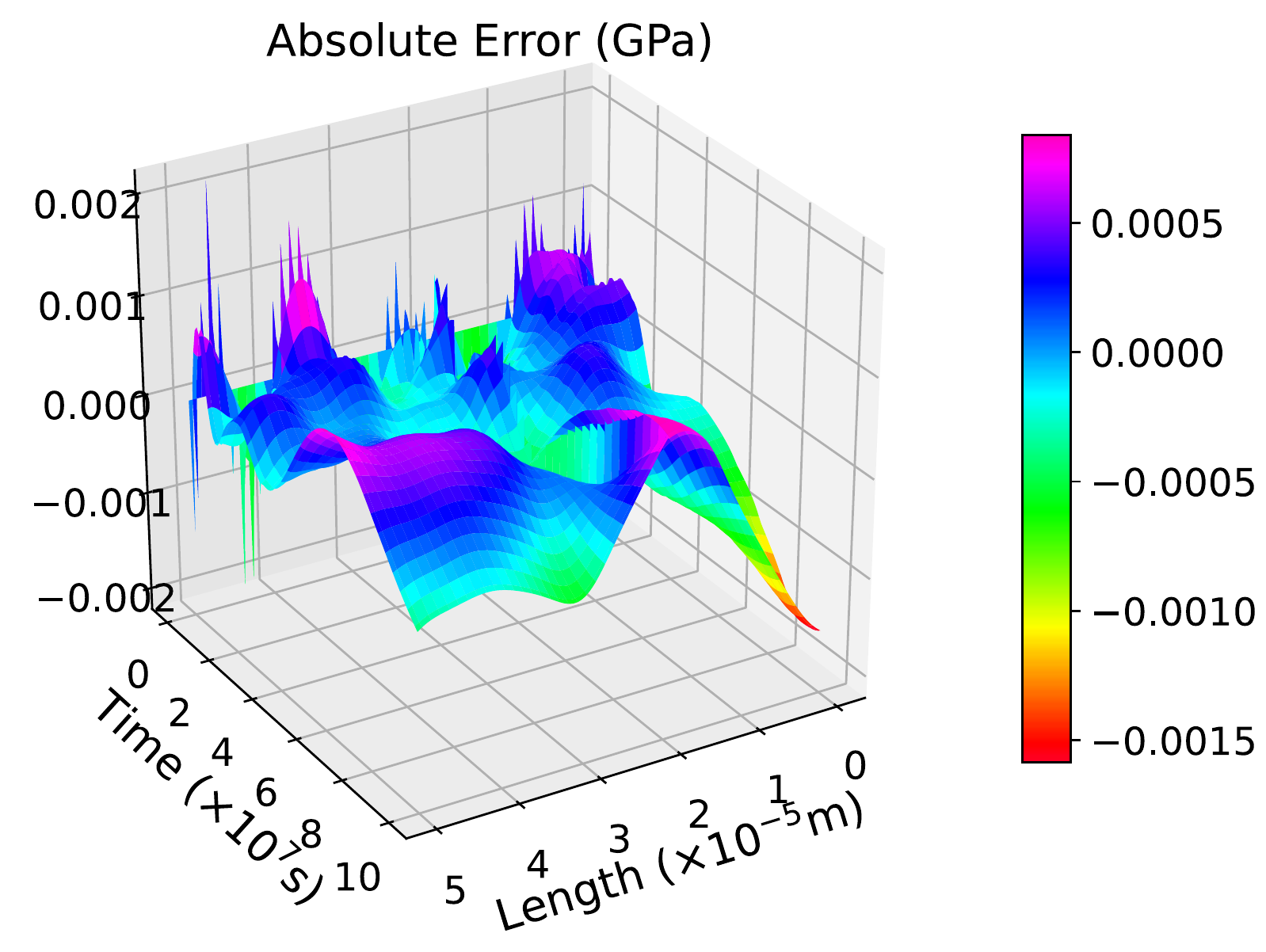}
		\caption{Absolute error of EM stress development under Case I between the proposed model and FEM along two-segment interconnect tree in the time range $0s \sim 10^8 s$.}
		\label{fig:consae}
	\end{figure}

% \begin{figure}[t]
% 	\centering 
% 	\subfigure[]{
% 	\includegraphics[width=0.445\columnwidth]{consacc.pdf}
% 	\label{fig:consacc}} 
% 	\subfigure[]{
% 		\includegraphics[width=0.52\columnwidth]{consae.pdf}
% 		\label{fig:consae}}
% 		\caption{(a) Comparisons of EM stress development under Case I between the proposed model and FEM along two-segment interconnect tree at different times. (b) Absolute error of EM stress development under Case I between the proposed model and FEM along two-segment interconnect tree in the time range $0s \sim 10^8 s$.}		
% 	\label{fig:gradient}
% \end{figure}

%In Case I, the constant temperature is set to be 350K. Based on the Korhonen's equations, we use a PINN consisting of 10 hidden layers with 40 neurons per layer to obtain the stress evolution under constant temperature.
% Based on the Korhonen's equations, when the objective function is optimized to the minimum value, stress evolution under the simplified constant temperature in Case I can be obtained by the PINN. 
\textcolor{black}{In Case I, the constant temperature is set to be 350K in FEM, PINN, and STPINN.} Based on the Korhonen's equation, we use a PINN consisting of 10 hidden layers with 40 neurons per layer for obtaining the stress evolution under constant temperature. It can be seen in Fig.~\ref{fig:consacc} that the approximate stress solutions along the segments at different times obtained by PINN agree well with the numerical results obtained by the FEM.
% less than $0.27\%$ error. 
Fig.~\ref{fig:consae} shows the absolute error defined by $|\sigma_{\rm FEM}-\sigma|$, illustrating that the absolute error is mainly in the range $-0.0015 GPa \sim 0.0005 GPa$ and the amplitude of error grows near the terminals with the increase of time. The distribution of error fits well with the trend of hydrostatic stress evolution. {\color{black} The results demonstrate the promising capability of PINN in solving stress evolution under constant temperature. However, PINN cannot well capture the EM-induced stress evolution under space-time related temperature, which will be discussed in Section~\ref{performance}.}
% (\textcolor{blue}{numbers are better put in \$\$})
%However, in system-level thermal reliability analysis, the constant temperature in Black's model~\cite{Black1969} eliminates the thermal impact in time and location~\cite{zlu2005:ieeem, XHuang2014:DAC}. 
%Research in~\cite{Chen2017:TDMR} has proposed an EM analysis model suitable for dynamic temperature condition.

\begin{figure}[t]
	\centering 
	%	\subfigure[]{
	\includegraphics[width=0.7\columnwidth]{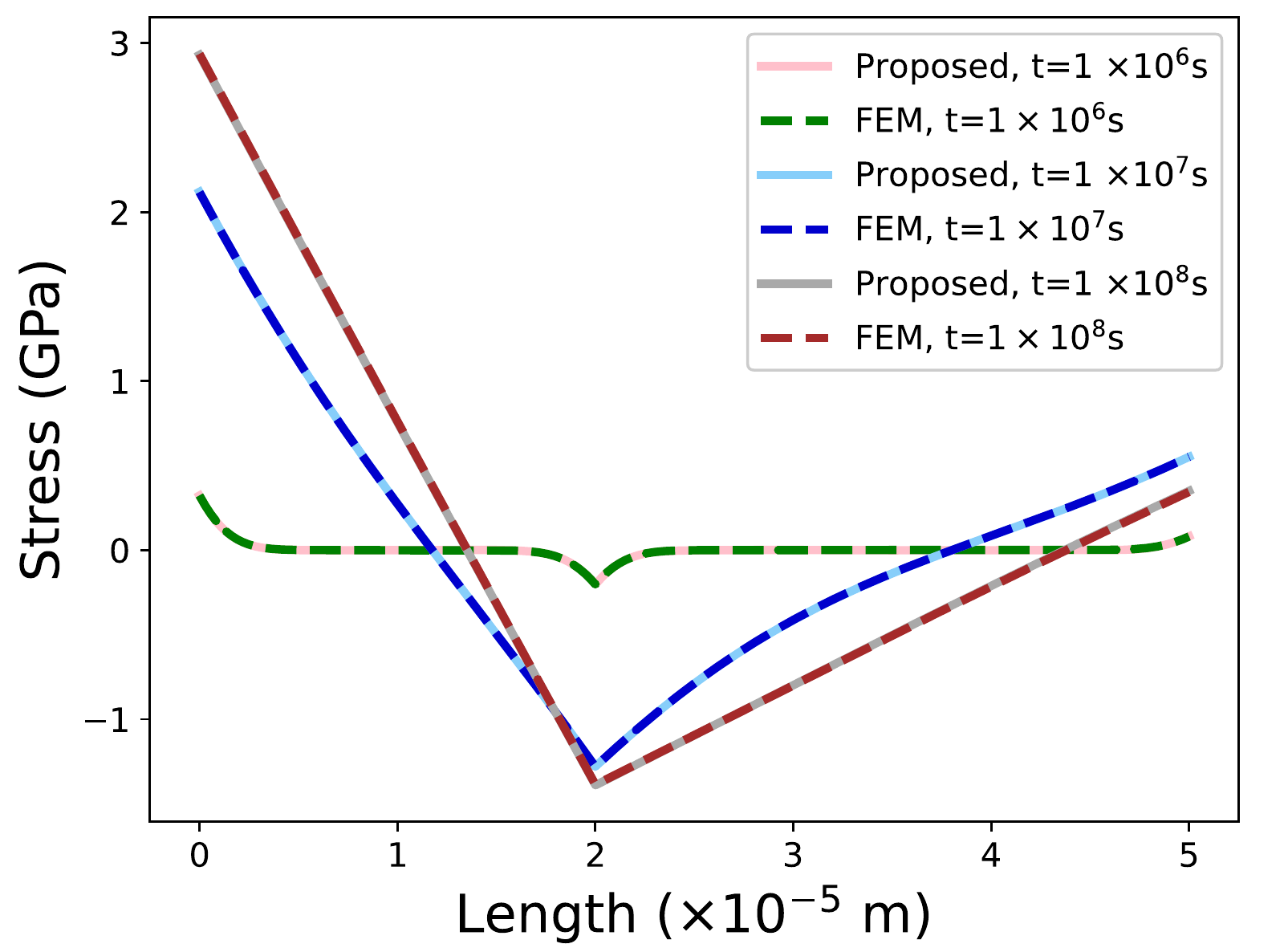}
	%	\subfigure[]{
	%		\includegraphics[width=0.45\columnwidth]{pictt.pdf}
	%		\label{fig:pictt}}
	\caption{Comparisons of EM stress development under Case II between the proposed model and FEM along two-segment interconnect tree at different time.}
		\label{fig:ktacc} 
\end{figure}

\begin{figure}[t]
	\centering 
	\includegraphics[width=0.8\columnwidth]{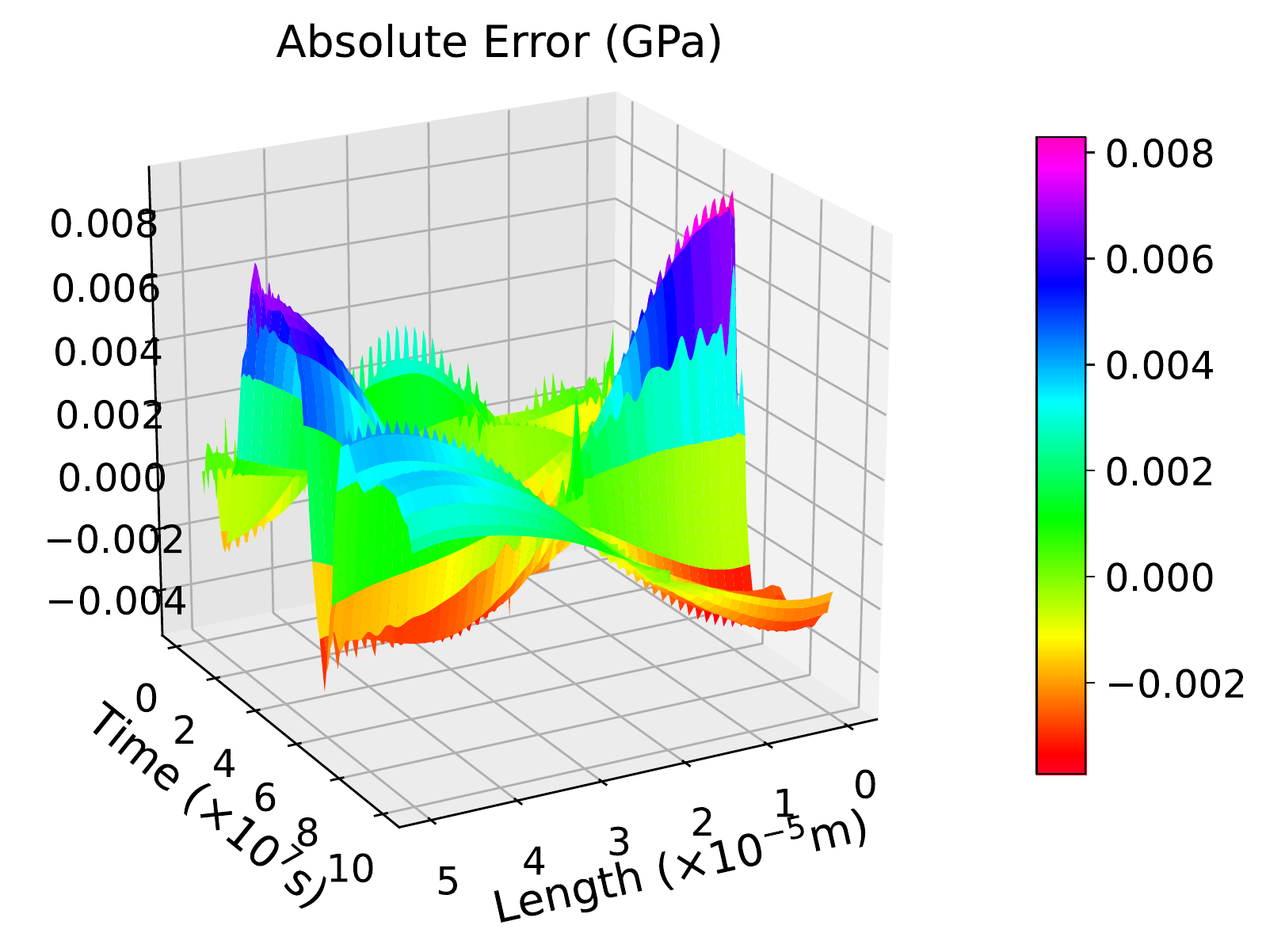}
	\caption{Absolute error of EM stress development under Case II between the proposed model and FEM along two-segment interconnect tree in the time range $0s \sim 10^8 s$.}
	\label{fig:ktae}
\end{figure}

%In Case II, the dynamic temperature is configured as $T=[350+30\times \sin(4\times10^{-8}\pi t)]K$. We employ a one-channel STPINN (1-STPINN) consisting of 1 hidden layer with 100 neurons per layer in $\mathcal{F}_t(t;\alpha)$ and 10 hidden layers with 40 neurons per layer in $\mathcal{F}(x,t;\beta)$ in the case test. Moreover, the first FNN $\mathcal{F}_t$ represents the integration operation of temporal variable $t$ to convert the time-dependent diffusivity towards constant diffusivity in the PDE. Both comparison and absolute error between results obtained by FEM and STPINN are shown in Fig.~\ref{fig:ktacc} and Fig.~\ref{fig:ktae}. The relative error is less than 1.22\% and the absolute error is mainly oscillating from -0.002 GPa to 0.008 GPa. The relatively large error is distributed near the terminals of the metal interconnect.

\textcolor{black}{In Case II, the dynamic temperature is configured as $T=[350+30\times \sin(4\times10^{-8}\pi t)]K$, which is used in the simulations of both FEM and STPINN.} We employ a one-channel STPINN (1-STPINN) consisting of 1 hidden layer with 100 neurons per layer in $\mathcal{F}_t(t;\alpha)$ and 10 hidden layers with 40 neurons per layer in $\mathcal{F}(x,t;\beta)$ for this case. The first FNN $\mathcal{F}_t$ represents the integration operation of temporal variable $t$ for converting the time-dependent diffusivity towards constant diffusivity in the stress evolution equation. The comparison results between the FEM and the proposed STPINN method are shown in Figs.~\ref{fig:ktacc} \&~\ref{fig:ktae}. The relative error is less than 1.22\% and the absolute error is mainly oscillating from $-0.002\ GPa$ to $0.008\ GPa$. The relatively large error is distributed near the terminals of the interconnect wire.

\begin{figure}[t]
	\centering 
		\includegraphics[width=0.8\columnwidth]{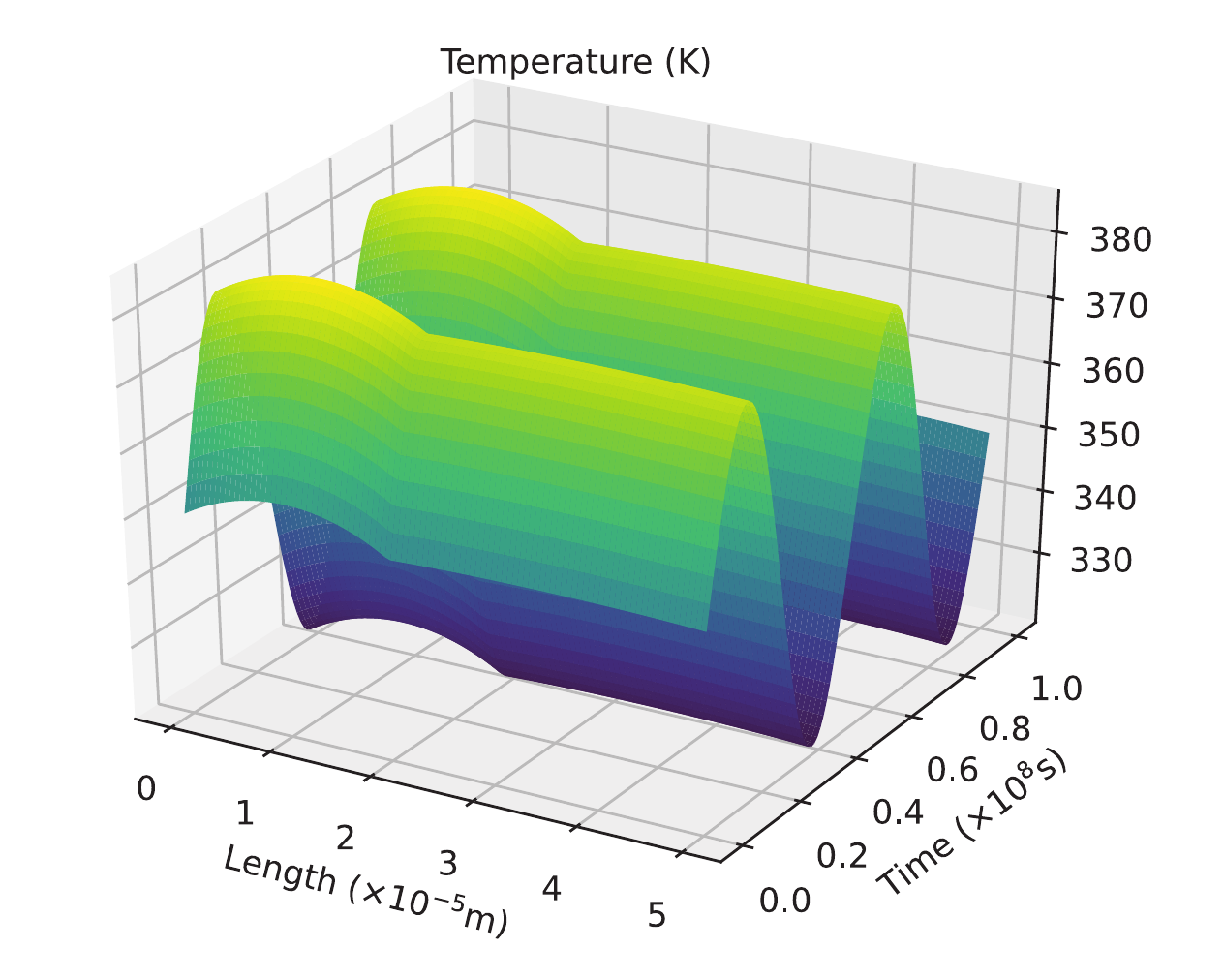}
	\caption{Temperature profile along Cu wires with $20 \mu m$ and $30 \mu m$ via separation under dynamic temperature (Case III).}
	\label{fig:temp}
\end{figure}

\begin{figure}[t]
	\centering 
	%	\subfigure[]{
	\includegraphics[width=0.7\columnwidth]{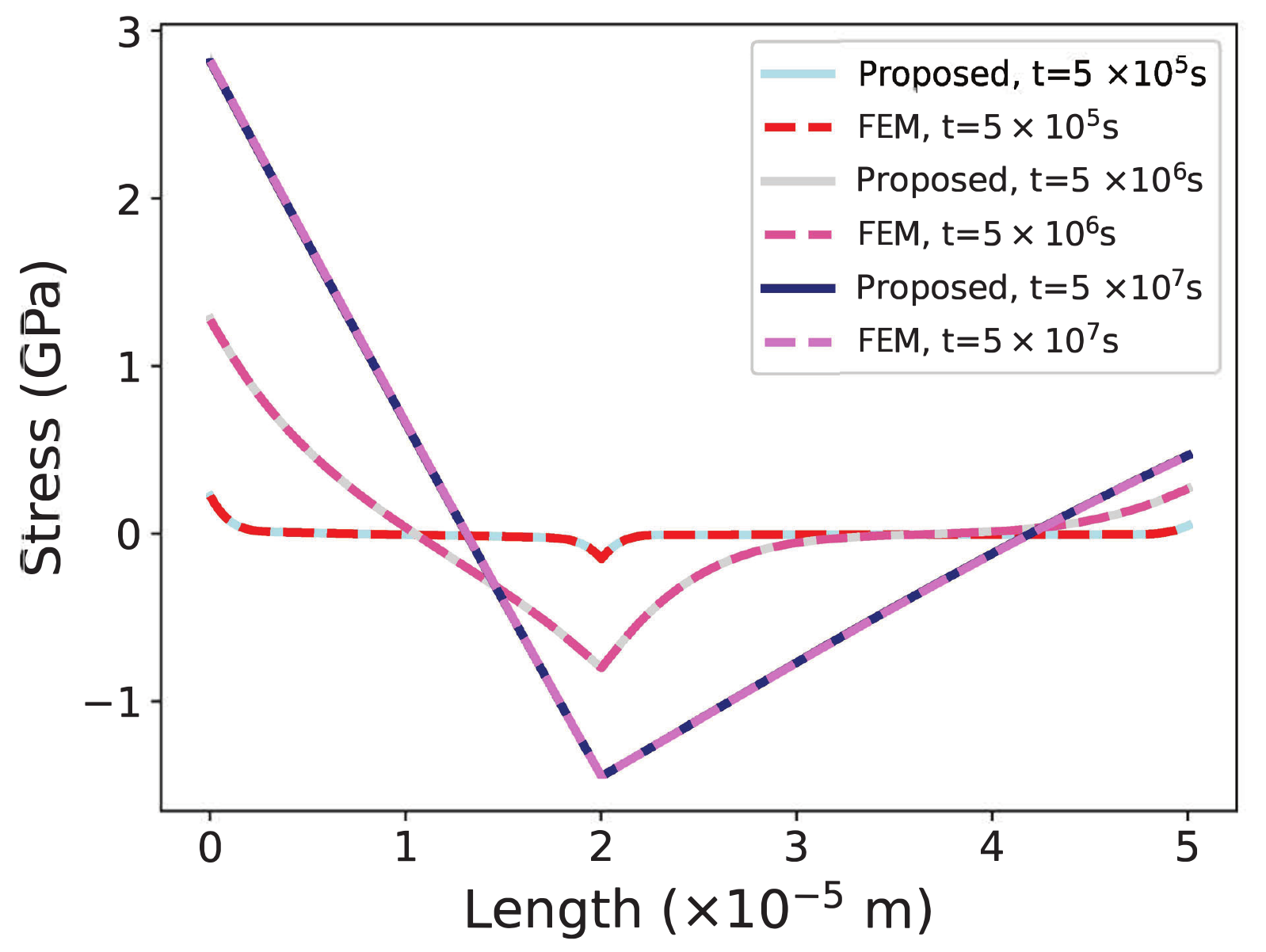}
	%	\subfigure[]{
	%		\includegraphics[width=0.45\columnwidth]{pict.pdf}
	%		\label{fig:pict}}
	\caption{Comparisons of EM stress development under Case III between the proposed model and FEM along two-segment interconnect tree at different time.}
	\label{fig:kxtacc}
\end{figure}

\begin{figure}[t]
	\centering 
	\includegraphics[width=0.8\columnwidth]{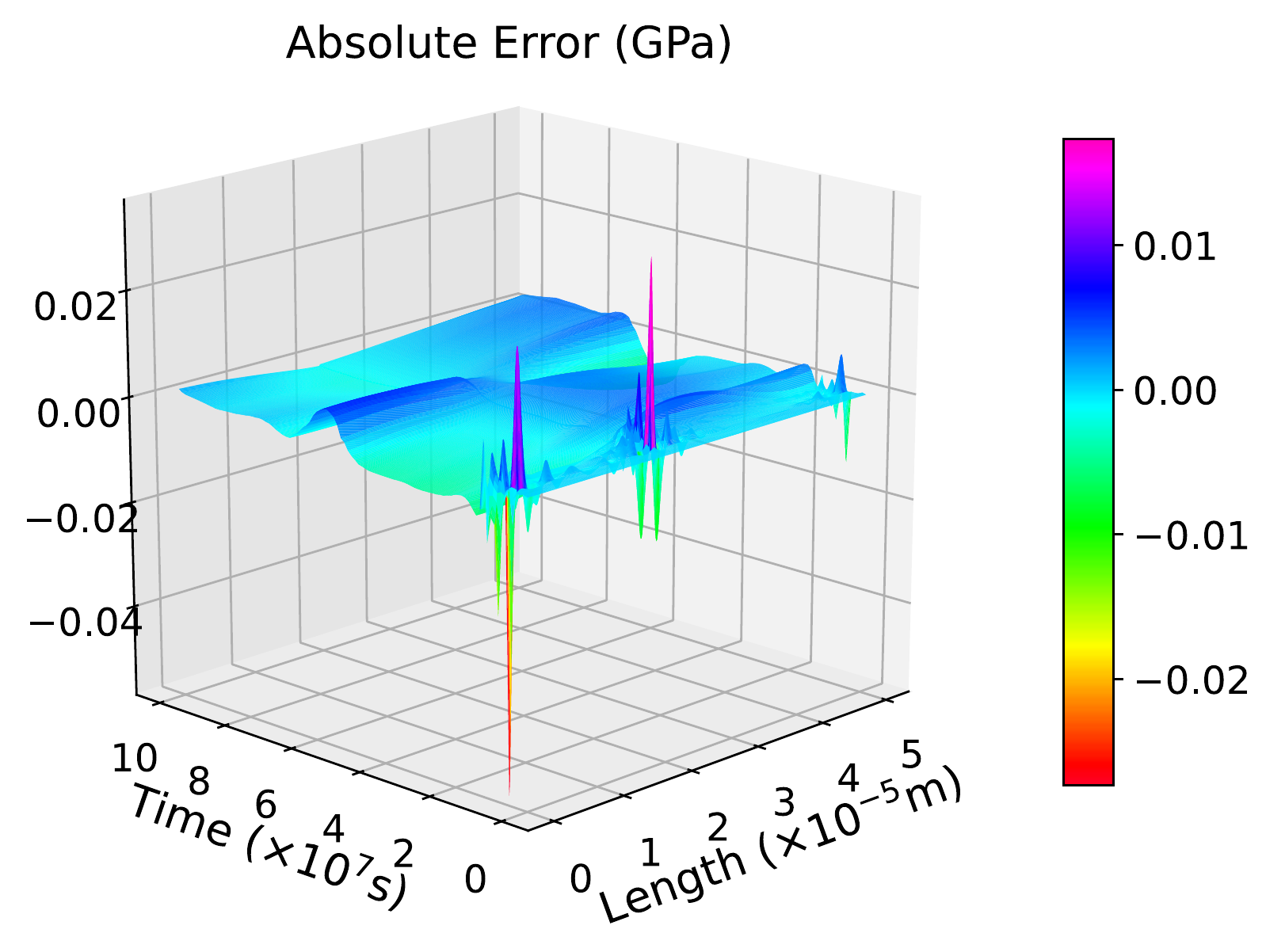}
	\caption{Absolute Error of EM stress development under Case III between the proposed model and FEM along two-segment interconnect tree in the time range $0s \sim 10^8 s$.}
	\label{fig:kxtae}
\end{figure}

%In Case III, we combine temperature configuration in Case II and thermal equation \eqref{eq:thermal}, where we set $H=t_{I\!L\!D}=0.8\mu m,\ w=d=0.3\mu m$. The temperature profile is shown in Fig.~\ref{fig:temp}. We assume low-k insulator ${\rm SiO_2}$ ($k_{oxide}=1.2W/mK$) as ILD, which is more deeply affected by via effect. The two-channel STPINN (2-STPINN) we used consists of 2 hidden layers with 50 neurons per layer in $\mathcal{F}_t(t;\alpha)$ and 10 hidden layers with 40 neurons per layer in $\mathcal{F}(x,t;\beta)$. It should be noted that PINN has poor approximating performance in Case III, which will be discussed in Section~\ref{performance}.
%The stress profile is shown in Fig.~\ref{fig:kxtacc}, illustrating good agreement with 0.80\% error at $t=5\times 10^5s,\ 5\times10^6s,\ 5\times10^7s$. The absolute error distribution shown in Fig.~\ref{fig:kxtae} indicates a sharp and large error fluctuation at junctions near the initial time. {\color{black} Due to the complex diffusivity, the accuracy of numerical solution decreases when configuring regular mesh grids and time steps in FEM.}

\textcolor{black}{In Case III, for the comparisons of FEM and STPINN, the space-time related temperature is configured as the thermal model \eqref{eq:thermal} where we set $H=t_{I\!L\!D}=0.8\mu m,\ w=d=0.3\mu m$ and $T_0(t)=[350+30\times \sin(4\times10^{-8}\pi t)]K$.} The temperature profile is shown in Fig.~\ref{fig:temp}. We assume low-k insulator ${\rm SiO_2}$ ($k_{oxide}=1.2W/mK$) as ILD, which is more deeply affected by via effect. For this case, we use the two-channel STPINN (2-STPINN) consisting of 2 hidden layers with 50 neurons per layer in $\mathcal{F}_t(t;\alpha)$ and 10 hidden layers with 40 neurons per layer in $\mathcal{F}(x,t;\beta)$. 
% It should be noted that PINN has poor approximating performance in Case III,
% {\color{black} It should be noted that PINN has unsatisfactory approximating performance as the complexity of the relationship between temperature and space-time variables increases,}
% which will be discussed in Section~\ref{performance}.
The stress evolution profile is shown in Fig.~\ref{fig:kxtacc}, illustrating good agreement with less than 0.80\% error at $t=5\times 10^5s,\ 5\times10^6s,\ 5\times10^7s$. The absolute error distribution shown in Fig.~\ref{fig:kxtae} indicates a relatively sharp and large error fluctuation at the junctions near the initial time. Due to the complex diffusivity, the accuracy of numerical solution decreases when configuring regular mesh grids and time steps in the FEM.

%Therefore, the error analysis involves the self deviations in the FEM and the proposed method.

%\begin{figure}[htbp]
%	\centering 
%	\includegraphics[width=0.9\columnwidth]{mt.pdf}
%	\caption{Absolute Error of EM stress development under Cases II between proposed model and COMSOL along two-segment interconnect tree in the time range $0s \sim 10^8 s$.}
%	\label{fig:mt}
%\end{figure}

\begin{figure}[t]
	\centering 
	\subfigure[]{
		\includegraphics[width=0.465\columnwidth]{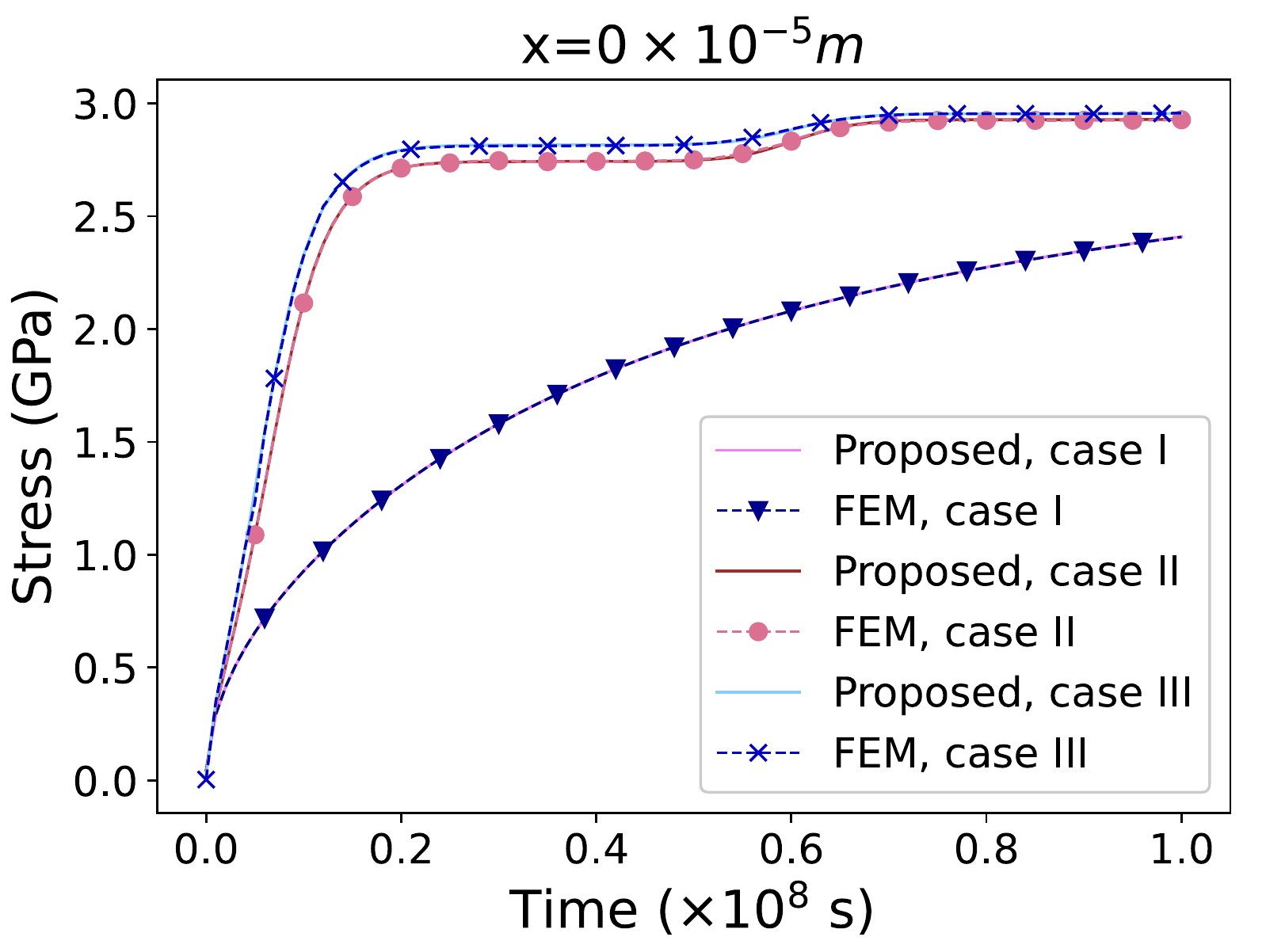}
		\label{fig:lt}} 
	\subfigure[]{
		\includegraphics[width=0.465\columnwidth]{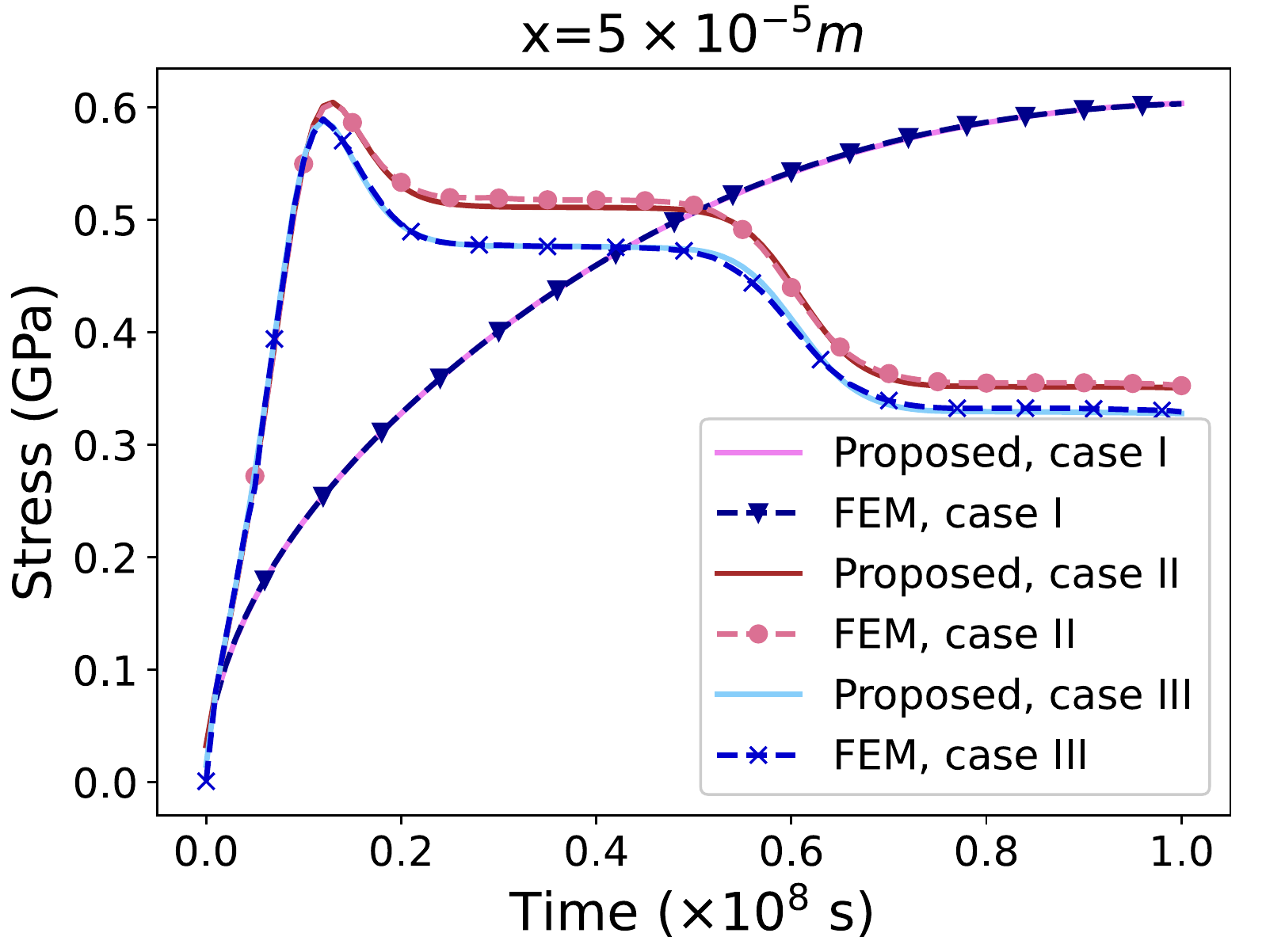}
		\label{fig:rt}}
	\caption{Comparisons of EM stress development under Cases I-III between the proposed model and FEM in time range from $0$s to $1\times10^8$s at: (a) left terminal, (b) right terminal.}
	\label{fig:t}
\end{figure}

\begin{table*}[h]
	\centering
	\caption{Accuracy and computation time comparison between STPINN, compact analytical solution method and FEM. W.S. represents Whole-Stress at 10 specified time points and J.S. represents Junction-Stress.
	}
\setlength{\tabcolsep}{3.5mm}{
\begin{tabular}{|c|c|c|c|c|c|c|c|c|c|c|c|}
\hline
\multirow{3}{*}{\textbf{Case}} & \multicolumn{4}{c|}{\textbf{STPINN}} & \textbf{FEM} & \multicolumn{4}{c|}{\textbf{Chen et al.(2016)}} & \multirow{3}{*}{\textbf{\begin{tabular}[c]{@{}c@{}}Speed\\ -up\\ Chen\end{tabular}}} & \multirow{3}{*}{\textbf{\begin{tabular}[c]{@{}c@{}}Speed\\ -up\\ FEM\end{tabular}}} \\ \cline{2-10}
 & \multirow{2}{*}{\textit{\textbf{Level}}} & \multicolumn{2}{c|}{\textit{\textbf{Error (\%)}}} & \textit{\textbf{\begin{tabular}[c]{@{}c@{}}Time (s)\end{tabular}}} & \multirow{2}{*}{\textit{\textbf{\begin{tabular}[c]{@{}c@{}}Time\\ (s)\end{tabular}}}} & \multirow{2}{*}{\textit{\textbf{Iteration}}} & \multicolumn{2}{c|}{\textit{\textbf{Error (\%)}}} & \textit{\textbf{Time (s)}} &  &  \\ \cline{3-5} \cline{8-10}
 &  & \textit{\textbf{W.S}} & \textit{\textbf{J.S}} & \textit{\textbf{W.S}} &  &  & \textit{\textbf{W.S}} & \textit{\textbf{J.S}} & \textit{\textbf{W.S}} &  &  \\ \hline
\multirow{2}{*}{II} & \multirow{2}{*}{1} & \multirow{2}{*}{1.2240} & \multirow{2}{*}{0.1703} & \multirow{2}{*}{0.3227} & \multirow{2}{*}{5} & 100 & 2.0749 & 1.1129 & 0.9595 & 2.97$\times$ & \multirow{2}{*}{15.59$\times$} \\ \cline{7-11}
 &  &  &  &  &  & 1000 & 0.9845 & 0.4943 & 1.0082 & 3.12$\times$ &  \\ \hline
III & 2 & 0.7976 & 0.1751 & 0.5027 & 14 & - & - & - & - & - & 27.84$\times$ \\ \hline
\end{tabular}}
\label{tab:solution}
\end{table*}

In order to illustrate the stress evolution under different temperature conditions, we compare the transient stress profiles at each junction under Cases I-III shown in Figs.~\ref{fig:lt}-\ref{fig:rt}, which also demonstrate good agreements between the proposed model and the FEM. It can be seen that when taking the thermal conditions with the same mean value into consideration, the stress of Cases II-III reach the steady-state earlier than the stress of Case I. As shown in Fig.~\ref{fig:lt}, the stress under Case III always maintains a higher value than the stress under Case II at the left terminal. Fig.~\ref{fig:rt} shows that at the right terminal of the interconnect wire, the stress under Cases II-III remains almost the same growth rate near the initial time and then decreases in a similar trend, while the stress under Case III commences falling earlier. The consideration of Joule heating provides a more effective estimation in EM-based reliability analysis during the void nucleation phase. The higher error in Case II is mainly due to the impact of time step selection in the FEM, which will be discussed in Section~\ref{performance}.

%Furthermore, STPINN shows better performance compared with existing methods, which requires meshing or iteration to solve PDEs at a specified space-time point. Table~\ref{tab:solution} compares the proposed STPINN method with the FEM and compact analytical solution model in the accuracy analysis. The compact analytical solution model only provides stress analysis in Case II in EM simulation~\cite{Chen2017:TDMR}. The table also illustrates the computation time of STPINN and existing methods. Here, time and the junction stress accuracy represent computation time and mean relative error between the proposed method and existing methods in 1k time steps at interconnect tree junction.
%where COMSOL is configured with $1\times 10^{-7}m$ mesh size and $1\times 10^5 s$ time step size. 

% The proposed STPINN method can achieve better performance compared with existing methods that require meshing or iteration for solving PDEs. 
The comparisons of the proposed STPINN method, FEM and the compact analytical method \cite{Chen2017:TDMR} are shown in Table~\ref{tab:solution} in terms of the accuracy and the computation time.
% In this table, the computation time and 
The error of STPINN represents the mean relative error of solutions between STPINN and FEM performing 1k iterations. To further demonstrate the effectiveness of the proposed STPINN, the solutions achieved from STPINN are compared against the results obtained by the compact analytical method under different numbers of iterations.
%of calculating the time period.
The junction stress accuracy represents the error at the junction of the interconnect tree in the time range $0\ s\sim10^8\ s$ and the whole stress accuracy presents the error at 10 specified time instances along the wire in the time range $10^5\ s\sim10^8\ s$. 
It should be noted that the compact analytical method mainly focuses on the EM-based stress analysis for Case II. 
When calculating the whole stress in the compact analytical method, the mean relative error reduces with increased iterations at the cost of computation time. 
% The junction locates at the cathodes of the two segments where the maximum tensile stress occurs. 
It can be seen from Table~\ref{tab:solution} that the proposed STPINN method can achieve a speedup of $2\times \sim 52\times$ over the existing methods with the mean relative error about $1.2\%$.

\subsection{Performance Analysis}\label{performance}
%To analyze the performance of STPINN, Table~\ref{tab:f} and Table~\ref{tab:ft} show the following systematic studies of network configuration, demonstrating the relative errors in the $\mathbb{L}_2$ norm between the STPINN and FEM method when architectures of STPINN ($\mathcal{F}(x,t;\beta)$ and $\mathcal{F}_t(t;\alpha)$) are set with different layers and neurons under Case III. As shown in Table~\ref{tab:f}, for network $\mathcal{F}(x,t;\beta)$ with little hidden layers, the increase of neurons per hidden layer will not reduce the relative error and not speed up the convergence. When the number of layers is added to 9 and 11, the error shows its downtrend with increased neuron number. On the contrary, raising the layer number in $\mathcal{F}_t(t;\alpha)$ does not achieve better performance. As shown in Table~\ref{tab:ft}, although error shows a decline as neurons per layer increase, the network with 2 hidden layers seems to perform best among the comparison of architectures with 1, 2, 3 layers. We summarize the results of the systematic studies above and fix the network architecture to 2 hidden layers with 50 neurons per layer in $\mathcal{F}_t(t;\alpha)$ and 10 hidden layers with 40 neurons per layer in $\mathcal{F}(x,t;\beta)$. 

For analyzing the performance of the proposed STPINN method, Tables~\ref{tab:f} \&~\ref{tab:ft} show the following systematic studies of the network configuration, demonstrating the relative errors in the $\mathbb{L}_2$ norm between the STPINN method and the FEM when the architectures of STPINN ($\mathcal{F}(x,t;\beta)$ and $\mathcal{F}_t(t;\alpha)$) are set with different layers and neurons under Case III. As shown in Table~\ref{tab:f}, for the network $\mathcal{F}(x,t;\beta)$ with few hidden layers, the increase of neurons per hidden layer will neither reduce the relative error nor accelerate the convergence. When the number of layers rises to 9 and 11, the relative error shows its downtrend with increased neurons. On the contrary, experimental results show that raising the number of layer in $\mathcal{F}_t(t;\alpha)$ does not achieve better performance when the number of layers is larger than two. 
% As shown in Table~\ref{tab:ft}, the relative error appears downtrend with the growth of the number of neurons per layer and the network with two hidden layers performs best in the architectures with one, two, three layers. 
{\color{black}As a result, we can fix the network architecture to 2 hidden layers with 50 neurons per layer in $\mathcal{F}_t(t;\alpha)$ and 11 hidden layers with 40 neurons per layer in $\mathcal{F}(x,t;\beta)$ for stress analysis in the two-segment wire. It can be observed from Tables~\ref{tab:f} \&~\ref{tab:ft} that changing the structure of $\mathcal{F}(x,t;\beta)$ has a greater impact on the accuracy compared with adjusting network structure of $\mathcal{F}_t(t;\alpha)$. The width and depth of the proposed STPINN architecture is problem dependent on various wire structures. For multi-segment wires, we mainly focus on adjusting the structure of $\mathcal{F}(x,t;\beta)$. We prefer to first change the width of $\mathcal{F}(x,t;\beta)$ and then the depth when adjusting $\mathcal{F}(x,t;\beta)$ in the experiments.}

{\color{black} In Table~\ref{tab:sample}, we report the relative $\mathbb{L}_2$ error of STPINN under different number of initial collocation points $N_0$, boundary collocation points $N_b$, $N_c$, and internal collocation points $N_f$. Here, we set $N_b=N_c=N_0$. It is the general trend that the error is decreased as $N_f$ is increased, given sufficient training points for initial and boundary conditions. 
% The number of collocation points $N_f$ should also be adjusted through the smoothness of the diffusivity so that the neural network can accurately learn the features of the spatial-temporal related diffusion coefficient. 
For various multi-segment wires, we mainly focus on adjusting the number of internal collocation points $N_f$ according to the segment structure as well as the space-time dependent stress diffusivity for achieving satisfactory loss. During the training phase, the cross-validation method can be employed to monitor whether the network is overfitting. }
\begin{table}[t]
	\caption{Relative prediction error with respect to FEM in the $\mathbb{L}_2$ norm for different number of hidden layers and neurons per layer in $\mathcal{F}(x,t;\beta)$. Here, the training step for L-BFGS is 20k and $\mathcal{F}_t(t;\alpha)$ is set as 3 hidden layers with 50 neurons.}
	\centering
	\setlength{\tabcolsep}{2.3mm}{
\begin{tabular}{|c|c|c|c|c|}
\hline
 \textbf{\diagbox {Layers}{Neurons}}& \multicolumn{1}{c|}{\textbf{10}} & \multicolumn{1}{c|}{\textbf{20}} & \multicolumn{1}{c|}{\textbf{30}} & \multicolumn{1}{c|}{\textbf{40}} \\ \hline
\textbf{5} & 1.949e-2 & 1.324e-2 & 1.583e-3 & 5.470e-3 \\ \hline
\textbf{7} & 2.749e-2 & 1.676e-3 & 6.566e-3 & 6.678e-3 \\ \hline
\textbf{9} & 5.700e-3 & 5.261e-3 & 4.441e-3 & 7.998e-4 \\ \hline
\textbf{11} & 1.025e-2 & 6.539e-3 & 3.876e-3 & 2.650e-4 \\ \hline
\end{tabular}}
	\label{tab:f}
\end{table}
\begin{table}[t]
	\caption{Relative prediction error with respect to FEM in the $\mathbb{L}_2$ norm for different number of hidden layers and neurons per layer in $\mathcal{F}_t(t;\alpha)$. Here, the training step for L-BFGS is 20k and $\mathcal{F}(x,t;\beta)$ is set as 9 hidden layers with 40 neurons.}
	\centering
		\setlength{\tabcolsep}{3.8mm}{
\begin{tabular}{|c|c|c|c|}
\hline
\textbf{\diagbox {Layers}{Neurons}} & \textbf{10} & \textbf{30} & \textbf{50} \\ \hline
\textbf{1} & 4.196e-3 & 5.710e-3 & 4.294e-3 \\ \hline
\textbf{2} & 2.871e-3 & 5.149e-3 & 1.230e-4 \\ \hline
\textbf{3} & 1.050e-3 & 6.411e-3 & 7.998e-4 \\ \hline
\end{tabular}}
	\label{tab:ft}
\end{table}

\begin{table}[t]
	\caption{{\color{black}The relative prediction error with respect to FEM in the $\mathbb{L}_2$ norm under different number of collocation points $N_0,\ N_b,\ N_c,\ N_f$. We set $N_b=N_c=N_0$ and the training step for L-BFGS is set to be 20k.}}
	\centering
		\setlength{\tabcolsep}{1.64mm}{
\begin{tabular}{|c|c|c|c|c|}
\hline
\textbf{\diagbox {$N_b,N_c,N_0$}{$N_f$}} & \textbf{2000} & \textbf{4000} & \textbf{6000} & \textbf{8000} \\ \hline
\textbf{300} & 5.453e-3 & 4.699e-3 & 5.524e-3 & 8.446e-3 \\ \hline
\textbf{500} & 5.471e-3 & 6.160e-4 & 2.070e-3 & 1.956e-3 \\ \hline
\textbf{700} & 4.287e-3 & 3.891e-3 & 2.919e-3 & 2.890e-4 \\ \hline
\end{tabular}}
	\label{tab:sample}
\end{table}

%We further evaluate the impact of the number of channels in STPINN in performance and compare STPINN against PINN for performance analysis of the same problem. The PINN is configured in the same 10-layer network structure with 40 neurons per layer as the $\mathcal{F}(x,t;\beta)$ in STPINN. The time mapping network $\mathcal{F}_t(t;\alpha)$ keeps the 2-layer architecture with 50 neurons per layer. Both methods employ the same optimization in training. Table~\ref {tab:channel analysis} shows the accuracy and runtime for stress evolution in Case III obtained by PINN, one-channel STPINN and two-channel STPINN after 30k steps of L-BFGS training.
%In the whole stress analysis, 5k equidistant points along segments are sampled in both network methods and FEM calculation while in junction stress analysis, 1k isometric time steps are recorded. Models in Table~\ref{tab:channel analysis} are trained for 30k iterations of L-BFGS.

\begin{figure}[t]
	\centering 
		\includegraphics[width=0.7\columnwidth]{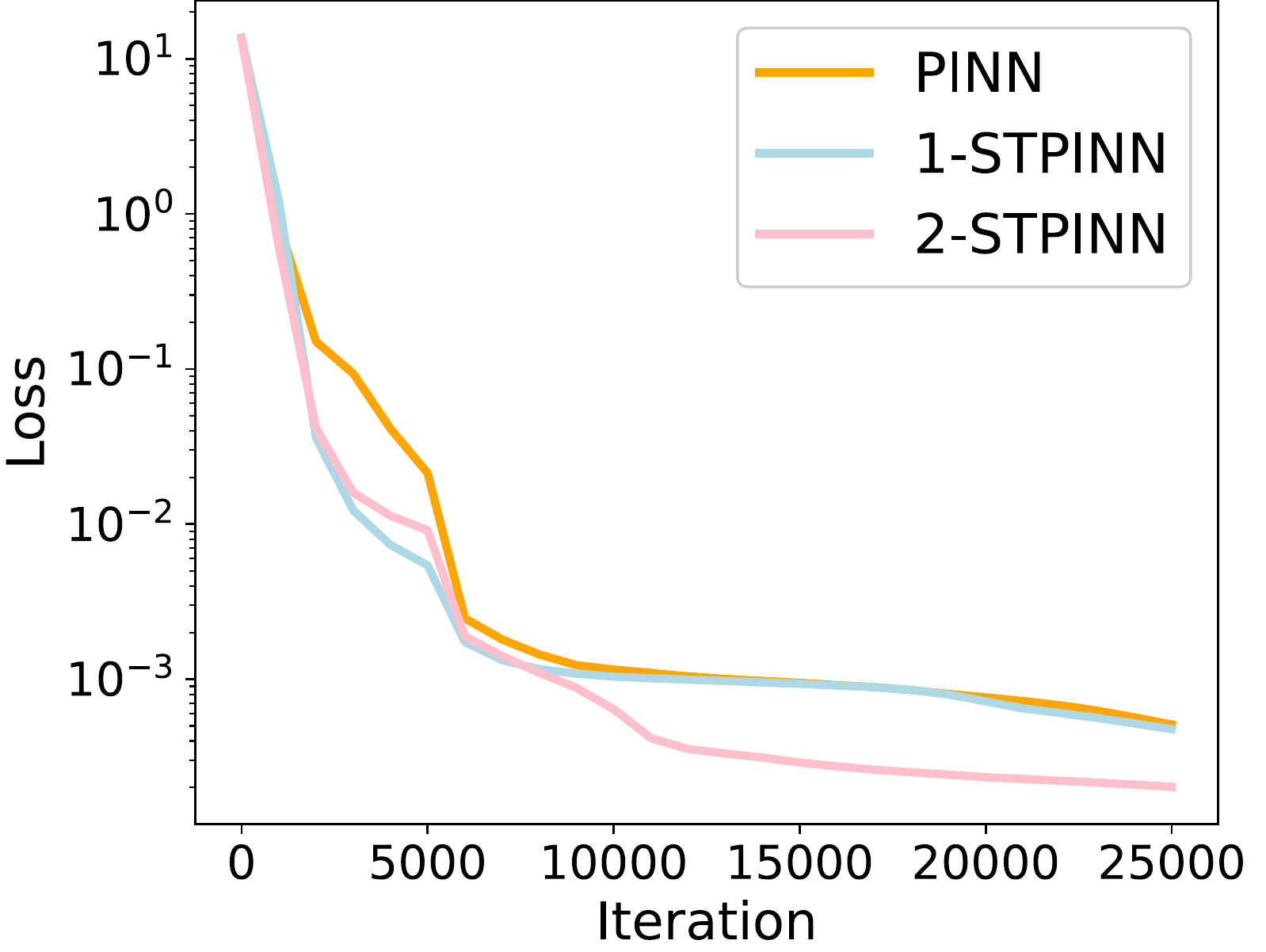}
	\caption{\color{black}Change process of the objective function over the number of training steps by PINN, 1-STPINN, 2-STPINN.}
	\label{fig:iteration}
\end{figure}

We further evaluate the impact of the number of channels in the proposed STPINN method for calculating the EM-based stress evolution. We also compare STPINN against PINN in terms of the accuracy and the computation time. 
{\color{black}The PINN method is configured in the 10-layer network structure with 40 neurons per layer, and $\mathcal{F}(x,t;\beta)$ in the proposed STPINN method is configured as an 8-layer network with 40 neurons per layer. The time mapping network $\mathcal{F}_t(t;\alpha)$ keeps the 2-layer architecture with 40 neurons per layer in the two-channel STPINN and 1-layer architecture with 80 neurons in the one-channel STPINN. Both the two methods employ the same optimization technique in the training phase.}
% The PINN method is configured in the same 10-layer network structure with 40 neurons per layer as the $\mathcal{F}(x,t;\beta)$ in the proposed STPINN method. The time mapping network $\mathcal{F}_t(t;\alpha)$ keeps the 2-layer architecture with 50 neurons per layer. Both the two methods employ the same optimization technique in the training phase. 
{\color{black}Table~\ref {tab:channel analysis} shows the accuracy and the runtime of calculating the stress evolution by PINN, one-channel STPINN and two-channel STPINN after 20k iterations of L-BFGS training. The temperature is configured as \eqref{eq:thermal} where $T_0(t)=[350+30\times \sin(8\times10^{-8}\pi t)]K$.}
{\color{black}The relative errors shown in Table~\ref {tab:channel analysis} imply that the proposed STPINN method can achieve higher accuracy than PINN for solving the space-time related diffusivity problem when the same number of neurons are employed. Fig.~\ref{fig:iteration} shows the change process of objective function based on PINN, 1-STPINN and 2-STPINN during the training phase. It can be observed that the loss of 2-STPINN is decreased to a lower value by employing the same number of neurons compared with 1-STPINN and PINN. Furthermore, 1-STPINN shows better generalization ability than PINN with similar loss.} 
It can be observed that, compared with the FEM, the mean relative error of the proposed STPINN decreases with an increased number of channels, whereas the time consumption increases with the broadening of the channel. 
For the proposed STPINN method, the number of channels should be determined by considering the accuracy and the time consumption.

\begin{table}[t]
	\caption{{\color{black}Performance analysis of stress evolution in the time range $10^5s\sim10^8s$ within segments and range $0s\sim10^8s$ at the interior junction by PINN, 1-STPINN, 2-STPINN. 
		%		FEM mesh size is $1\times 10^{-8}m$, time step size is $1\times 10^5s$. 
		%W.S. represents Whole-Stress at 10 %specified time points and J.S. represents %Junction-Stress.
		} }
	\centering
	\setlength{\tabcolsep}{4.6mm}{
		\begin{tabular}{|c|c|c|c|c|}
			\hline
			\multirow{2}{*}{\textbf{Method}} & \multicolumn{2}{c|}{\textbf{Error (\%)} } & \multicolumn{2}{c|}{\textbf{Time (s)}} \\ \cline{2-5} 
			& \textit{\textbf{W.S.}} & \textit{\textbf{J.S.}} & \textit{\textbf{W.S.}} & \textit{\textbf{J.S.}} \\ \hline
			PINN & 11.026 & 1.628 & 0.2117 & 0.0210 \\ \hline
			1-STPINN & 8.886 & 0.842 & 0.2808 & 0.0598 \\ \hline
			2-STPINN & 5.647 & 0.547 & 0.5127 & 0.0734 \\ 
			\hline
	\end{tabular}}
	\label{tab:channel analysis}
\end{table}

%In traditional numerical methods, we adjust the spatial grid density according to the interconnection scale. Specifically, higher grid density means less relative error and more computational resource consumption.
% The spatial meshing density is adjusted according to the interconnect structure scale. A more significant grid density means more computing resource consumption with more accurate solutions. 
%Fig.~\ref{fig:speedup} shows the improvement of runtime for the STPINN method compared to FEM under Case III with $1000\sim15000$ meshing points along the interconnect. It can be seen from Fig.~\ref{fig:speedup} that the speedup keeps rising when the number of meshing points is less than 9k. The STPINN achieves a steady speedup around $40\times$ when the number of meshing points is larger than 9k. 
%As a result, the mesh-free solver STPINN, not constrained by the number of meshing points in accuracy, shows better computing performance on more sophisticated stress analysis.
\begin{figure}[h]
	\centerline{\includegraphics[width=0.65\columnwidth]{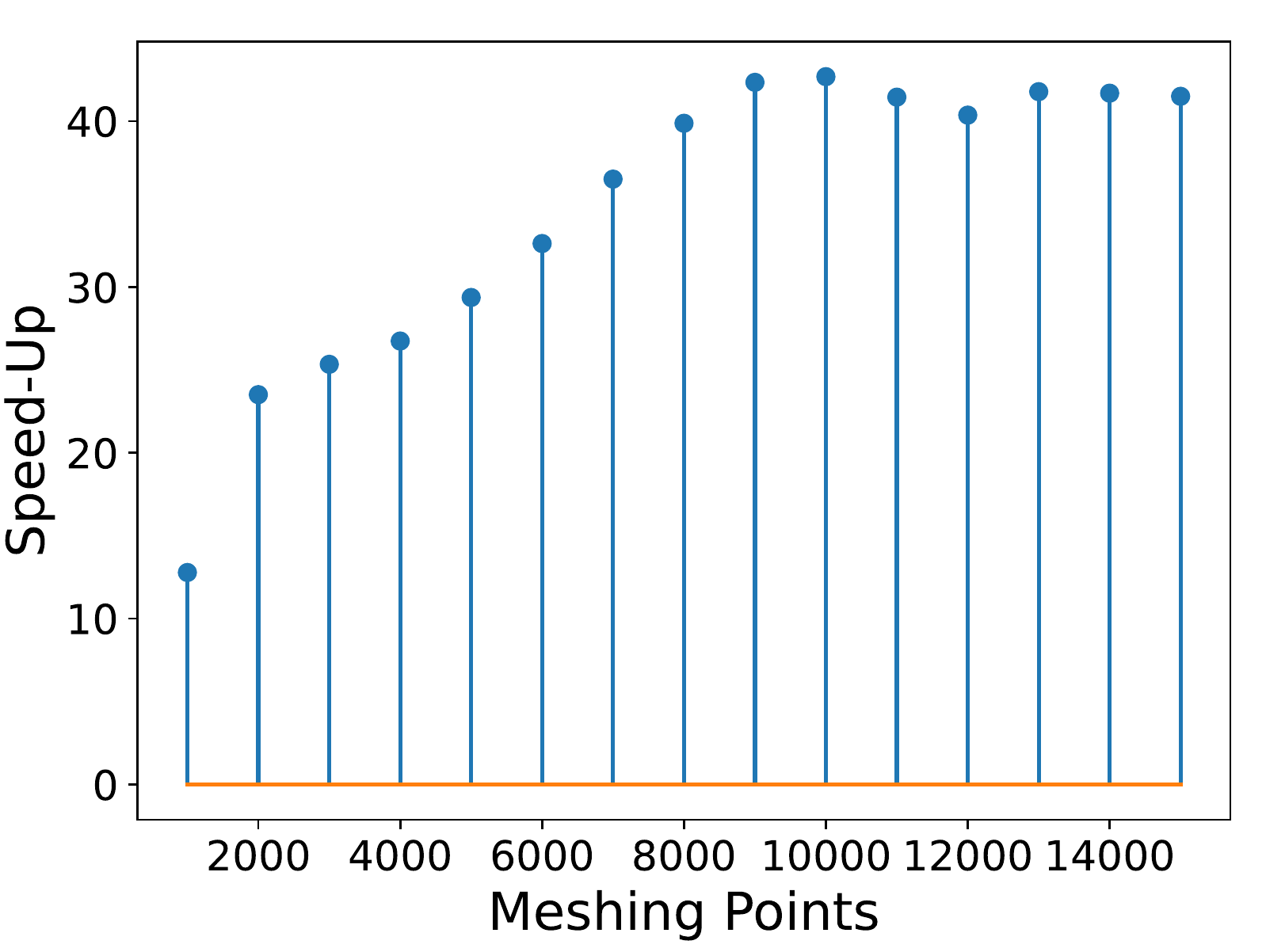}}
	\caption{Improvement of the runtime of STPINN compared to FEM in Case III.}
	\label{fig:speedup}
\end{figure} 
In the traditional numerical methods, we need to adjust the spatial grid density according to the interconnect scale. Specifically, higher grid density means less relative error and more computational resources. Fig.~\ref{fig:speedup} shows the improvement of computation time for the proposed STPINN method compared with the FEM under Case III with 1k$\sim$15k meshing points along the interconnect wire. It can be seen from Fig.~\ref{fig:speedup} that the speedup keeps rising when the number of meshing points is fewer than 9k. The STPINN method can achieve a speedup around $40\times$ when the number of meshing points is larger than 9k. As a result, the proposed mesh-free solver STPINN, not constrained by the number of meshing points for achieving higher accuracy, shows better computing performance on more sophisticated stress analysis of VLSI interconnect wires.

%We also compare the computation time between STPINN and FEM under similar accuracy. In order to obtain an accurate numerical solution by FEM, the narrower time step size is required to perform iterative operations, especially under complex temperature conditions. Stress evolutions in time range $0s\sim10^8s$ at junction in Case II obtained by 1-STPINN are compared with results of FEM after different iterations, as shown in Fig.~\ref{fig:step}. It can be seen that the predictions provided by STPINN are most similar to the numerical solutions obtained after 10k iteration steps. 
% More computational time is required in FEM. 
%The STPINN method shows the speedup of $9.30\times,\ 18.59\times,\ 52.68\times$ over FEM after 100, 1k, 10k iteration steps in solving the whole stress evolution. It should be noted that the STPINN method can achieve a $52.69\times$ speedup while keeping a similarly high accuracy over FEM.
\begin{figure}[t]
	\centerline{\includegraphics[width=0.7\columnwidth]{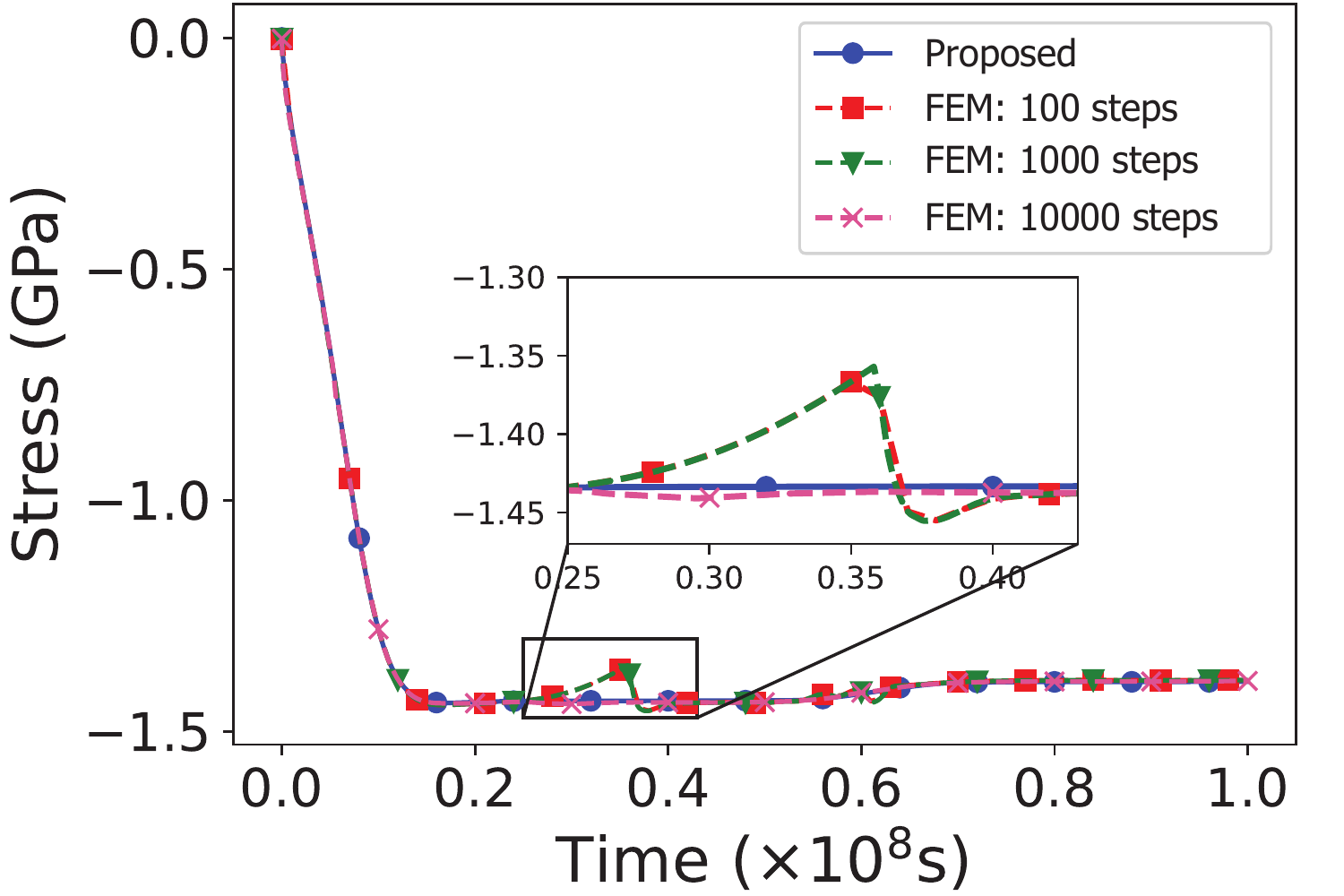}}
	\caption{Stress evolution comparison in Case II between results obtained by 1-STPINN and FEM after 100, 1k, 10k iteration steps. 
		%		FEM mesh size is $1\times 10^{-8}m$. 
		Total runtimes are 3s, 6s, 17s for 100, 1k, 10k iteration steps, respectively.}
	\label{fig:step}
\end{figure}

We also compare the computation time between the proposed STPINN and FEM with almost the same accuracy. 
\textcolor{black}{The smaller time step size is required for obtaining a more accurate numerical solution by the FEM, especially under complex thermal conditions. }
The stress values in the time range $0s\sim10^8s$ at the junction in Case II obtained by 1-STPINN are compared with the ones obtained by the FEM with different iterations, as shown in Fig.~\ref{fig:step}. It can be seen that the results obtained by the proposed STPINN method can agree well with the numerical solutions obtained by the FEM with 10k iteration steps. Moreover, the proposed STPINN method shows the speedup of $9.30\times,\ 18.59\times, \ 52.68\times$ over the FEM with 100, 1k, 10k iteration steps for calculating the whole stress values, respectively. It should be also noted that the proposed STPINN method can achieve a $52.69\times$ speedup while keeping a similarly high accuracy over the FEM.

% \begin{figure}[htb]
% 	\centerline{\includegraphics[width=0.8\columnwidth]{cd.pdf}}
% 	\caption{Current density configuration of a multi-segment interconnect tree.}
% 	\label{fig:cd}
% \end{figure} 

\subsection{Multi-segment Analysis}\label{old multi}
% \begin{figure}[htb]
% 	\centerline{\includegraphics[width=0.9\columnwidth]{3.pdf}}
% 	%	\caption{Stress evolution in 7-segment tree in ibmpg3 under dynamic temperature at $t=1\times 10^8 s$.}
% 	\caption{EM based stress evolution for a multi-segment interconnect tree under dynamic temperature.}
% 	\label{fig:7segment}
% \end{figure} 
\begin{figure}[t]
	\centering 
	\subfigure[]{
		\includegraphics[width=0.45\columnwidth]{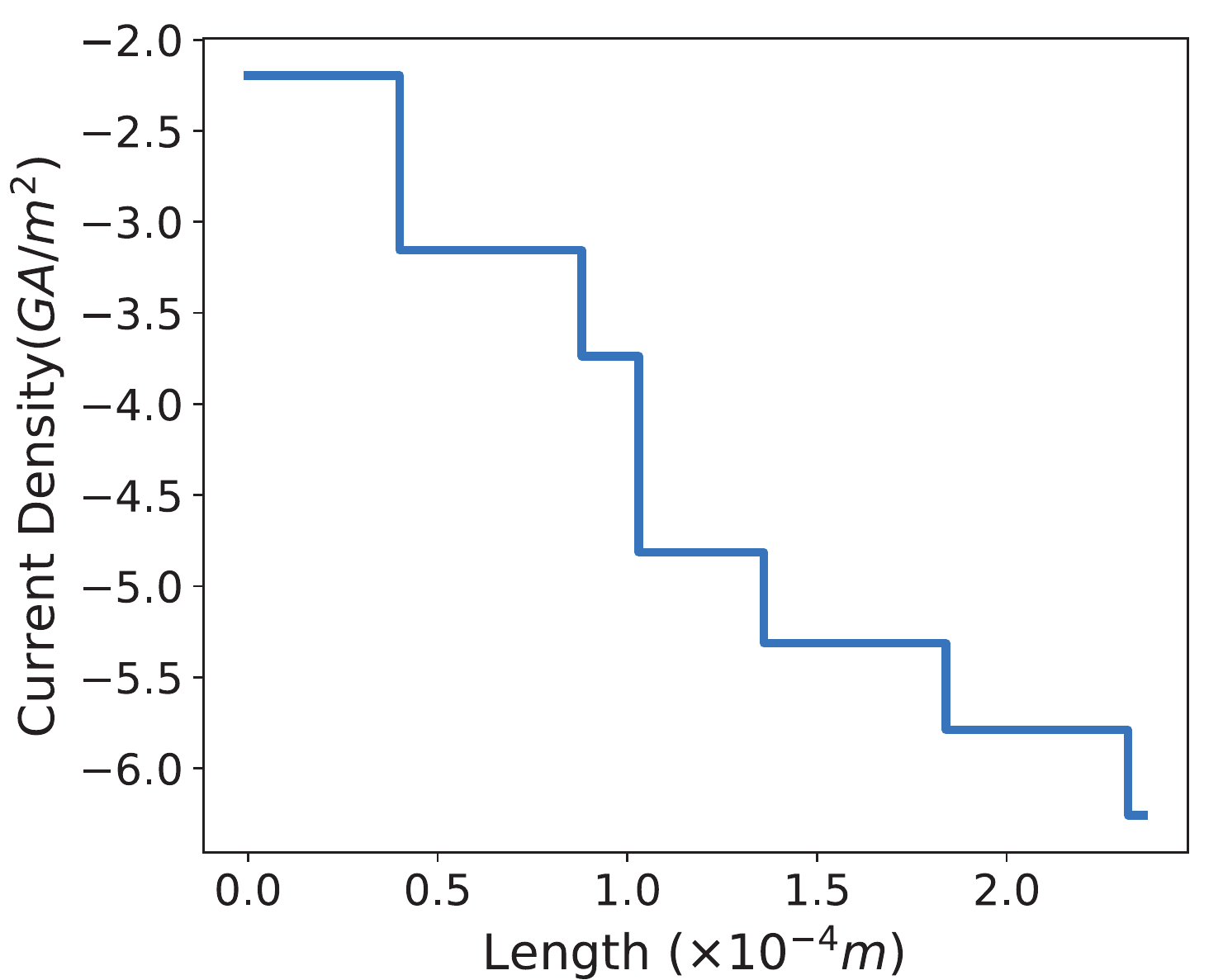}
		\label{fig:cd}} 
	\subfigure[]{
		\includegraphics[width=0.47\columnwidth]{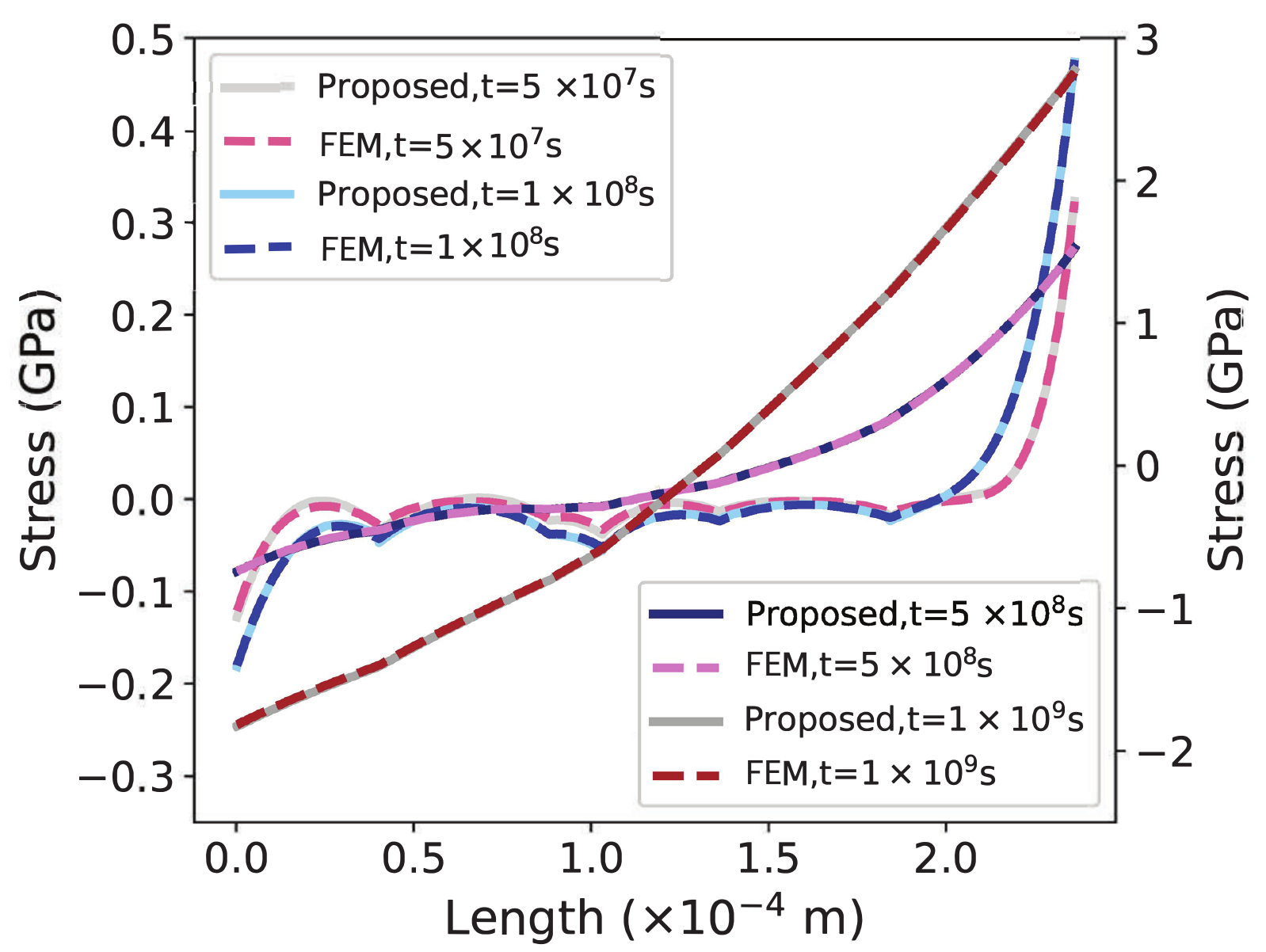}
		\label{fig:7segment}}
	\caption{{\color{black}(a) Current density configuration of a 7-segment interconnect tree, (b) EM based stress evolution for a 7-segment interconnect tree under dynamic temperature by using STPINN. The proposed STPINN takes the location $x$ and the time $t$ as the inputs.
%	with 850 neurons
	}}
\end{figure}
We now extend the proposed STPINN method to the stress evolution calculation of interconnect wires with multi-segments. \textcolor{black}{For multi-segment analysis, the original self-diffusion coefficient $D_0$ is kept the same in each segment, which is used to calculate the diffusion coefficient $D_a$ combining with space-time related temperature.} \textcolor{black}{For the comparisons of FEM and STPINN, we choose a straight 7-segment interconnect wire and calculate the stress evolution under the dynamic temperature condition (Case II) by the proposed 1-STPINN.} We employ a 1-hidden layer network with 100 neurons per layer for $\mathcal{F}_t(t;\alpha)$ and a 15-hidden layer network with 50 neurons per layer in $\mathcal{F}(x,t;\beta)$ for calculating the stress values. {\color{black}Fig.~\ref{fig:cd} shows the current density distribution of this 7-segment interconnect tree.} Fig.~\ref{fig:7segment} shows the comparison results of the stress evolution in the time range from $5\times 10^7s$ to $1\times 10^9s$ between the proposed method and the FEM. It can be demonstrated that the proposed STPINN method can accurately capture the stress evolution and the discontinuity through vias on the multi-segment tree at different time instances. 

%{\color{blue}It is supposed that the stress diffusivity is the same in each segment under the same temperature.} 

% \begin{figure}[htb]
% 	\centerline{\includegraphics[width=0.9\columnwidth]{cross.pdf}}
% 	%	\caption{Stress evolution in 7-segment tree in ibmpg3 under dynamic temperature at $t=1\times 10^8 s$.}
% 	\caption{{\color{black}EM based stress evolution for a cross-shaped 5-terminal interconnect tree under Case III.}}
% 	\label{fig:cross}
% \end{figure} 

% {\color{black}Then we choose a cross-shaped 5-terminal interconnect for the simulation of the proposed 2-STPINN. We employ a 2-hidden layer network with 50 neurons per layer for $\mathcal{F}_t(t;\alpha)$ and a 10-hidden layer network with 40 neurons per layer in $\mathcal{F}(x,t;\beta)$ for calculating the stress values under Case III. The length and current density of each segment is configured as $L_1=20\ \mu m,\ L_2=30\ \mu m,\ L_3=10\ \mu m,\ L_4=20\ \mu m,\ j_1=1\times 10^{10}A/m^2,\ j_2=2\times 10^{10}A/m^2,\ j_3=-3\times 10^{10}A/m^2,\ j_4=4\times 10^{10}A/m^2$. The experiment result is shown in Fig.~\ref{fig:cross}, which demonstrates good agreement of the results between the proposed STPINN and FEM.}

Experimental results show that deepening and widening of the network architecture in $\mathcal{F}_t(t;\alpha)$ has little impact on the accuracy of stress evolution, which means that excessive calculation of time transformation is not necessary for solving EM-base stress evolution equation under the complex thermal condition. It can be also observed that for more complex topological structures governed by coupled stress evolution equations, a larger network structure for $\mathcal{F}(x,t;\beta)$ is required to approximate the stress evolution process.
{\color{black}
\subsection{Additional Input to the Proposed STPINN}
%Parameterized STPINN for Multi-segment %Analysis
\label{sec:multi}

\begin{figure}[t]
	\centering 
	\subfigure[]{
		\includegraphics[width=0.45\columnwidth]{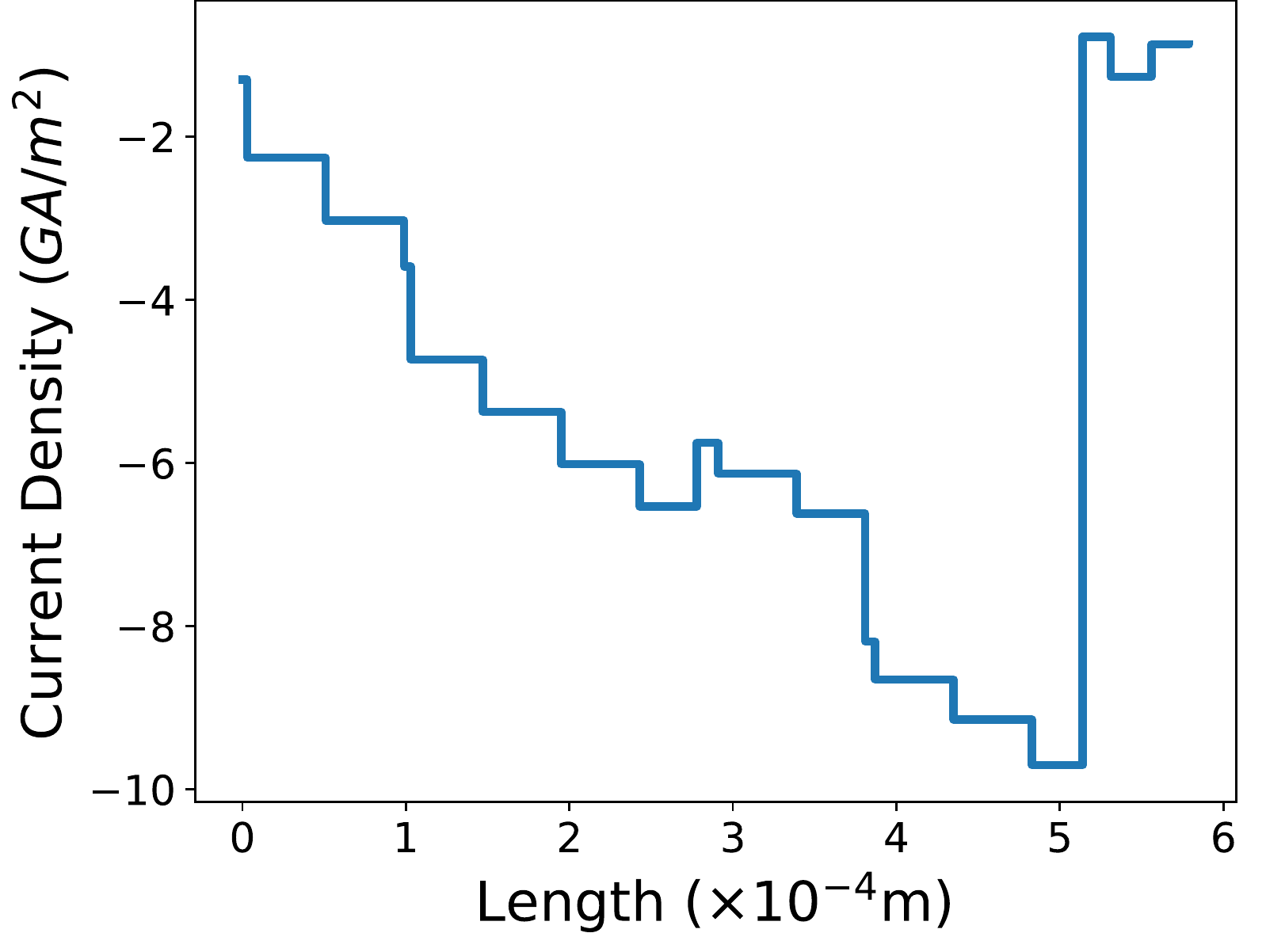}
		\label{fig:19cd}} 
	\subfigure[]{
		\includegraphics[width=0.46\columnwidth]{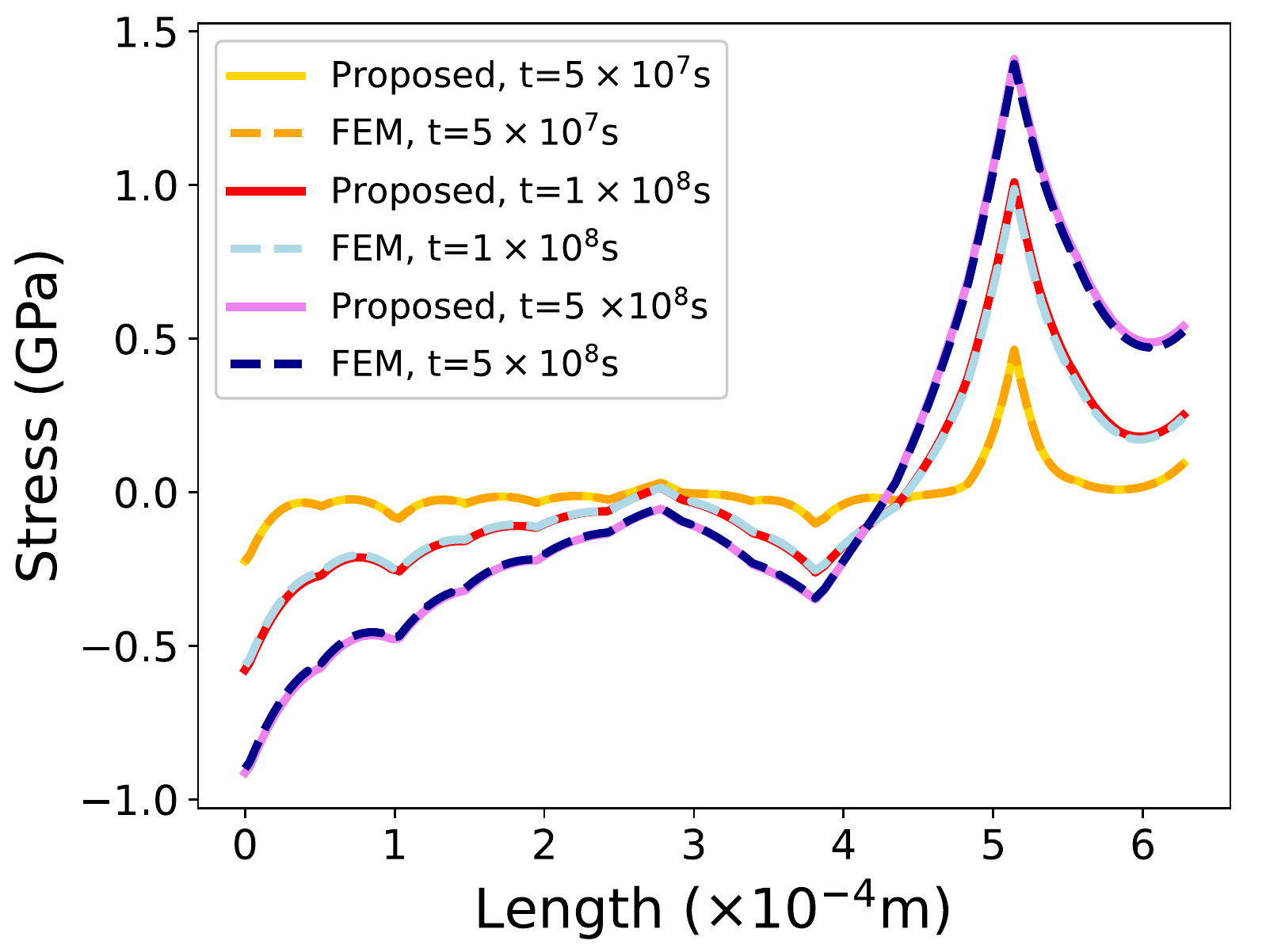}
		\label{fig:19seg}}
	\caption{{\color{black}(a) Current density configuration of a 19-segment interconnect tree, (b) EM based stress evolution for the 19-segment interconnect tree under Case II by using STPINN. 
	%with 420 neurons. 
	The proposed STPINN takes the location $x$, the time $t$ and the EM driving force as the inputs.}}
	\label{fig:19}
\end{figure}

We employ the EM driving force at the sample point as an additional input to STPINN. The solution of STPINN with an additional input can be expressed as $\mathcal{F}_c(\mathcal{F}(x,G,\mathcal{F}_t(t;\alpha);\beta);\theta)$.
% The new parameter mitigates the demand for the number of neurons in more complex interconnect structures.
In the simulation of 1-STPINN, we choose a straight 19-segment interconnect tree and calculate the stress evolution under Case II. We set $\mathcal{F}_t(t;\alpha)$ as the 1-hidden layer network with 100 neurons per layer and $\mathcal{F}(x,t;\beta)$ as the 8-hidden layer network with 40 neurons per layer. Fig.~\ref{fig:19cd} shows the current density distribution of the interconnect tree and Fig.~\ref{fig:19seg} shows the comparison results of the stress evolution in the time range from $5\times 10^7s$ to $5\times 10^8s$ between 1-STPINN and FEM.} 

\begin{figure}[t]
	\centerline{\includegraphics[width=0.75\columnwidth]{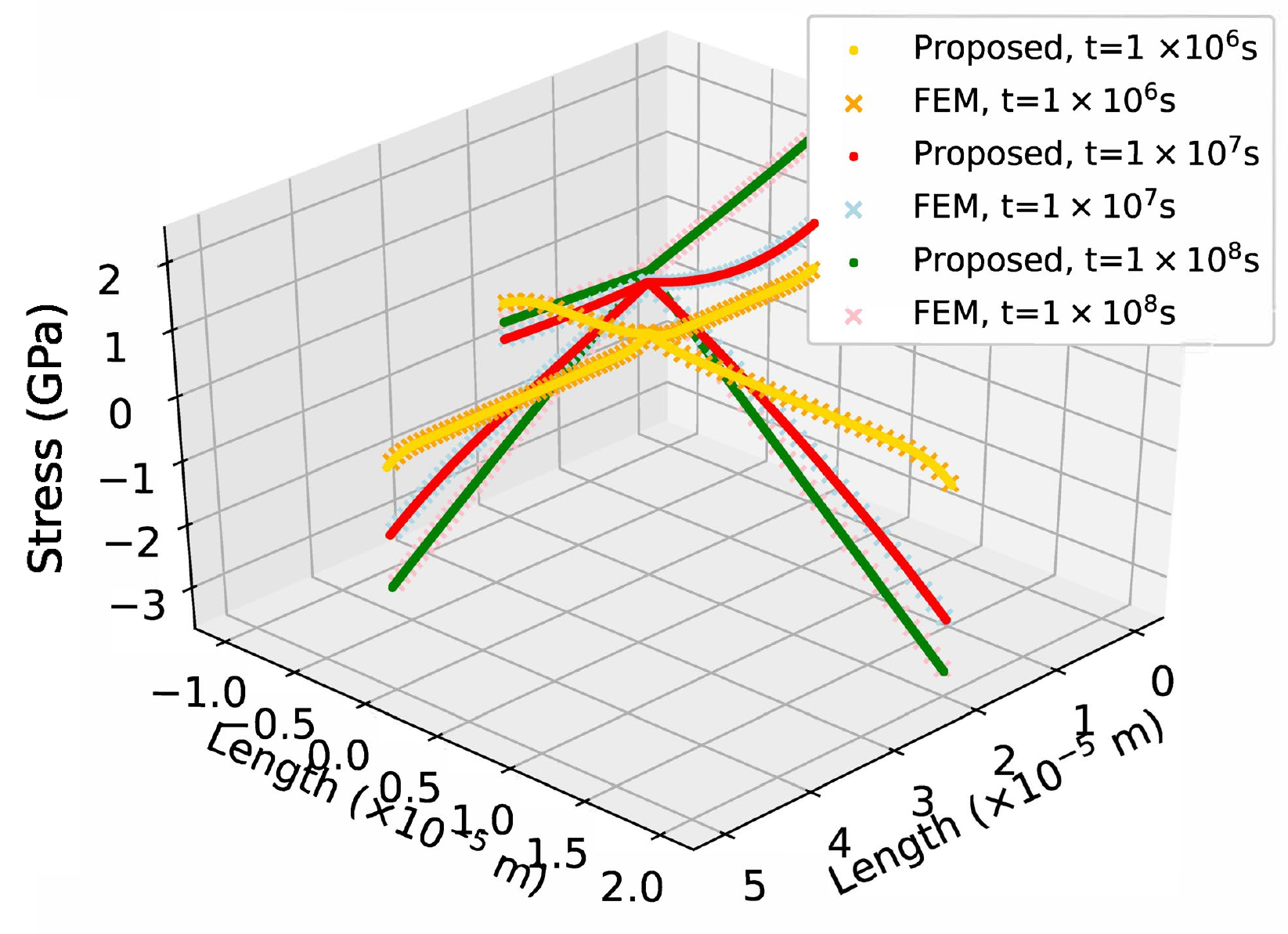}}
	\caption{{\color{black}EM based stress evolution for a cross-shaped 5-terminal interconnect tree under Case III by using STPINN.
%	with 300 neurons. 
	The proposed STPINN takes the location $x$, the time $t$ and the EM driving force as the inputs.}}
	\label{fig:cross}
\end{figure}

{\color{black}For obtaining the stress evolution of a cross-shaped 5-terminal interconnect, we employ a 2-hidden layer network with 50 neurons per layer in $\mathcal{F}_t(t;\alpha)$ and a 5-hidden layer network with 40 neurons per layer in $\mathcal{F}(x,t;\beta)$ for calculating the stress evolution under Case III. The location input of the proposed 2-STPINN is a two-dimensional coordinate for the cross-shaped interconnect tree. The length and the current density of each segment are configured as $L_1=20\ \mu m,\ L_2=30\ \mu m,\ L_3=10\ \mu m,\ L_4=20\ \mu m,\ j_1=1\times 10^{10}\ A/m^2,\ j_2=2\times 10^{10}\ A/m^2,\ j_3=-3\times 10^{10}\ A/m^2,\ j_4=4\times 10^{10}\ A/m^2$. The experimental results are shown in Fig.~\ref{fig:cross}, which demonstrates good agreement between 2-STPINN and FEM.} {\color{black}Compared with the number of neurons used in the 7-segment wire of Section~\ref{old multi}, the additional input of STPINN decreases the demand for neurons in more complex interconnect structures. It demonstrates the importance of segment features in STPINN based interconnect stress evolution analysis.}

% Experimental results show that deepening and widening of the network architecture in $\mathcal{F}_t(t;\alpha)$ has little impact on improving accuracy performance for obtaining stress evolution, which means that excessive computational allocation for time transformation is not necessary for solving EM-base stress evolution equation with complex thermal condition. 

% {\color{black}When the interconnect structure becomes more complicated, it is difficult for the neural network to provide the global approximation.There will be plenty of collocation points within the interconnect tree for network training, and the physics-informed constraints on such a large number of collocation points will make neural networks hard to fit. In our future work, we will build a trial function respecting the diffusion process on internal interconnect and employ the neural network as the nonlinear component to approximate the solutions respecting boundary conditions. In this way, the neural network only needs to provide an approximation for the boundaries instead of the global interconnect.}

% It can be also observed that for more complex topological structures governed by coupled stress evolution equations, a larger network structure for $\mathcal{F}(x,t;\beta)$ is required to approximate the stress evolution process.

\section{Conclusion}\label{5}
In this paper, we propose a novel composite neural network to compute stress evolution along multi-segment interconnect trees during the void nucleation phase considering complex thermal conditions. {\color{black}The proposed STPINN method aims at obtaining mesh-free solutions of PDEs with space-time related diffusivity.} We first construct the interconnect thermal model for Joule heat spreading incorporating via effect and dynamic temperature. We then solve stress evolution equations under different temperature conditions by the STPINN method. To enhance the learning ability of neural network-based solver, space-time conversion and multi channels are used in the proposed STPINN method for obtaining more accurate solution. {\color{black}The interior junction node constraints and the large gradients in the EM model are settled by the virtual distance and preprocessing techniques.} Finally, we compare the stress evolution in configured interconnects under three different temperature configurations, which achieves mean relative errors $<1.22\%$ vs FEM and $<0.44\%$ vs the analytical model while delivering a $2\times \sim 52\times$ speedup over the competing schemes. {\color{black}Of particular interest, the neural network based large-scale EM stress approximation and the EM stochasticity assessment need further study. To obtain the stress evolution on large-scale interconnect trees through neural network based method, the key problem is how to efficiently reduce the total number of collocation points used for the physics-informed constrains.
For EM stochasticity assessment which supposes that the self-diffusion coefficient in each segment is different, a more generalized learning based model should be developed.}

\ifCLASSOPTIONcaptionsoff
  \newpage
\fi



% \input{reliability.bbl}
\bibliographystyle{IEEEtran}
\bibliography{reliability}

\newpage
\vspace{-10 mm}

\begin{IEEEbiography}
[{\includegraphics[width=1in,height=1.25in,clip,keepaspectratio]{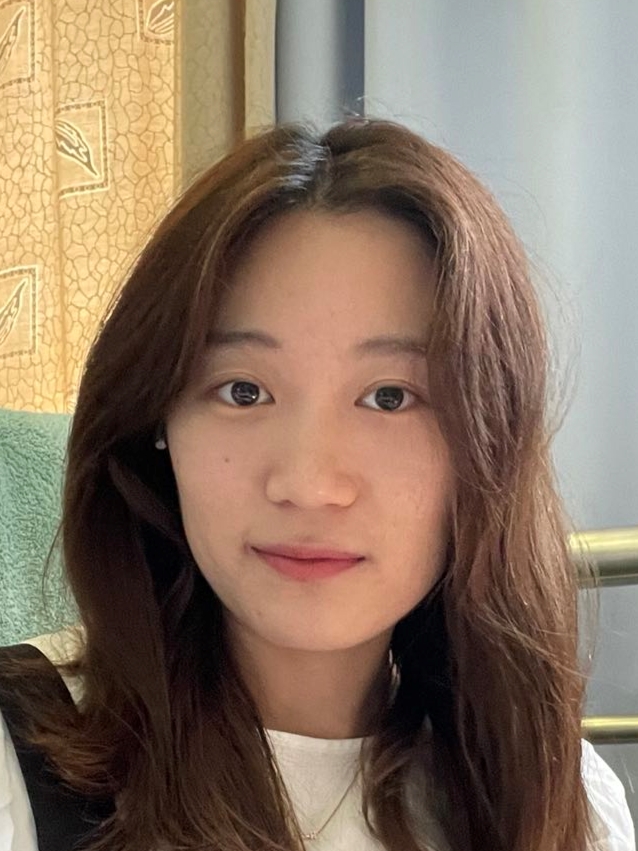}}]
{Tianshu Hou} received the B.S. degree in electronic information science and technology from Sichuan University, Sichuan, China in 2019. She is currently pursuing a Ph.D degree in the Department of Micro/Nano-electronics, Shanghai Jiao Tong University, Shanghai, China. Her current research interests include electromigration reliability modeling, assessment and optimization.
\end{IEEEbiography}

\vspace{-15 mm}
\begin{IEEEbiography}
[{\includegraphics[width=1in,height=1.25in,clip,keepaspectratio]{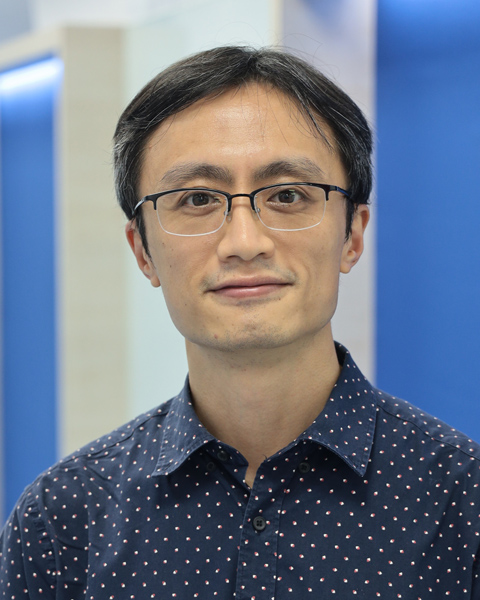}}]{Ngai Wong}  (SM, IEEE) received his B.Eng and Ph.D. in EEE from The University of Hong Kong (HKU), and he was a visiting scholar with Purdue University, West Lafayette, IN, in 2003. He is currently an Associate Professor with the Department of Electrical and Electronic Engineering at HKU. His research interests include electronic design automation (EDA), model order reduction, tensor algebra, linear and nonlinear modeling \& simulation, and compact neural network design. 
\end{IEEEbiography}

\vspace{-15 mm}

\begin{IEEEbiography}
[{\includegraphics[width=1in,height=1.25in,clip,keepaspectratio]{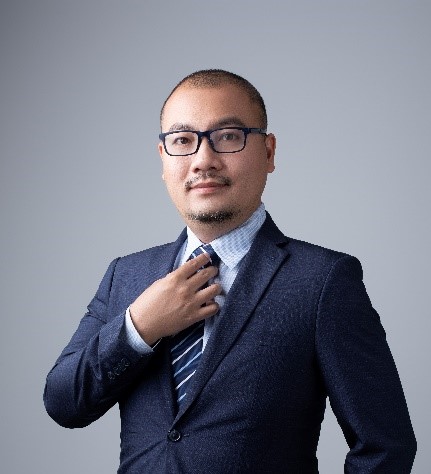}}]{Quan Chen} (S'09-M'11) received his B.S. degree in Electrical Engineering from the Sun Yat-Sen University, China, in 2005 and the M.Phil. and Ph.D. degree in Electronic Engineering from The University of Hong Kong, Hong Kong, in 2007 and 2010. From 2010-2011 he was postdoctoral fellow at the department of Computer Science and Engineering of the University of California, San Diego (UCSD). In 2012-2018, he was a research assistant professor at the department of Electrical and Electronic Engineering, The University of Hong Kong (HKU). He joined the Southern University of Science and Technology (SUSTech) in Shenzhen, China in 2019, where he is an assistant professor now. 
His research interests include ultra-large-scale circuit simulation and multi-physics analysis in the field of electronic design automation (EDA), as well as EDA techniques for emerging technologies such as sub-10nm devices, memristors, and quantum computing. He also has years of experience in technical transformation and commercialization.

\end{IEEEbiography}

\vspace{-15 mm}

\begin{IEEEbiography}
  [{\includegraphics[width=1in,height=1.25in,clip,keepaspectratio]{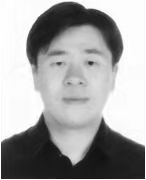}}]
  {Zhigang Ji} received his Ph.D. degree from Liverpool John Moores University (LJMU), Liverpool, U.K., in 2010. He currently holds the position as the professor in Nanoelectronics in Shanghai Jiaotong University (SJTU). His current research interests focus on characterization, modeling and design of reliable, low-power, and high-performance computation systems, motivated by both evolutionary and revolutionary advances in nanoelectronics.
\end{IEEEbiography}

\vspace{-15 mm}

\begin{IEEEbiography}
[{\includegraphics[width=1in,height=1.25in,clip,keepaspectratio]{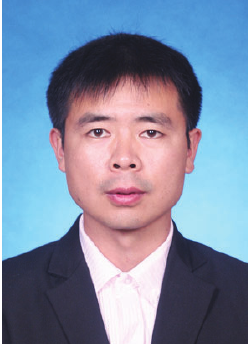}}]
{Hai-Bao Chen} received the B.S. degree in information and computing sciences, and the M.S. and Ph.D. degrees in applied mathematics from Xian Jiaotong University, Xian, China, in 2006, 2008, and 2012, respectively. He then joined Huawei Technologies, where he focused on cloud computing and big data. He was a Post-Doctoral Research Fellow with Electrical Engineering Department, University of California, Riverside, CA, USA, from 2013 to 2014. He is currently an Associate Professor in the Department of Micro/Nano-electronics, Shanghai Jiao Tong University, Shanghai, China. His current research interests include VLSI reliability, machine learning and neuromorphic computing, numerical analysis and modeling for VLSIs, integrated circuit for signal and control systems. Dr. Chen has authored or co-authored about 70 papers in scientific journals and conference proceedings. He received one Best Paper Award nomination from Asia and South Pacific Design Automation Conference (ASP-DAC) in 2015. Since 2016, Dr. Chen serves as an Associate Editor for Integration-the VLSI Journal.
\end{IEEEbiography}

\end{document}